\documentclass[prd,amsmath,amssymb,floatfix,nofootinbib,superscriptaddress]{revtex4}

\usepackage{graphicx}
\usepackage{bm}
\usepackage{amsmath,amssymb}
\usepackage{epsfig}
\usepackage{color}
\usepackage{natbib}
\usepackage{enumerate}
\usepackage[hypertex]{hyperref}
\usepackage{ifthen}

\newcommand{\quickedit}[1]{#1}
\newcommand{\lens}{{\rm{lens}}}
\newcommand{\aml}[1]{ \emph{(AML: #1)}}
\newcommand{\dmh}[1]{} 

\newcommand{\fixme}[1]{} 

\renewcommand{\vec}{\text{vec}}
\newcommand{\tB}{\underline{\boldvec{B}}}
\newcommand{\tA}{\underline{\boldvec{A}}}
\newcommand{\tD}{\underline{\boldvec{D}}}
\newcommand{\mCov}{\text{\bf{Cov}}}

\newcommand{\tC}{\tilde{C}}
\newcommand{\tCtot}{\tC_{\text{tot}}}
\newcommand{\tTemp}{\tilde{T}}
\newcommand{\Cgtwo}{C_{\rm{gl},2}}
\newcommand{\var}{{\rm var}}

\def\eprinttmp@#1arXiv:#2 [#3]#4@{
\ifthenelse{\equal{#3}{x}}{\href{http://arxiv.org/abs/#1}{#1}}{\href{http://arxiv.org/abs/#2}{arXiv:#2} [#3]}}

\providecommand{\eprint}[1]{\eprinttmp@#1arXiv: [x]@}
\newcommand{\adsurl}[1]{\href{#1}{ADS}}
\providecommand{\bibinfo}[2]{\ifthenelse{\equal{#1}{isbn}}{
\href{http://cosmologist.info/ISBN/#2}{#2}}{#2}}

\newcommand{\cov}{{\rm{Cov}}}
\newcommand{\Xstwo}{{{}_{s_2}X}}
\newcommand{\XstwoL}{{{}_{s_2}\tilde{X}}}

\newcommand{\vecp}{\text{vecp}}

\newcommand{\mBp}{\mB_{nks}}

\newcommand{\ra}{\rangle}
\newcommand{\la}{\langle}
\renewcommand{\d}{\text{d}}
\newcommand{\grad}{\nabla}

\newcommand{\begm}{\begin{pmatrix}}
\newcommand{\enm}{\end{pmatrix}}
\newcommand{\threej}[6]{{\begm #1 & #2 & #3 \\ #4 & #5 & #6 \enm}}

\newcommand{\edth}{\,\eth\,}
\renewcommand{\beth}{\,\overline{\eth}\,}

\newcommand{\ud}{{\rm d}}
\providecommand{\clo}{{\mathcal{O}}}
\providecommand{\cla}{{\mathcal{A}}}
\providecommand{\clf}{{\mathcal{F}}}

\newcommand{\boldvec}[1]{\mathbf{#1}}
\newcommand{\mgamma}{\bold{\gamma}}

\newcommand{\mA}{\bm{A}}
\newcommand{\mB}{\bm{B}}
\newcommand{\mC}{\bm{C}}
\newcommand{\mD}{\bm{D}}

\newcommand{\mF}{\bm{F}}
\newcommand{\mG}{\bm{G}}
\newcommand{\mH}{\bm{H}}
\newcommand{\mI}{\bm{I}}

\newcommand{\mM}{\bm{M}}

\newcommand{\vL}{\boldvec{L}}

\newcommand{\vW}{\boldvec{W}}

\newcommand{\va}{\boldvec{a}}

\newcommand{\ve}{\boldvec{e}}

\newcommand{\vk}{\boldvec{k}}
\newcommand{\vl}{\boldvec{l}}

\newcommand{\vn}{\boldvec{n}}

\renewcommand{\vr}{\boldvec{r}}

\newcommand{\vx}{\boldvec{x}}

\newcommand{\valpha}{{\boldsymbol{\alpha}}}
\newcommand{\vgrad}{{\boldsymbol{\nabla}}}

\newcommand{\vnhat}{{\hat{\vn}}}

\newcommand{\CMBFAST}{\textsc{cmbfast}}
\newcommand{\CAMB}{\textsc{camb}}

\newcommand{\chire}{\chi_{\mathrm{re}}}
\newcommand{\etare}{\eta_{\mathrm{re}}}

\newcommand{\muK}{\mu\rm{K}}

\newcommand{\lmax}{l_{\text{max}}}

\newcommand{\fnl}{f_{\rm{NL}}}

\newcommand\ba{\begin{eqnarray}}
\newcommand\ea{\end{eqnarray}}
\newcommand\be{\begin{equation}}
\newcommand\ee{\end{equation}}
\newcommand{\hide}[1]{}

\newcommand{\temp}{T}
\newcommand{\Ltemp}{\tilde{T}}

\newcommand{\Cgrads}{\tilde{C}^{\temp\grad\temp}}
\begin{document}


\title{The shape of the CMB lensing bispectrum}

\author{Antony Lewis}
\address{Department of Physics \& Astronomy, University of Sussex, Brighton BN1 9QH, UK}

\author{Anthony Challinor}
\address{Institute of Astronomy and Kavli Institute for Cosmology, Madingley Road, Cambridge, CB3 0HA, UK}
\address{DAMTP, Centre for Mathematical Sciences, University of Cambridge, Wilberforce Road, Cambridge CB3 OWA, UK}

\author{Duncan Hanson}
\address{Institute of Astronomy and Kavli Institute for Cosmology, Madingley Road, Cambridge, CB3 0HA, UK}
\address{Jet Propulsion Laboratory, California Institute of Technology, 4800 Oak Grove Drive, Pasadena CA 91109, USA}

\begin{abstract}
Lensing of the CMB generates a significant bispectrum, which should be detected by the
Planck satellite at the 5-sigma level and is potentially a non-negligible source of bias for $\fnl$ estimators of local
non-Gaussianity.
We extend current understanding of the lensing bispectrum in several directions:
(1) we perform a non-perturbative calculation of the lensing bispectrum which is $\sim~10\%$ more accurate than previous, first-order calculations;
(2) we demonstrate how to incorporate the signal variance of the lensing bispectrum into estimates of its amplitude, providing a good analytical explanation for previous Monte-Carlo results; and
(3) we discover the existence of a significant lensing bispectrum in polarization, due to a previously-unnoticed correlation between the lensing potential and $E$-polarization as large as $30\%$ at low multipoles.
We use this improved understanding of the lensing bispectra to re-evaluate Fisher-matrix predictions, both for Planck and cosmic variance limited data.
We confirm that the non-negligible lensing-induced bias for estimation of local non-Gaussianity should be robustly treatable,
and will only inflate $\fnl$ error bars by a few percent over predictions where lensing effects are completely ignored (but note that lensing must still be accounted for to obtain unbiased constraints).
We also show that the detection significance for the lensing bispectrum itself is ultimately limited to 9 sigma by cosmic variance.
The tools that we develop for non-perturbative calculation of the lensing bispectrum are directly relevant to other calculations,
and we give an explicit construction of a simple non-perturbative quadratic estimator for the lensing potential and relate its cross-correlation power spectrum to the bispectrum.
Our numerical codes are publicly available as part of CAMB and LensPix.
\end{abstract}

\date{\today}

\maketitle

\pagenumbering{arabic}

\section{Introduction}

The large-scale CMB temperature anisotropy has a contribution from the blue- and red-shifting of photons as they fall in and out of potential wells between the last-scattering surface and our observation. This integrated-Sachs-Wolfe (ISW) effect is not present during matter domination, but becomes important at redshift $z\alt 2$ at which dark energy starts to affect the evolution of the matter perturbations. The CMB is also gravitationally lensed by structures along the line of sight, with most of the effect also coming from $z\alt 2$, so there is a correlation between the ISW signal and the CMB lenses. The effect of an overdensity is to magnify the last-scattering surface, effectively locally shifting the scale of the acoustic peaks. The variance over some range of scales is therefore changed by the magnification if the spectrum is not flat. This leads to a correlation between the small-scale CMB power and the large-scale lenses, and hence a correlation between the large-scale CMB temperature and the small-scale power. This corresponds to a `squeezed' bispectrum shape --- it is the correlation of one large scale with two much smaller scales. The lensing bispectrum falls off rapidly as the largest scale decreases, since the ISW contribution to the temperature falls rapidly on smaller scales.
The existence of a significant temperature bispectrum is well known, and must be modeled when trying to detect small levels of local primordial non-Gaussianity~\cite{Smith:2006ud,Serra:2008wc,Hanson:2009kg,Mangilli:2009dr}. It can also be used as a probe of the perturbation growth and expansion history of the universe at low redshift, and hence help to constrain the dark energy and curvature~\cite{Seljak:1998nu,Goldberg:1999xm,Hu:2001fb,Giovi:2003ri,Giovi:2004te,Gold:2004ee}.

Calculations of the temperature lensing bispectrum have until now been calculated at lowest order in the lensing effects, although some simulation work has also been done to verify that the effect of
higher-order terms is small \cite{Hanson:2009kg}. In this work, we demonstrate how to extend these calculations non-perturbatively to higher-order by working in an `unlensed short-leg' approximation, where we take one large-scale mode of the CMB temperature to be unlensed. This produces higher-order corrections to the usual lensing result which may be accurately reproduced simply by replacing the unlensed power spectra which appear in the lowest-order calculation of the lensing bispectrum with their lensed counterparts. This results in $\clo(10\%)$ corrections to the lensing bispectrum which we verify using Monte-Carlo simulations.

The lensing bispectrum should be detected soon at high-significance (e.g. $\sim 5\sigma$ in the data of the recently-launched Planck satellite \cite{Hanson:2009kg}). In this regime, the cosmic variance of the lensing signal can have large effects on the expected error of the bispectrum amplitude. Calculation of the increase in error at first appears daunting as it involves a six-point function in the non-Gaussian, lensed CMB, however we will show how a heuristic interpretation of the lensing bispectrum estimator as a cross-correlation between the observed CMB temperature and a quadratic reconstruction of the lensing effects can be used to intuit an accurate approximation to the signal variance. This method also generalizes straightforwardly to a calculation of the increase in variance for other estimators of non-Gaussianity, where the bias due to lensing represents an additional effective source of noise. This increase has already been investigated numerically by Ref.~\cite{Hanson:2009kg} under the assumption that the amplitude of the lensing bispectrum is well constrained and may simply be subtracted from the data. We are able to reproduce this result analytically, as well as extend it to the case where the amplitude of the lensing bispectrum is treated as a free parameter and marginalized over directly from the data. Our discussion also leads to improvements to standard bispectrum estimators, which incorporate the signal variance appropriate to the lensing bispectrum.

Discussion of the CMB lensing bispectrum in the literature has focused on the temperature anisotropies since there is no direct analogue of the ISW effect in
polarization.
However, as we will show here, the large-scale $E$ polarization from reionization is also directly correlated with the $z\alt 3$ matter distribution, giving a correlation between the $E$-polarization and lensing potential at up to the $30\%$ level. This generates a significant polarized lensing bispectrum, detectable at $\sim 2.5\sigma$ with cosmic-variance limited data. We present the first calculations of these effects, and generalize our analytical non-perturbative bispectrum and variance calculations to the polarization case. Including this effect in a fit for the amplitude of the lensing bispectrum would increase the significance with which it is detected from $3.8\sigma$ to $4.5\sigma$ for Planck, or from $5.3\sigma$ to $8.3\sigma$ for an experiment which is cosmic-variance limited to $\lmax=2000$.

The outline for this paper closely follows the description above. In Section~\ref{potentials} we review the quantitative description of lensing effects as a remapping by the gradient of a lensing potential $\psi$,
and derive the cross-correlation between the lensing potential and the CMB temperature and polarization. In Section~\ref{flat-sky} we then present calculations of the lensing bispectrum on the flat-sky, both at first order in the lensing potential as well as in the short-leg approximation which is effectively accurate at higher order as well. Use of the flat-sky expressions makes it straightforward to gain an intuition for the terms involved. In Section~\ref{full-sky} we proceed to give full-sky results for both temperature and polarization, which generalize straightforwardly from the flat-sky limit. In Section~\ref{estimators-variance-bias} we discuss the variance of estimators for the lensing bispectrum, and the increased variance for other non-Gaussian bispectra which occurs when marginalizing or subtracting the lensing contribution to avoid biases. Our conclusions are summarized in Section~\ref{conclusions}, and the details of several more involved calculations are contained in
appendices. Throughout we assume a standard $\Lambda$CDM cosmology, and for numerical examples use a  constant spectral index spatially-flat model with
 $\Omega_b h^2=0.0226$, $\Omega_c h^2=0.112$, $h=0.7$, $A_s=2.1\times 10^{-9}$, $n_s=0.96$, $\tau=0.09$, and approximate the three neutrinos as massless.

\section{The lensing potential and its cross-correlation with temperature
anisotropy and polarization}
\label{potentials}

The effect of gravitational lensing is to alter the direction of propagation
of photons such that when we look in direction $\vnhat$ we are actually seeing photons that originate from $\vnhat+\valpha$ on the last-scattering surface, where $\valpha$ is a deflection angle.
Using the Born approximation, the deflection angle of a source at conformal distance $\chi_*$ is given in terms of the Weyl potential $\Psi$ (i.e.\ the average of the Newtonian-gauge potentials) by the line-of-sight integral
\begin{equation}
\valpha= -2 \int_0^{\chi_*} \d\chi \frac{f_K(\chi_*-\chi)}{f_K(\chi_*)f_K(\chi)}\grad_{\vnhat}\Psi(\chi \vnhat; \eta_0 -\chi),
\end{equation}
where $\grad_\vnhat$ represents the angular derivative, equivalent to the covariant derivative on the sphere defined by $\vnhat$.
The quantity $\eta_0 -\chi$ is the conformal time at which the photon was at position $\chi \vnhat$, and $f_K(\chi)$ is the comoving angular-diameter distance.
It is convenient to define the \emph{lensing potential},
\begin{equation}
\psi(\vnhat) \equiv -2 \int_0^{\chi_*} \ud \chi\,
\frac{f_K(\chi_*-\chi)}{f_K(\chi_*)f_K(\chi)} \Psi(\chi \vnhat; \eta_0 -\chi),
\label{psin}
\end{equation}
so that the deflection angle is given by $\grad_\vnhat \psi$. From now on we write this simply as $\grad\psi$.
For full derivations and review see Refs.~\cite{Lewis:2006fu,Hanson:2009kr}.

\begin{figure}
\includegraphics[width=8cm]{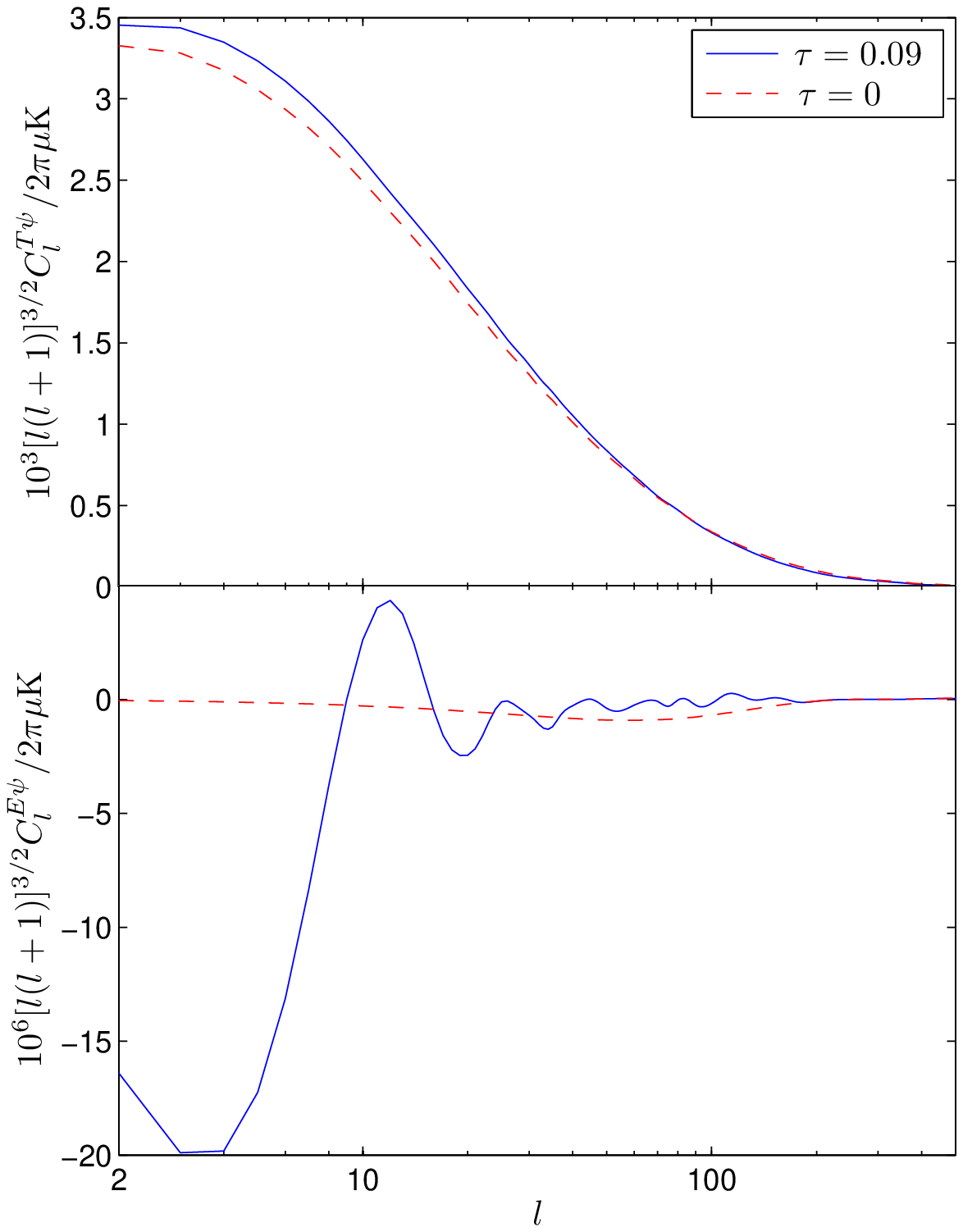}
\includegraphics[width=8cm]{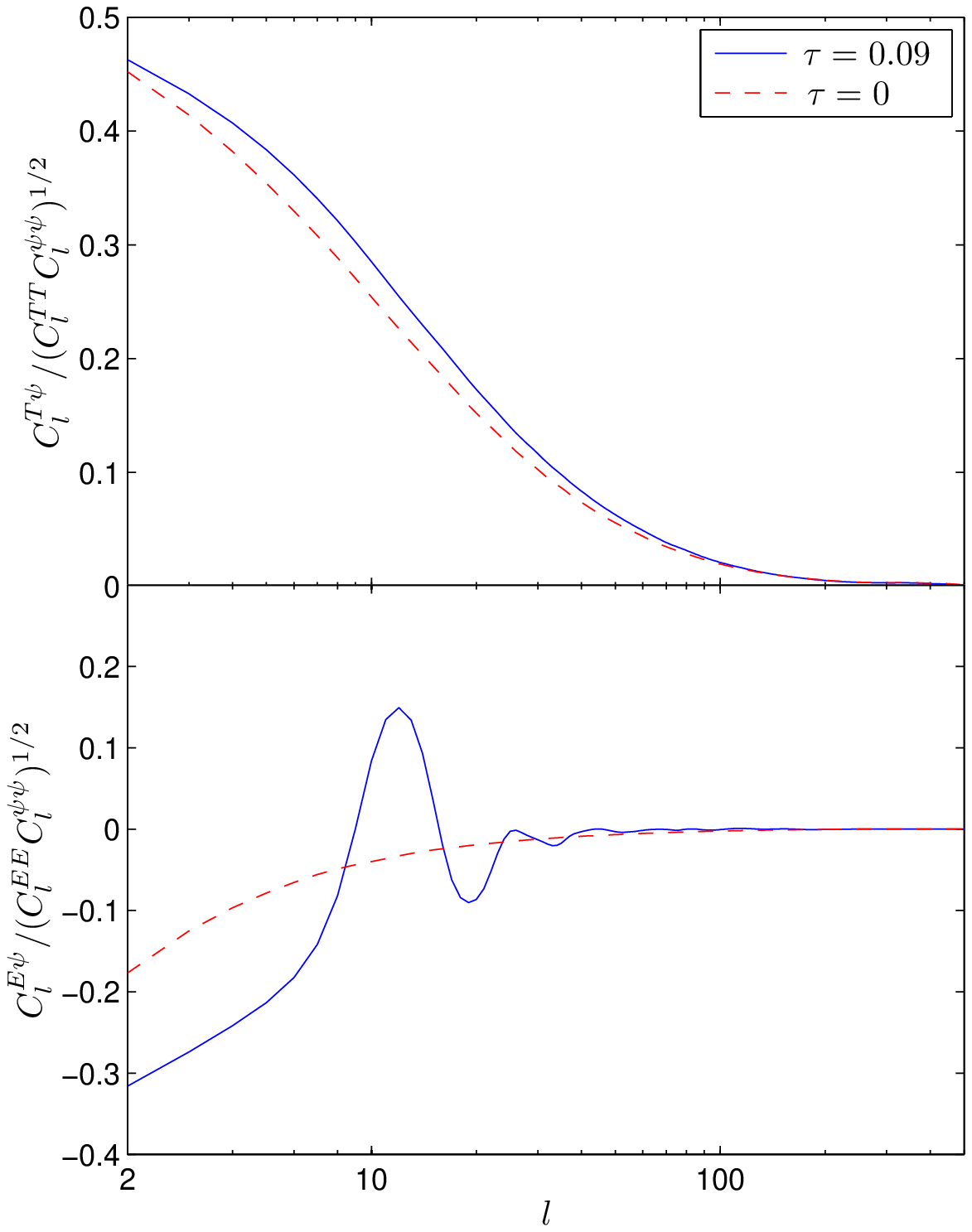}
\caption{The power spectrum (left) and dimensionless correlation coefficient (right) for the correlation of the CMB lensing potential with the CMB temperature anisotropy (top) and $E$-mode polarization (bottom) for a standard $\Lambda$CDM cosmology with an optical depth to reionization $\tau=0.09$ (solid lines) and $\tau=0$ (dashed lines). Note that without reionization the $E$-polarization power spectrum on large scales is very small, so although the correlation is still significant the actual size of the signal is very tiny. The $T$--$\psi$ correlation is due mainly to the ISW effect; the $E$--$\psi$ correlation to the reionization signal.
}
\label{tpecorrelations}
\end{figure}

Since the lensing potential is a weighted integral of the Weyl potential along the line of sight, it is correlated to the ISW contribution to the CMB temperature given by
\be
\Delta T_{\rm{ISW}}(\vnhat) = 2 \int_0^{\chi_*} \d\chi \dot{\Psi}(\chi \vnhat; \eta_0 -\chi),
\ee
where the dot denotes a conformal time derivative. In concordance $\Lambda$CDM models, $\Delta T_{\rm{ISW}}$ and $\psi$ are highly correlated (at above the 90\% level) due to the similarity of their redshift kernels, which leads directly to a correlation between the total CMB anisotropy and the lensing potential. The full result for the angular power spectrum $C_l^{T\psi}$ can easily be calculated numerically, and typical results are shown later in Fig.~\ref{tpecorrelations}; in total the correlation is nearly $50\%$ at $l=2$, but decreases rapidly with scale as the ISW contribution to the total diminishes, giving only a few percent correlation by $l=100$.

\quickedit{
\begin{figure}
\includegraphics[width=8cm]{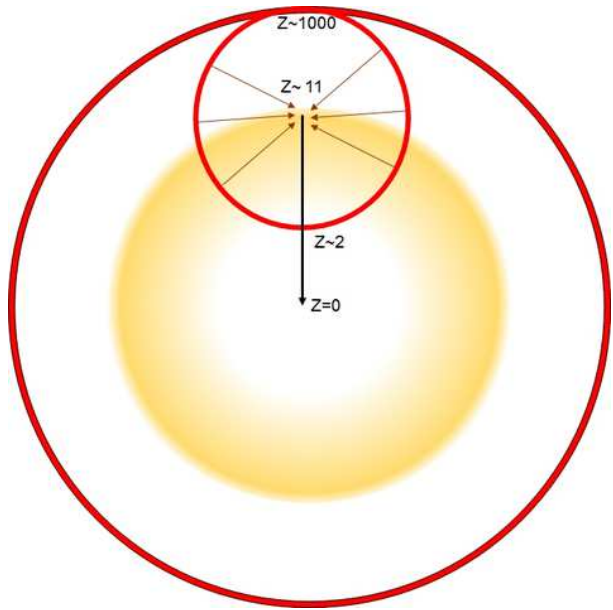}
\caption{The geometry of the polarization signal generated by scattering at reionization: the figure is roughly to scale in comoving distance, with reionization (shading) starting at $z\sim 11$, about 2/3 of the way to the last-scattering surface (dark red outer shell). After the universe has reionized, the probability of scattering falls off as the universe expands, so most scattering occurs between $\sim 1/2$ and $\sim 2/3$ of the distance to recombination. An electron at reionization sees its own last scattering surface as indicated by the red shell, and Thomson scattering of the quadrupolar component of the distribution of photons originating from this surface generates $E$-polarization. For an electron at the start of reionization its last scattering shell extends from our last-scattering surface down to a redshift of about $z\sim 2$.
Perturbations that generate a large-scale polarization signal will be correlated on large-scales, and therefore be correlated to
perturbations at $z\alt 2$ (see Fig.~\ref{waves}).
}
\label{LSS}
\end{figure}
\begin{figure}
\includegraphics[width=5.3cm]{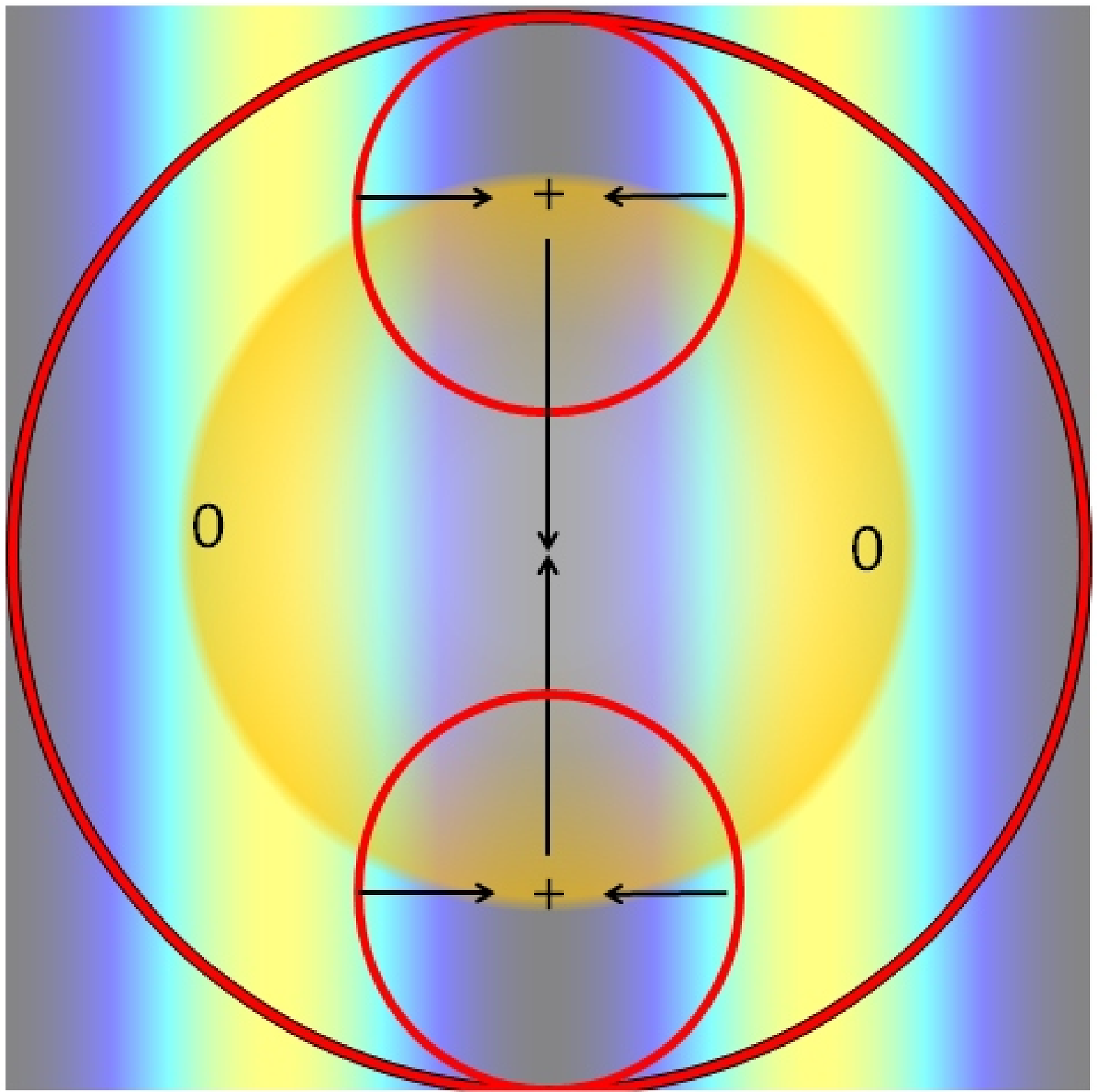}
\includegraphics[width=5.3cm]{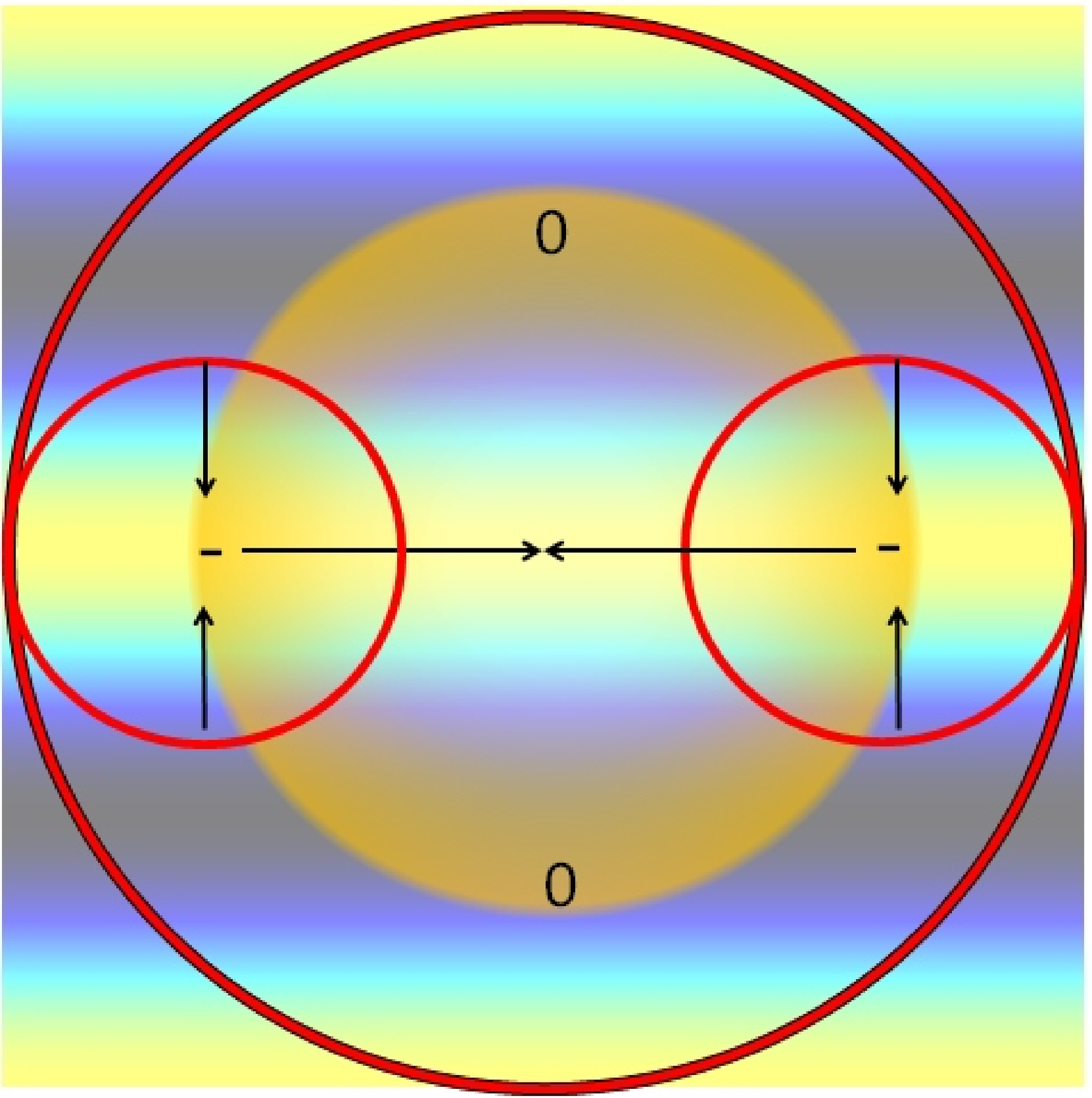}
\includegraphics[width=5.3cm]{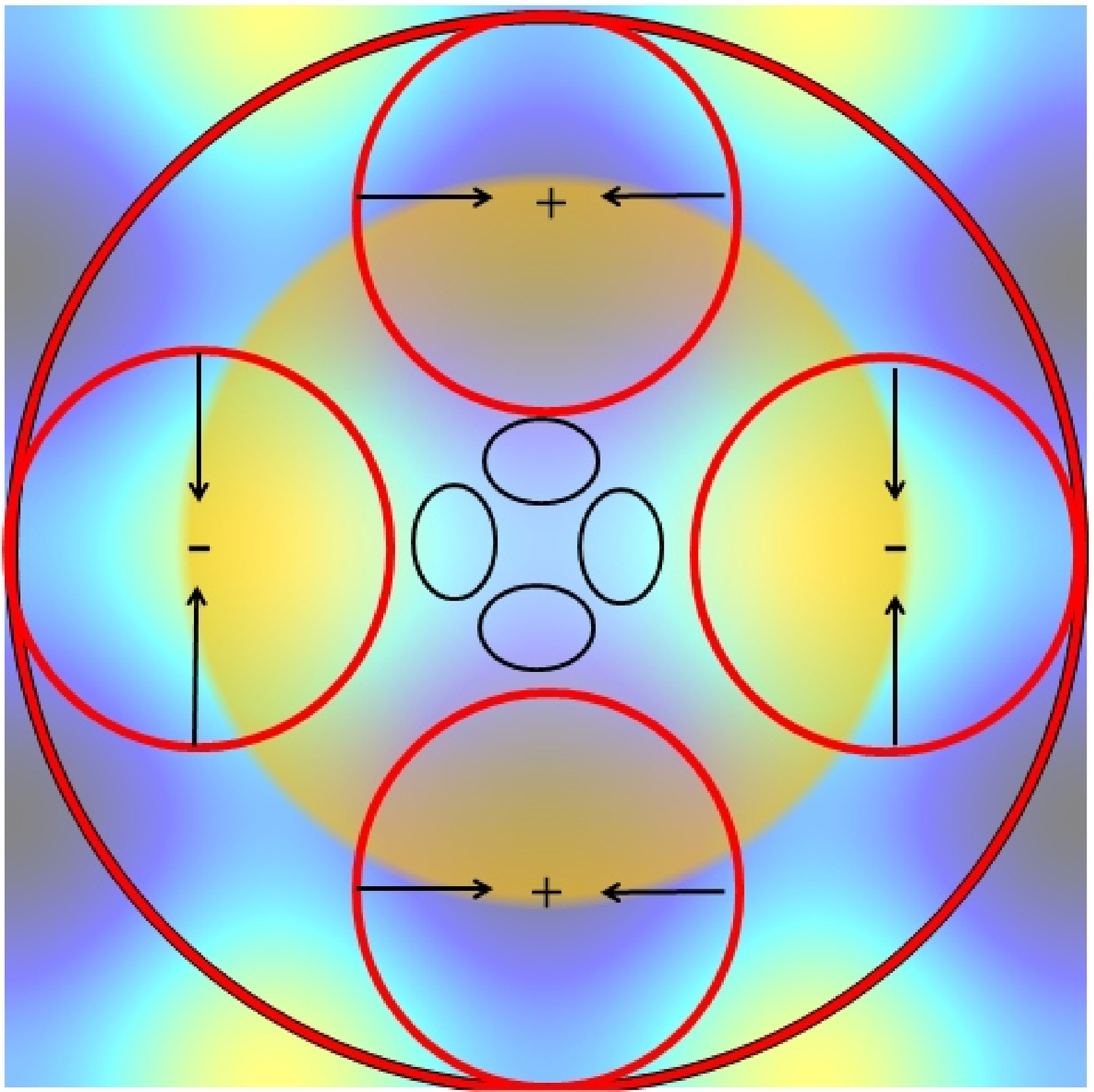}
\caption{
An illustration of the correlation between the quadrupole component of the polarization from reionization and the local density quadrupole, here generated by a superposition of two orthogonal plane waves.
}
\label{waves}
\end{figure}
}

The story with the polarization is rather different. The temperature quadrupole generated by the ISW can re-scatter leading to a correlated polarization
signal; however as shown in Ref.~\cite{Cooray:2005yj} this signal is tiny because there is little scattering at the low redshifts where the ISW signal
becomes significant. The dominant correlation is actually between the lensing potential and the large-scale polarization $E$-modes generated by
scattering at reionization. At redshift $z\sim 11$ where reionization occurs, $E$-mode polarization is generated by Thomson scattering of the local radiation quadrupole.
This quadrupole has contributions from a wide range of redshifts (for the observer), overlapping
 with the region $1 \lesssim z \lesssim 6$ from which the CMB lensing potential is sourced, and is correlated over long distances.
This is illustrated more concretely in Figs.~\ref{LSS} and~\ref{waves}, and plots of the cross-spectra
and correlation coefficient are given in Fig.~\ref{tpecorrelations}.
The large-angle $E$--$\psi$ correlation is negative and so produces radial
polarization around large-scale overdensities.
Further discussion of the $E$--$\psi$ correlation is given in
Appendix~\ref{app:Epsi} where a simple analytic model which reproduces the
main features of Fig.~\ref{tpecorrelations} is developed for the
case of instantaneous reionization.
With cosmic-variance limited full-sky $\psi$ and $E$, $C_l^{\psi E}$ could be detected as non-zero at approximately 2.5 sigma (compared to nearly 8 sigma for $C_l^{\psi T}$ from $\psi$ and $T$).

Although the ISW effect does not directly generate the correlated $E$-polarization signal (reionization occurs well before dark energy becomes dynamically
important), there is nonetheless a significant indirect correlation between the ISW and $E$ because, as we have noted, the lensing potential is highly correlated to the ISW signal. Indeed the $C^{TE}_l$ correlation at large scales is suppressed by about $1/5$ due to the (anti-)correlation between the ISW signal and the $E$ polarization. Note that the latter has the same sign as the $C_l^{E\psi}$
correlation.
Accurate numerical calculations of both $C_l^{T\psi}$ and $C_l^{E\psi}$ are now included in CAMB\footnote{\url{http://camb.info}}~\cite{Lewis:1999bs}.

\section{Flat-sky CMB temperature lensing bispectrum}
\label{flat-sky}
To understand the basic shape of the lensing bispectrum it is useful to start by considering the simple case of the CMB temperature in the flat-sky approximation.
We follow the flat-sky notation and conventions of Ref.~\cite{Lewis:2006fu}.
Assuming statistical isotropy and that parity invariance holds in the mean, the reduced bispectrum $b_{l_1 l_2 l_3}$ can be defined as
\be
\la  \tTemp(\vl_1)\tTemp(\vl_2)\tTemp(\vl_3) \ra = \frac{1}{2\pi} \delta(\vl_1+\vl_2+\vl_3)b_{l_1 l_2 l_3},
\ee
where $\tTemp(\vl)$ is the Fourier transform of the lensed temperature, and the delta-function ensures the triangle constraint. The reduced bispectrum is
symmetric in its arguments and it is therefore convenient to restrict the values of $l_1$, $l_2$ and $l_3$ such that $l_1 \le l_2 \le l_3$, with other combinations obtainable by permutation.

\subsection{Leading perturbative result}

Lensing remaps the temperature anisotropies so that the lensed
temperature field $\tilde{T}(\vx)$ is related to the unlensed field
$T(\vx)$ by $\tilde{T}(\vx) = T(\vx+\vgrad\psi)$.
Fourier transforming and performing a series expansion to first order in $\psi$ gives\footnote{We are assuming the unlensed CMB is a single source plane at recombination governed by a single lensing potential. This is not quite correct on large scales because the ISW contributions are more local; however the bispectrum is only significant for small-scales of the lensed field, so we can neglect this complication to good accuracy.}
\be
\tilde{\temp}(\vl) =\temp(\vl) -\int \frac{\ud^2 \vl_1}{2\pi}\frac{\ud^2 \vL}{2\pi} \temp(\vL) \psi(\vl_1)\vl_1\cdot\vL (2\pi)\delta(\vl_1+\vL-\vl).
\ee
Assuming Gaussianity of the lensing potential and the CMB temperature anisotropies, to first order in $\psi$
we then obtain the three-point correlation~\cite{Zaldarriaga:2000ud,Hu:2000ee}
\be
\la  \tTemp(\vl_1)\tTemp(\vl_2)\tTemp(\vl_3) \ra \approx -\frac{1}{2\pi}\delta(\vl_1+\vl_2+\vl_3)
\left[(\vl_1\cdot \vl_2) C_{l_1}^{\temp\psi} C^{\temp\temp}_{l_2} + \mbox{5 perms.}\right].
\ee
This is the standard first-order result for the lensing bispectrum; as we shall see higher-order corrections result in corrections at the $10\%$ level.

\subsection{Unlensed short-leg approximation and non-perturbative result}

As we have described, the physical origin of the bispectrum signal is the small-scale power changing due to (de)magnification and shearing by large-scale lenses. If we consider a lensed CMB sky, and add an additional large-scale lens, it will look substantially similar, but re-sized. We would therefore expect the power spectrum of the small-scale fluctuations over the extent of the large-scale lens to be determined by a shifted version of a \emph{lensed} power spectrum. In the lowest-order result we calculated above, the expression for the bispectrum involved the \emph{unlensed} small-scale temperature spectrum, but since the result is only lowest order we can expect higher-order corrections on small scales.

\begin{figure}
\includegraphics[width=8cm]{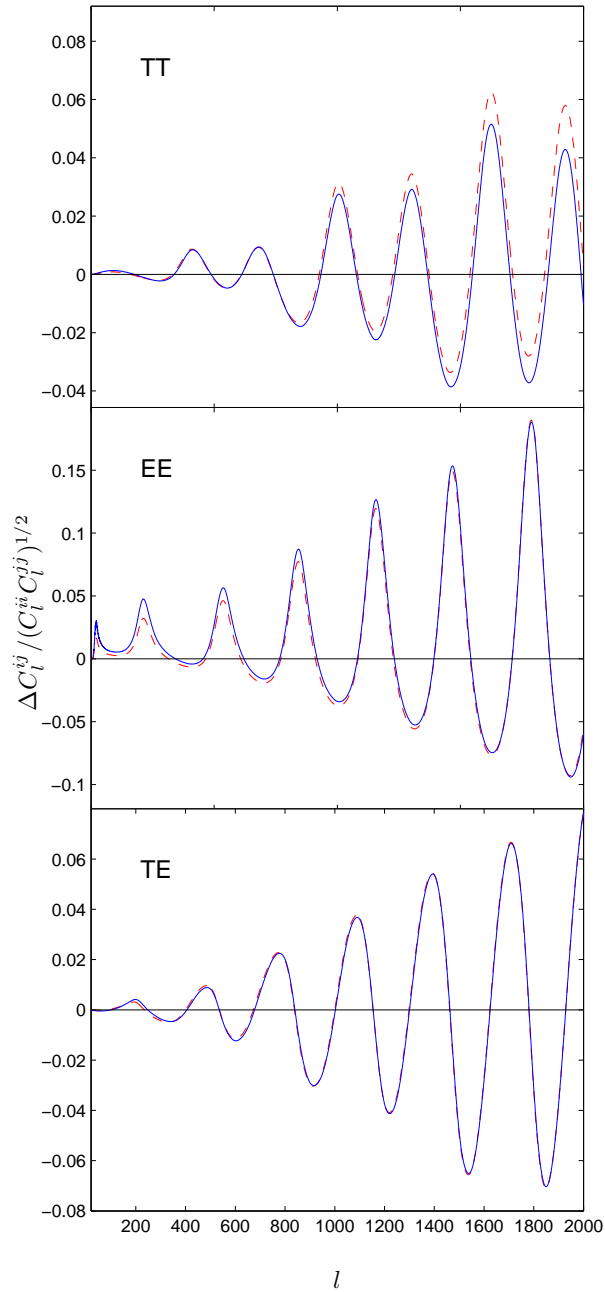}
\caption{The fractional difference between the flat-sky lensed temperature power spectra and the unlensed power spectra for $\tilde{C}^{\temp\temp}_l/C^{\temp\temp}_l-1$ (top; dashed), $\Cgrads_l/C^{\temp\temp}_l-1$ (top; solid) and analogously for the polarization in the lower two plots (see Appendix~\ref{Cgards_correlation}). The lensing bispectrum depends on $\Cgrads_l$, with $\Cgrads_l \approx \tilde{C}^{\temp\temp}_l$ to within about one percent.
}
\label{Cgrads}
\end{figure}

Since the lensing potential correlations fall off rapidly at high $l$, all of the bispectrum signal is at small $l_1$ ($l_1\alt 100$, where we restrict to $l_1\le l_2\le l_3$).
Since the lensing effect on the temperature at $l\alt 100$ is very small, to a good approximation we can calculate the bispectrum neglecting the lensing of the short leg $l_1$, i.e.
$\la \tilde{T}(\vl_1) \tilde{T}(\vl_2) \tilde{T}(\vl_3)\ra \approx
\la T(\vl_1) \tilde{T}(\vl_2) \tilde{T}(\vl_3)\ra$.
As we will show, many of the results in this paper may be verified numerically with Monte-Carlo simulations, and in all squeezed-shape cases we have checked, short legs have proved more than adequate. A more general approximation may be required to accurately assess the lensing bias on non-squeezed bispectra.

Using the fact that the lensed temperature is linear in $T$, and integrating the Gaussian expectation by parts, we have
\be
\la \temp(\vl_1)\Ltemp(\vl_2)\Ltemp(\vl_3)\ra = C^{\temp\psi}_{l_1} \left\la \frac{\delta}{\delta \psi(\vl_1)^*} \left(\Ltemp(\vl_2)\Ltemp(\vl_3)\right)\right\ra,
\label{eqn:short_leg_bispec}
\ee
where the ensemble average is taken over realizations of both the CMB and the lensing potential. The only approximation here is that $T(\vl_1)$ is uncorrelated
with the unlensed temperature modes that contribute to $\tilde{T}(\vl_2)$
and $\tilde{T}(\vl_3)$.
Using
\be
 \frac{\delta}{\delta \psi(\vl_1)^*}  \Ltemp(\vl) = -\frac{i}{2\pi} \vl_1\cdot \widetilde{\vgrad\temp}(\vl+\vl_1),
\ee
where we introduced the lensed temperature gradient, $\widetilde{\vgrad\temp}(\vx) = (\vgrad\temp)[\vx+\vgrad\psi]$ and its Fourier transform,
we then have
\ba
\left\la  \frac{\delta}{\delta \psi(\vl_1)^*} \left(\Ltemp(\vl_2)\Ltemp(\vl_3)\right)\right\ra
&=& -\frac{i}{2\pi} \vl_1\cdot \left\la \widetilde{\vgrad\temp}(\vl_1+\vl_2)\Ltemp(\vl_3)\right\ra + (\vl_2\leftrightarrow \vl_3) \\
&=& -\frac{1}{2\pi} \delta(\vl_1+\vl_2+\vl_3) \left[ (\vl_1\cdot \vl_2)  \Cgrads_{l_2} + (\vl_1\cdot \vl_3)  \Cgrads_{l_3}\right].
\label{eqn:lensedcovresponse}
\ea
Here we have defined the power spectrum $\Cgrads_l$ by
\be
\la \widetilde{\vgrad T}(\vl) \tilde{T}(\vl') \ra = i \vl \Cgrads_l \delta(\vl + \vl') ,
\ee
so that
\be
-i l^{-2}\left\la \vl\cdot \widetilde{\vgrad\temp}(\vl)\Ltemp(\vl')\right\ra  = \Cgrads_l \delta(\vl+\vl').
\ee
The expression for $\la \temp(\vl_1)\Ltemp(\vl_2)\Ltemp(\vl_3)\ra $ then follows simply from Eq.~\eqref{eqn:short_leg_bispec}.
In the absence of lensing, $\Cgrads_l$ reduces to the usual temperature power spectrum. With lensing, to the extent that gradients and lensing commute, $\Cgrads_l$ is reasonably well approximated by the lensed power spectrum.
Indeed, in Fig.~\ref{Cgrads} we show numerically that approximating $\Cgrads_l\approx \tC^{\temp\temp}_l$ is correct to about the percent level.
For the temperature bispectrum this then gives
\ba
b_{l_1 l_2 l_3} &\approx&
- C_{l_1}^{\temp\psi}\left[ (\vl_1\cdot \vl_2) \tC^{\temp\temp}_{l_2} + (\vl_1\cdot \vl_3) \tC^{\temp\temp}_{l_3}\right].
\ea
In Appendix~\ref{third_order} we show explicitly that this non-perturbative relation agrees with a direct perturbative calculation to third order in $\psi$. Figure~\ref{flat_squeeze} shows the effect of the higher-order corrections, effectively smoothing out the lensing bispectrum at the 10\% level; this may be important to estimate correctly the contribution of CMB lensing to estimators for other forms of non-Gaussianity, and also for using the lensing bispectrum to obtain cosmological constraints.

Note that Eq.~\eqref{eqn:lensedcovresponse} for the response of the lensed CMB covariance to a mode of the lensing potential differs from
that which is usually derived at lowest order in the lensing potential, e.g. for quadratic estimators \cite{Okamoto03}, in which the
\textit{unlensed} spectra appear rather than (effectively) the lensed spectra.
The neglect of these higher-order contributions leads to a bias in standard quadratic lensing estimators, which is more rigorously
calculated in Ref.~\cite{Hanson:2010rp}. The non-perturbative response of the lensed covariance to a mode of the lensing potential
which we present here provides a faster, more intuitive way to arrive at the same result.

Finally we can easily construct an accurate approximation for the lensing bispectrum which is non-perturbatively correct if the short-leg approximation holds, and also agrees with the perturbative result to leading order even if it is violated:
\be
b_{l_1 l_2 l_3} \approx -
\left[(\vl_1\cdot \vl_2) C_{l_1}^{\temp\psi} \Cgrads_{l_2} + \mbox{5 perms.}\right].
\ee

\subsection{Squeezed limit}

Since $C_{l_1}^{\temp\psi}$ rapidly becomes small on small scales, the bispectrum is nearly zero unless $l_1$ is small. However the lensing deflection angles are small, a few arcminutes, so the lensing only has a significant effect on $T(\vl_2)$ on small scales ($l_2\gg 1$). Hence almost all of the bispectrum signal is in squeezed triangles with $l_1 \ll l_2 \approx l_3$.
If we consider the ultra-squeezed limit we can define $\vl\equiv (\vl_2-\vl_3)/2 = \vl_2+\vl_1/2=-\vl_3-\vl_1/2$ and
expand in the small quantity $\vl_1/l$ giving the leading terms for the reduced bispectrum,
\begin{figure}
\includegraphics[width=8cm]{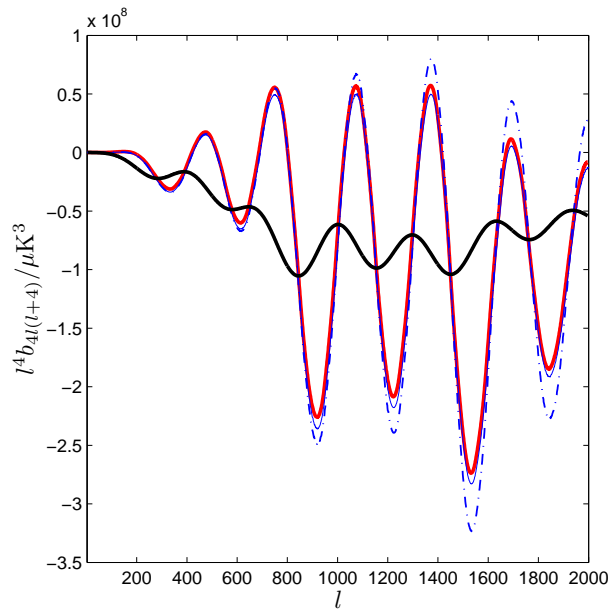}
\caption{The CMB temperature lensing reduced bispectrum $b_{4,l,l+4}$ using the squeezed-limit approximation of Eq.~\eqref{diff_approx} with the unlensed (thin blue dot-dashed) and lensed (thin blue solid) small-scale power spectrum, compared to the full spherical result of Eq.~\eqref{fullspherbispec} using the lensed small-scale spectrum (thick red). For comparison the thick black line shows the result for a local-model primordial bispectrum with $\fnl=10$ (which has not itself been lensed here; see Ref.~\cite{Hanson:2009kg}); note the difference in phase and scale-dependence.}
\label{flat_squeeze}
\end{figure}
\ba
b_{l_1 l_2 l_3} &\approx&
- C_{l_1}^{\temp\psi}\left[ (\vl_1\cdot \vl_2) \tC^{\temp\temp}_{l_2} + (\vl_1\cdot \vl_3) \tC^{\temp\temp}_{l_3}\right] \nonumber \\
&\approx&  l_1^2C_{l_1}^{\temp\psi}\biggl[ \frac{ (\vl_1\cdot \vl)^2}{l_1^2 l^2} \frac{ \ud  \tC^{\temp\temp}_{l}}{\ud \ln l} + \tC^{\temp\temp}_{l} + \clo(l_1^2/l^2)\biggr]
\nonumber\\
&\approx&  l_1^2C_{l_1}^{\temp\psi}\frac{1}{2}\biggl[ \cos 2\phi_{l_1 l} \frac{ \ud  \tC^{\temp\temp}_{l}}{\ud \ln l} + \frac{1}{l^2} \frac{\ud (l^2\tC^{\temp\temp}_{l})}{\ud \ln l} + \clo(l_1^2/l^2)\biggr].
\label{diff_approx}
\ea
The partly quadrupolar dependence on the angle $\phi_{l_1 l}$ between the large-scale and small-scale modes is very different from the isotropic squeezed limit expected from primordial modulations (e.g. the local $\fnl$ model), making the quadrupole part of the lensing signal orthogonal. For $\vl_1$ and $\vl$ parallel, the signal is proportional to $\tilde{C}^{\temp\temp}_{l} \ud \ln (l\tilde{C}^{\temp\temp}_{l})/\ud \ln l$, reflecting the change in small-scale power due to shifting of scales by lensing (de)magnification and shearing. Since the spectrum has acoustic oscillations, the derivative term oscillates in $l$, with a phase shift compared to the power spectrum. For $\vl_1\cdot \vl\approx 0$ (i.e.\ $l_3 = l_2[1+\mathcal{O}(l_1/l_2)^2]$)
the derivative term is small and the bispectrum is generally of much smaller amplitude and has the same phase of acoustic oscillations as the power spectrum.
The phase shift of the dominant lensing bispectrum signal compared to the phase of the acoustic oscillations is rather distinctive, and different from that expected for any primordial bispectrum of adiabatic perturbations. The strong scale-dependence (very little signal for $l_1\agt 100$) is also different from standard local non-Gaussianity models; see Figs.~\ref{flat_squeeze} and~\ref{threedcontour}.
However as we shall see the isotropic part of the lensing bispectrum signal does have significant overlap with the local $\fnl$ model,
so although it is easily distinguished it is also important to model when studying local primordial non-Gaussianity.
 \quickedit{
%
\begin{figure}
\includegraphics[width=8cm]{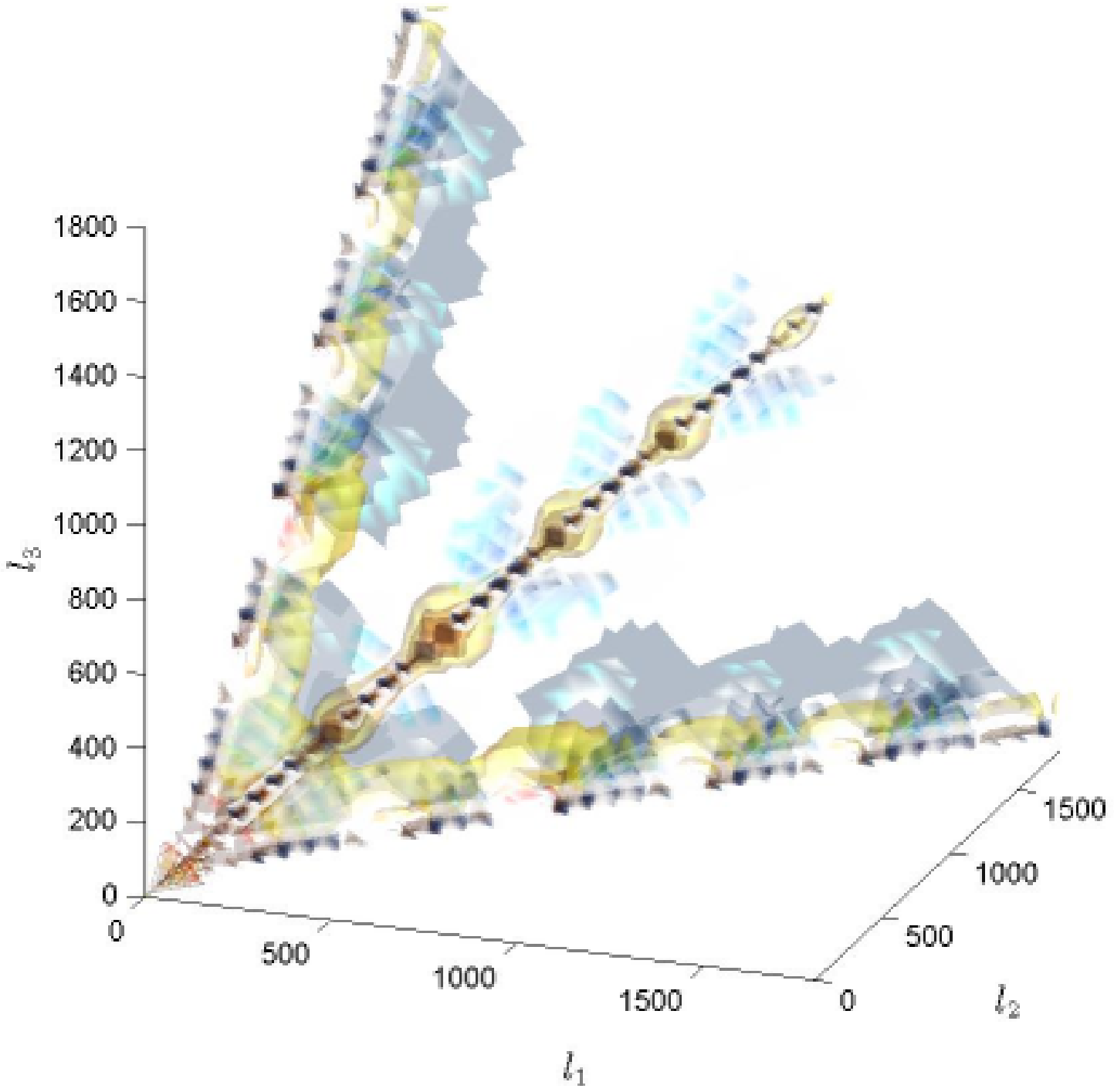}
\includegraphics[width=8cm]{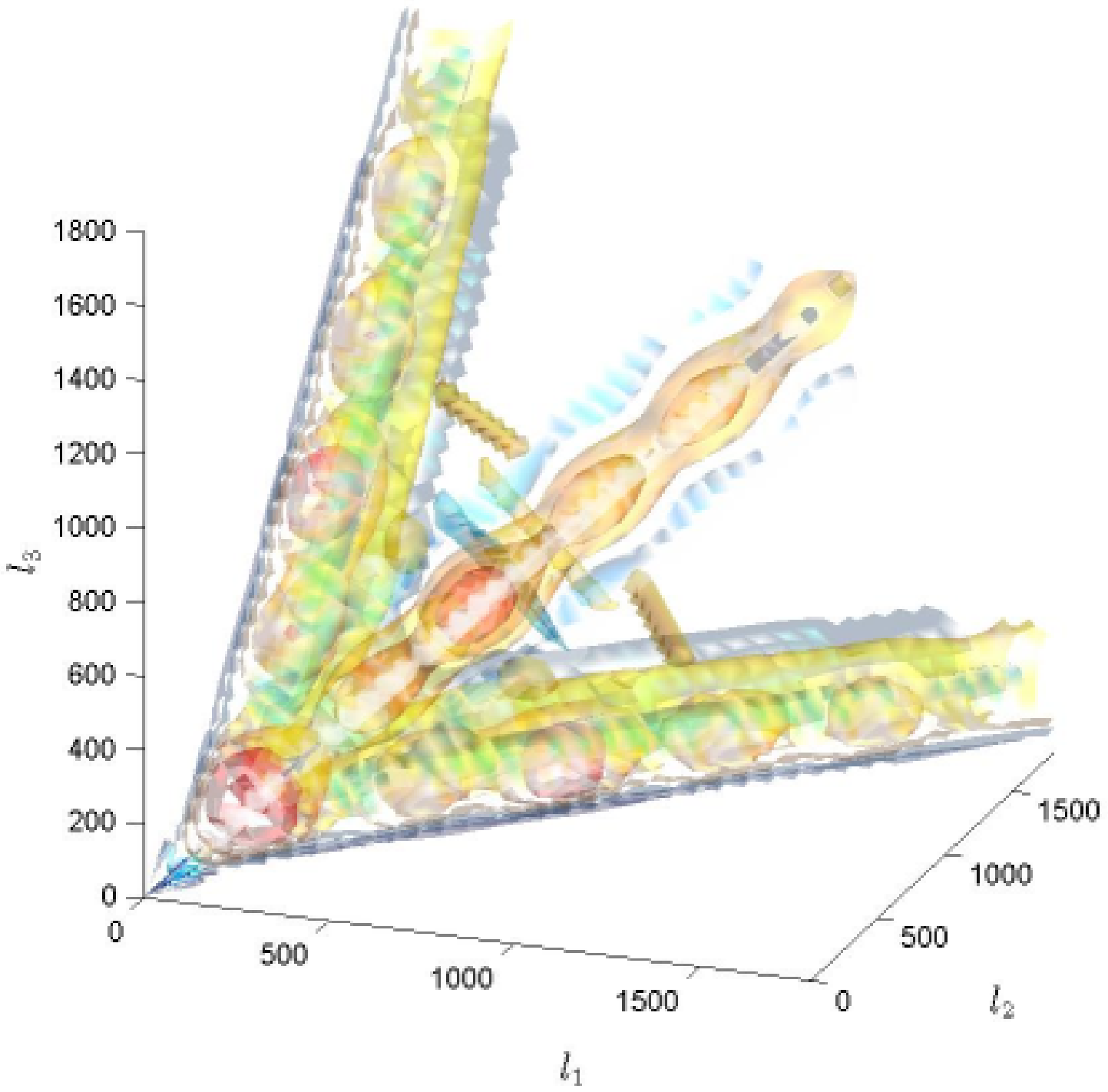}
\caption{Contour plot of $l_1 l_2 l_3(l_1+l_2+l_3)b_{l_1 l_2 l_3}$ (with non-linear intervals) for ISW-lensing (left) and local-model primordial non-Gaussianity (right; different overall scale). Although both are peaked for squeezed configurations, there are large phase and shape differences.
}.
\label{threedcontour}
\end{figure}
}

We can also derive the squeezed limit following an argument similar to Refs.~\cite{Maldacena:2002vr,Creminelli:2004yq,Creminelli:2004pv} by considering one fixed very large-scale lensing mode of the magnification matrix $\mA$, where
\be
A_{ij} \equiv \delta_{ij} + \frac{\partial \alpha_j}{\partial x^i}
= (1-\kappa) \delta_{ij} - \gamma_{ij}.
\ee
Here, $\kappa$ is the convergence and $\gamma_{ij}$ is the symmetric, trace-free
shear.
Since for a local displacement $\zeta$ we have $\tilde{\temp}(\zeta) = \temp(\mA \zeta)$, it follows that taking the average with fixed $\mA$ we have
\be
\langle \tilde{T}(\vl_2) \tilde{T}(\vl_3) \rangle =
C^{\temp\temp}_{|\mA^{-1}\vl_2|}
\frac{\delta(\vl_2+\vl_3)}{|\mA|}.
\ee
Expanding to first order in the convergence $\kappa$ and shear matrix $\gamma$ gives~\cite{Bucher:2010iv}
\be
\langle \tilde{T}(\vl_2) \tilde{T}(\vl_3) \rangle = C^{\temp\temp}_{l_2}
\delta(\vl_2+\vl_3) \left[1+ \kappa \frac{\ud \ln(l_2^2 C^{\temp\temp}_{l_2})}{\ud\ln l_2}
+ \hat{\vl}_2^T
\mgamma
 \hat{\vl}_2 \frac{\ud \ln C^{\temp\temp}_{l_2} }{\ud\ln
l_{2}} \right].
\label{squeezed_mag}
\ee
If there are also small-scale lensing modes then approximately the same result is obtained, with the power spectra replaced by the lensed CMB power spectra, in agreement with Eq.~\eqref{diff_approx} when correlated with $T(\vl_1)$. Equation\eqref{squeezed_mag} makes clear the different shape dependence of the convergence and shear effects: a scale-invariant spectrum looks the same under uniform magnification, but shear introduces observable distortion to the hot and cold spots (only the $\gamma$ term contributes if $l^2 C_l^{\temp\temp}=\text{const.}$); a white-noise spectrum looks the same after shearing, but the noise amplitude is changed under magnification (only the $\kappa$ term contributes if $C_l^{\temp\temp}=\text{const.}$)\footnote{The squeezed-limit form of the bispectrum here disagrees with Ref.~\cite{Creminelli:2004pv} which has incorrect $l^2$ factors in the anisotropic term. The result given in Ref.~\cite{Boubekeur:2009uk} is in agreement in the matter-dominated Sachs-Wolfe limit.}.

\section{General full-sky CMB lensing bispectra}
\label{full-sky}
We now present the generalization of the lensing bispectrum calculation of the previous section to the full-sky and polarization. Further details are contained in Appendix~\ref{Cgards_correlation}.
Following Ref.~\cite{Okamoto03} the lensed field is given by $\tilde{a}^i_{lm} = a^i_{lm} + \delta a^i_{lm}+\dots$ where the leading-order lensing correction for $\va_{lm} = (T_{lm},E_{lm},B_{lm})$ is
\ba
\delta T_{lm} &=& \sum_{LM}\sum_{l'm'} \threej{l}{L}{l'}{m}{M}{m'} F^0_{lLl'}
\psi_{LM}^* T_{l'm'}^* \\
\delta E_{lm} &=& \sum_{LM}\sum_{l'm'} \threej{l}{L}{l'}{m}{M}{m'}
\psi_{LM}^* \left[F^{+2}_{lLl'} E_{l'm'}^* - i F^{-2}_{lLl'} B_{l' m'}^*\right] \\
\delta B_{lm} &=& \sum_{LM}\sum_{l'm'} \threej{l}{L}{l'}{m}{M}{m'}
\psi_{LM}^* \left[F^{+2}_{lLl'} B_{l'm'}^* + i F^{-2}_{lLl'} E_{l' m'}^*\right],
\ea
where
\be
F^{\pm s}_{l L l'} \equiv \frac{1}{4}\left[L(L+1) + l'(l'+1)-l(l+1)\right]\sqrt{\frac{(2l+1)(2L+1)(2l'+1)}{4\pi}}
\left[\threej{l}{L}{l'}{s}{0}{-s}\pm \threej{l}{L}{l'}{-s}{0}{s}\right].
\ee
Note that $F^{+s}_{l_1 l_2 l_3}$ is only non-zero for $l_1 + l_2 +
l_3$ even and is then symmetric in $l_2$ and $l_3$; $F^{-s}_{l_1 l_2
  l_3}$ is only non-zero for $l_1 + l_2 + l_3$ odd and is then
antisymmetric in $l_2$ and $l_3$.
We shall assume there are no unlensed $B$ modes, so that $B_{lm}$ is due entirely to lensing.
The non-perturbative flat-sky derivation generalizes directly to the curved-sky case; we implement the result here
by simply using the lowest-order series-expansion result and then replacing the unlensed power spectra with their lensed counterparts.
 The leading-order lensing-induced 3-point function, using the lensed power spectra for the small scales to reproduce accurately the non-perturbative calculation, is then given by
\be
\la a^i_{l_1 m_1} a^j _{l_2 m_2} a^k_{l_3 m_3} \ra \approx
\left[C^{a^j \psi}_{l_2} \tilde{C}_{l_3}^{a^i a^k} F^{s_i}_{l_1 l_2 l_3} +
i C^{a^j \psi}_{l_2} \tilde{C}^{\bar{a}^ia^k}_{l_3} F^{-s_i}_{l_1 l_2 l_3}
\right] \threej{l_1}{l_2}{l_3}{m_1}{m_2}{m_3} + \mbox{5 perms.} \, ,
\label{fullspherbispec}
\ee
where $s_E=s_B=2$ and $s_T=0$, and $\bar{E}=-B$, $\bar{B}=E$ and $\bar{T}=0$.
Equation~(\ref{fullspherbispec}) includes a sum over all six permutations of
$i(lm)$. The bispectrum then follows from
\be
B^{ijk}_{l_1 l_2 l_3} \equiv \sum_{m_1 m_2 m_3} \threej{l_1}{l_2}{l_3}{m_1}{m_2}{m_3} \la a^i_{l_1 m_1}  a^j_{l_2 m_2} a^k_{l_3 m_3}\ra .
\ee
Note that under interchange of a pair of arguments, e.g.\ $il_1 \leftrightarrow
j l_2$, the bispectrum changes by a factor $(-1)^{l_1+l_2+l_3}$. If the
bispectrum involves a parity-odd combination of fields, e.g.\
$\la T B E \ra$, parity-invariance in the mean requires non-zero bispectra
to have $l_1+l_2+l_3$ odd and hence to change sign under interchange of
a pair of arguments. Furthermore, non-zero bispectra with $l_1+l_2+l_3$
odd are necessarily imaginary.

\begin{figure}
\includegraphics[width=8cm]{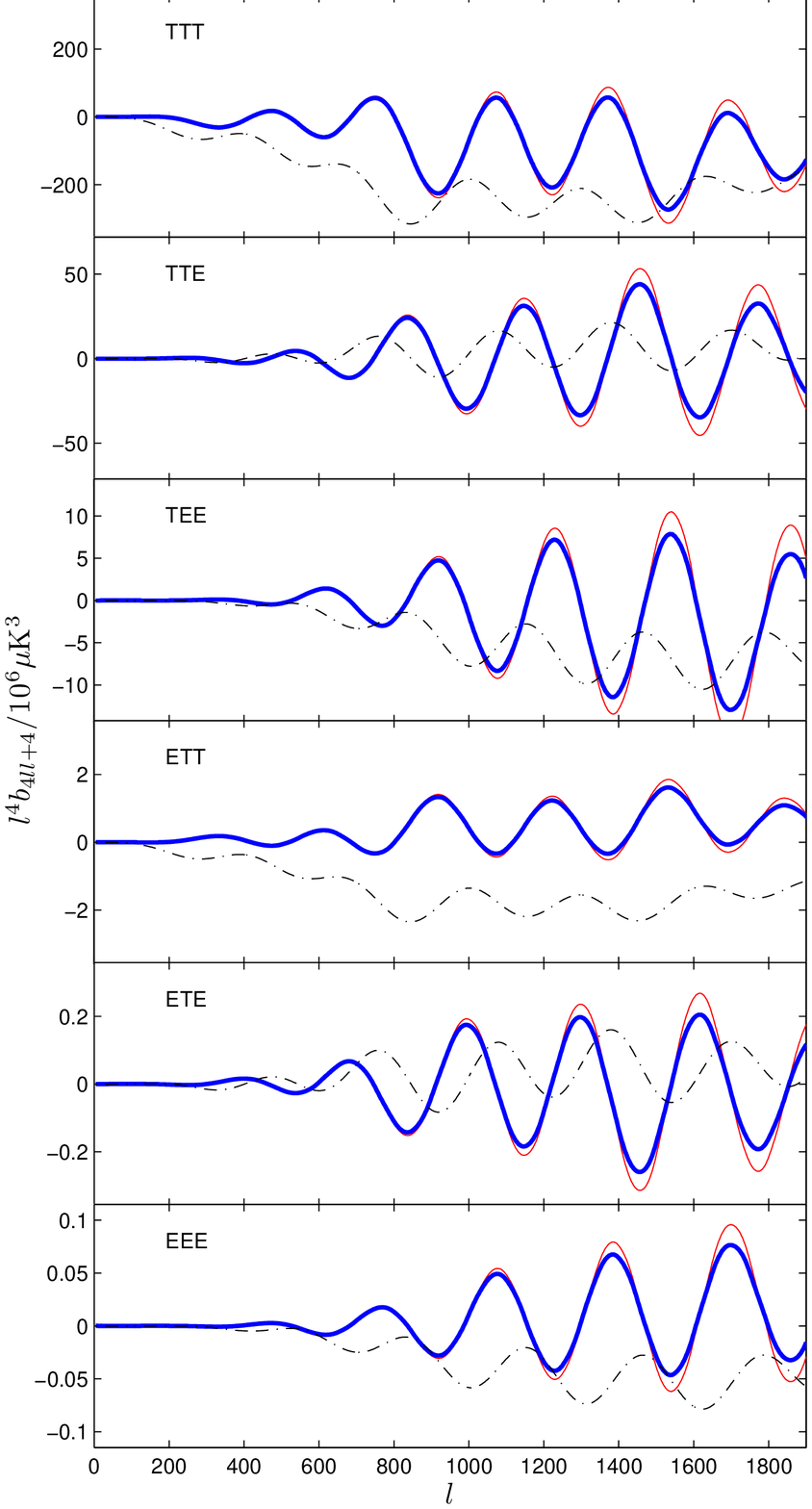}
\includegraphics[width=8cm]{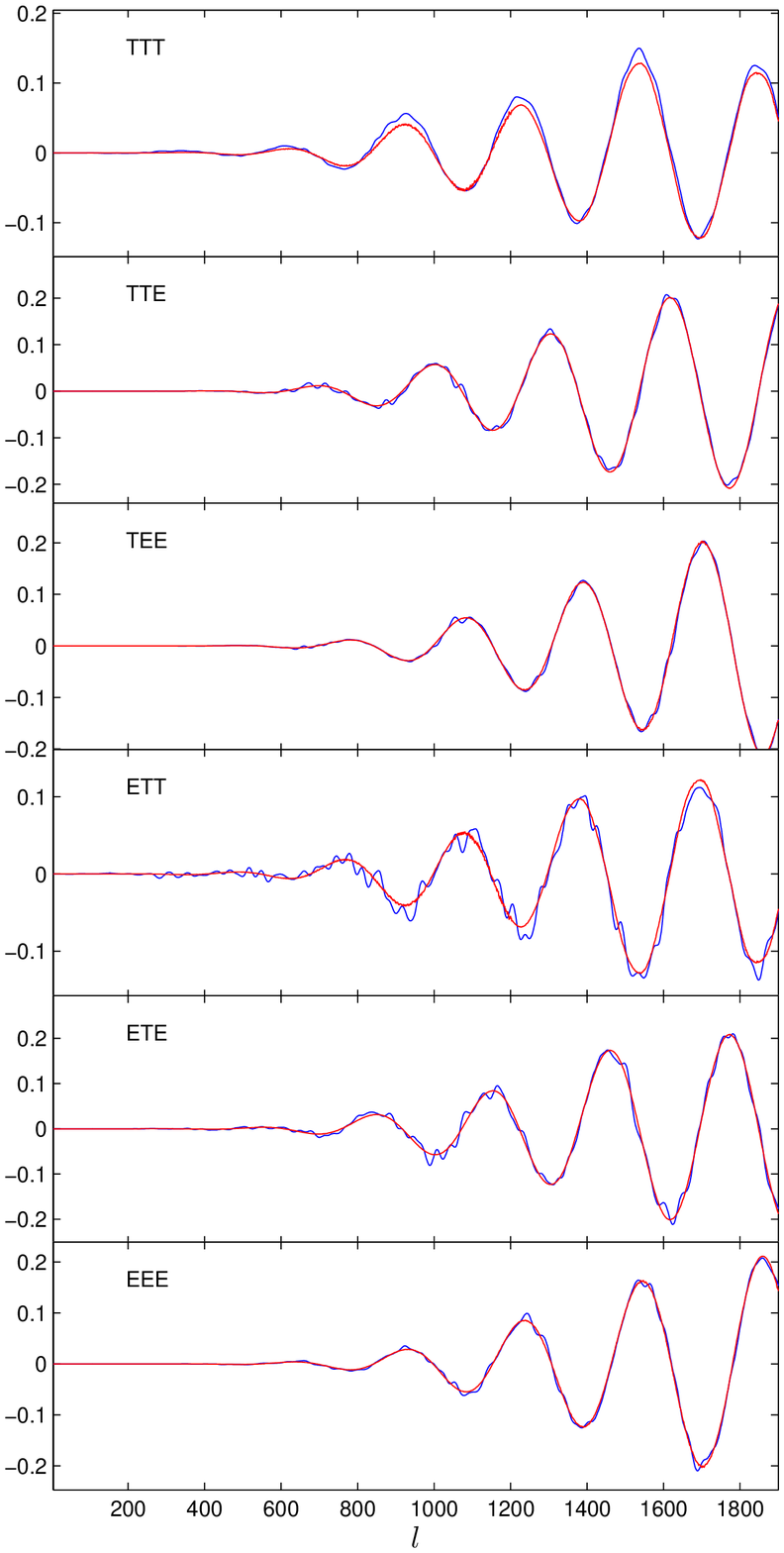}
\caption{The reduced bispectra $b^{ijk}_{4 l l+4}$ for various temperature and $E$-polarization combinations.
\emph{Left:} Thin solid (red) lines show the lowest-order result for the CMB lensing bispectrum; thick solid (blue) show the approximate non-perturbative result using the lensed CMB power spectrum. Dash-dotted (black) lines for comparison show the result for a local-model primordial bispectrum with $\fnl=30$. \emph{Right:} The difference between the non-perturbative bispectrum and the lowest-order result, in units of the maximum of the absolute value of each bispectrum. Smooth (red) lines show the theoretical approximation of Eq.~(\ref{fullspherbispec});
noisy (blue) lines are results from 1000 simulations, smoothed over $\Delta l=10$.
}
\label{correlations}
\end{figure}

\begin{figure}
\includegraphics[width=8cm]{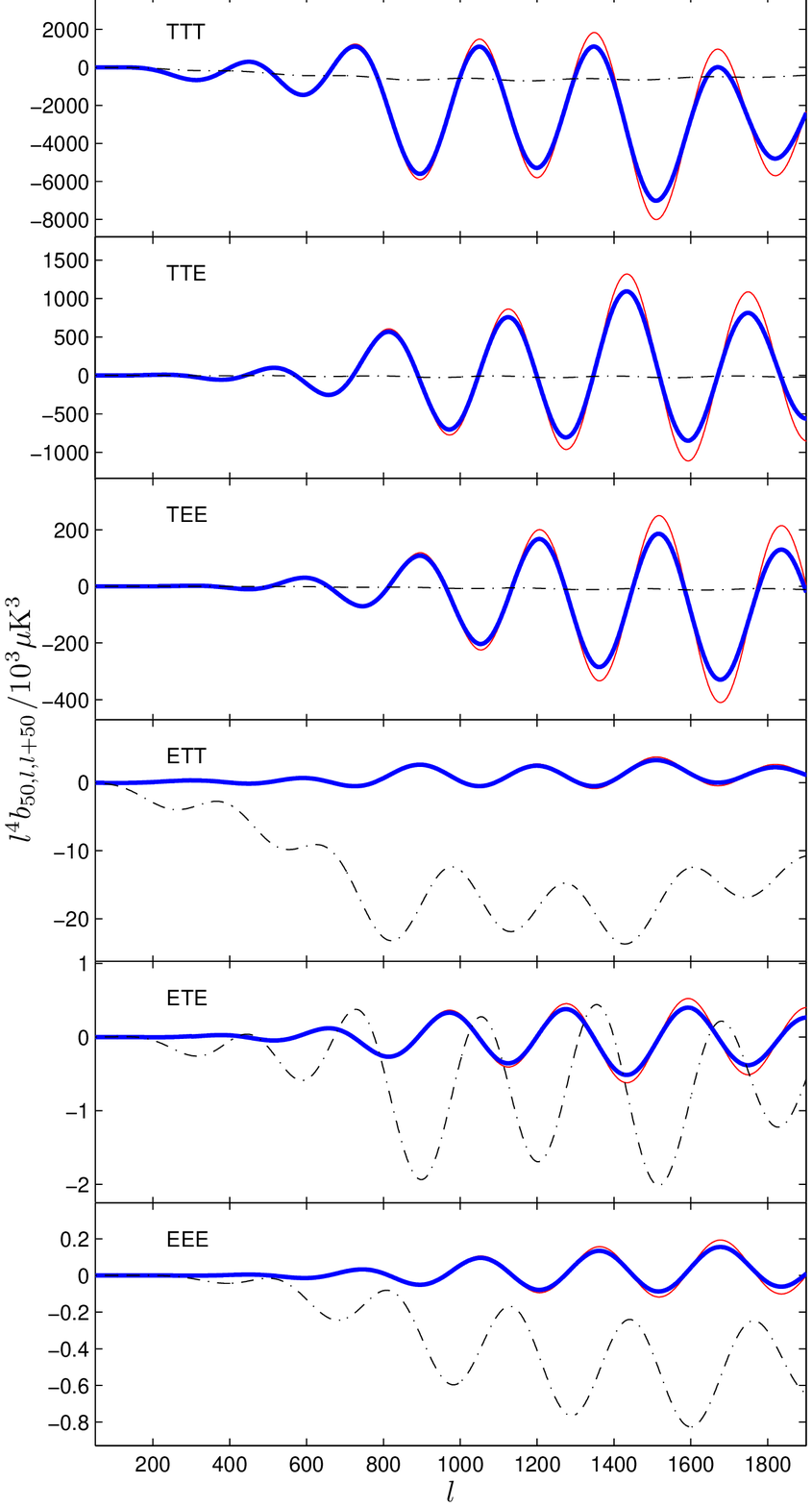}
\includegraphics[width=8cm]{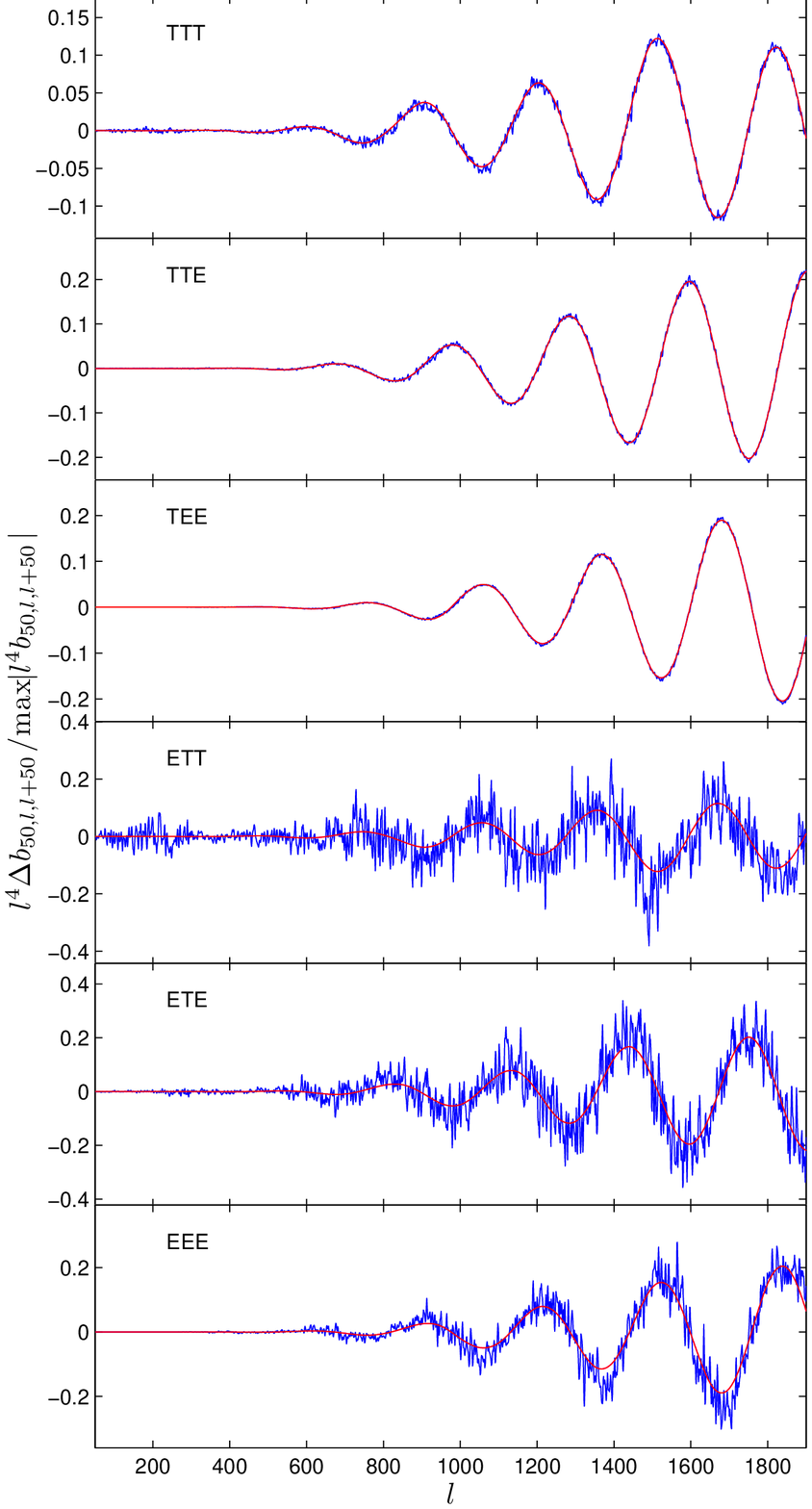}
\caption{Same as Fig.~\ref{correlations} but now  showing $b^{ijk}_{50 l l+50}$, where results on the right are from 700 simulations and are unsmoothed.
}
\label{correlations50}
\end{figure}

\begin{figure}
\includegraphics[width=8cm]{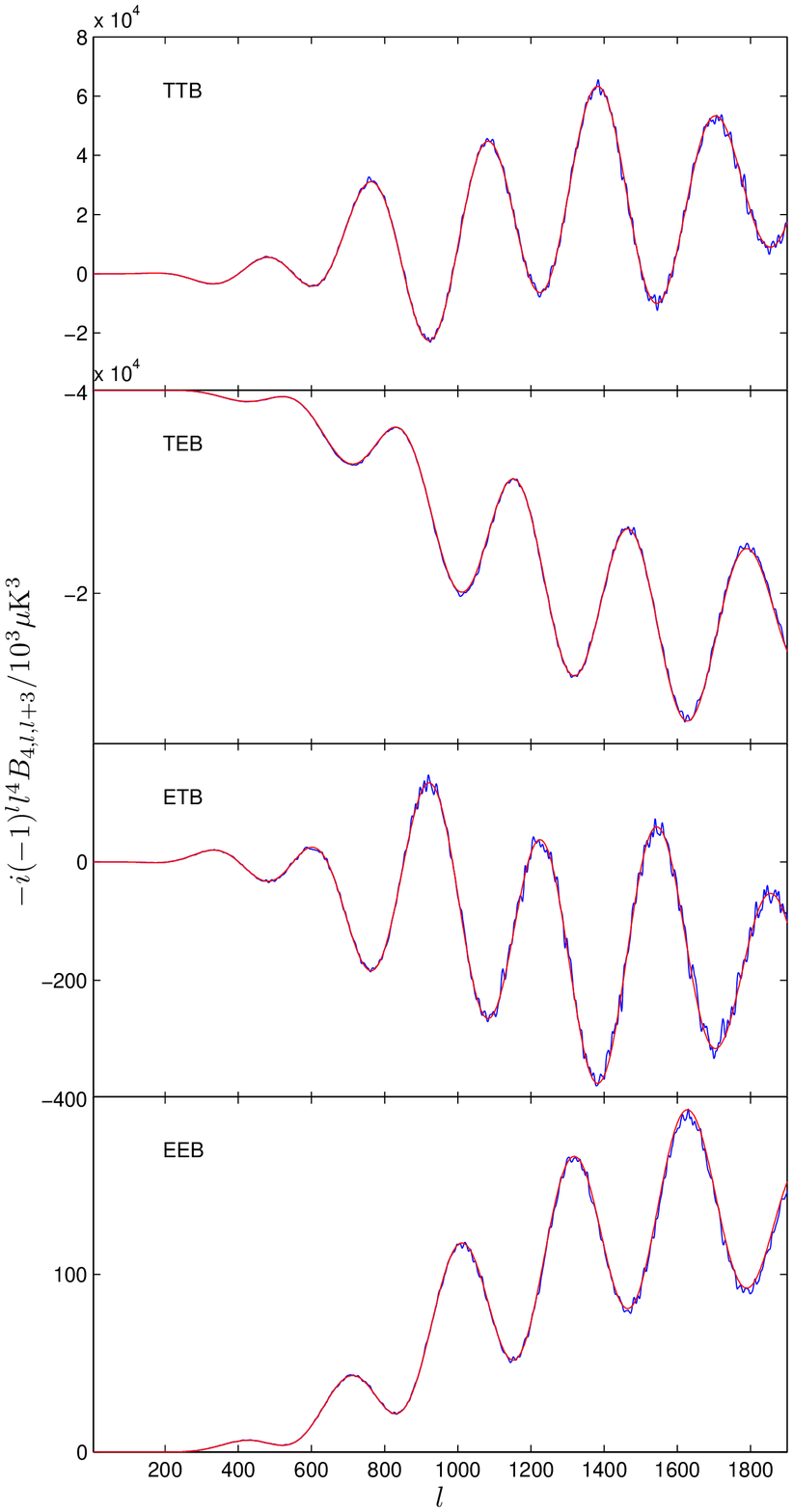}
\includegraphics[width=8cm]{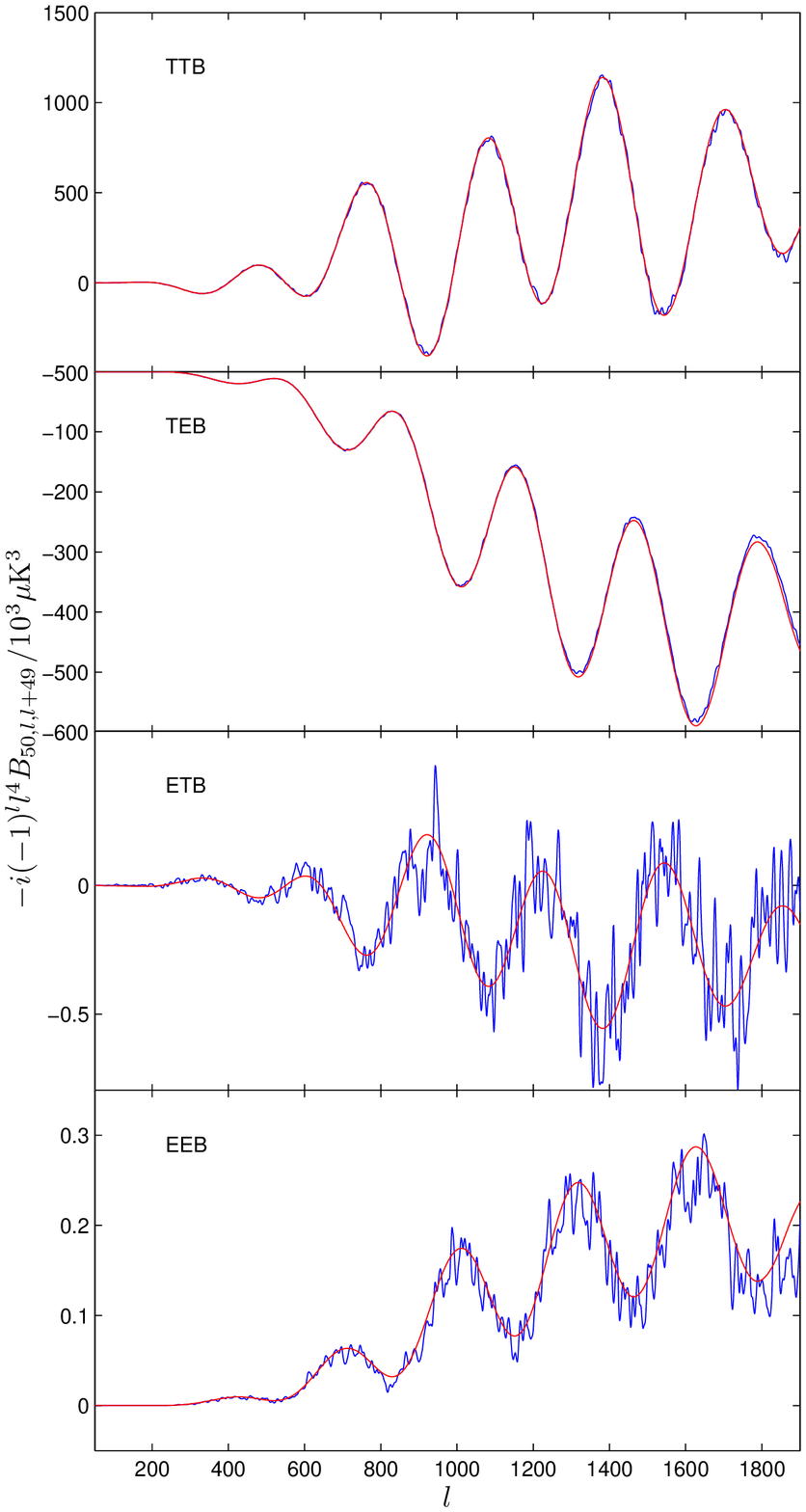}
\caption{The ``odd-parity'' bispectra $-i (-1)^l B^{ijk}_{4,l,l+3}$ (left) and $-i (-1)^l B^{ijk}_{50,l,l+49}$ (right) comparing the theoretical result
of Eq.~(\ref{fullspherbispec}) (smooth line; red) to the result (noisy line;
blue) of 700 (left) and 150 (right) simulations.
}
\label{correlationsodd}
\end{figure}

Results for the polarization bispectra have been derived before (e.g.~Ref.~\cite{Hu:2000ee}), however previous calculations have invariably set the large-scale $C^{E\psi}_l=0$, missing a signal detectable at several sigma with cosmic-variance limited data, and the power spectra have usually been the unlensed ones, giving a systematic error of $\clo(10\%)$. We show several slices through the temperature and polarization bispectra in Figs.~\ref{correlations}, \ref{correlations50}, \ref{correlationsodd}, using both analytical calculations as well as simulations using the Monte-Carlo procedure outlined in Appendix~\ref{simulation}, testing the accuracy of the unlensed short-leg approximation and the use of lensed power spectra rather than e.g.\ $\Cgrads_l$. We demonstrate in Appendix~\ref{Cgards_correlation} that for the polarization case, the non-perturbative calculation
involves a new spectrum $\tilde{C}_l^{PP\perp}$ as well as $\tilde{C}_l^{E\nabla
E}$ and $\tilde{C}_l^{B\nabla B}$. Terms involving this spectrum are missed
in the approximation of replacing e.g.\ $\tilde{C}_l^{E\nabla E}$ by
$\tilde{C}_l^{EE}$. However, this is harmless since $\tilde{C}_l^{PP\perp}$ is
of similar magnitude to $\tilde{C}_l^{BB}$ and the error from neglecting
such terms is small compared to the change in $C_l^{EE}$ due to lensing
(which is the dominant correction to the leading-order bispectra).

For the temperature, the reduced bispectrum $b_{l_1 l_2 l_3}$ is defined so that
\be
B^{TTT}_{l_1 l_2 l_3} =\sqrt{\frac{(2l_1+1)(2l_2+1)(2l_3+1)}{4\pi}}
\threej{l_1}{l_2}{l_3}{0}{0}{0} b_{l_1 l_2 l_3},
\ee
where $b_{l_1 l_2 l_3}$ is taken to be zero for $l_1+l_2+l_3$ odd. This
generalizes straightforwardly to bispectra involving only $T$ and/or $E$
since parity-invariance forces the bispectra to vanish for $l_1+l_2+l_3$ odd.
However, there does not appear to be a standard equivalent definition for
the bispectra involving a product of fields with net odd parity.
For sufficiently sensitive data, these ``odd-parity'' bispectra are well measured because of the expected absence of small-scale primordial $B$-modes; this is equivalent to the lensing reconstruction from $E$--$B$ correlations having lowest statistical noise, and hence correlating well with the large-scale temperature (and polarization). Since gravity waves decay on sub-horizon scales, the squeezed ``odd-parity'' CMB lensing bispectra are yet another way in which the CMB bispectra are very different from any primordial source. With sufficiently low noise, the large-scale lensing potential can be reconstructed very well using the small-scale $E$ and $B$ polarization~\cite{Okamoto03,Hirata:2003ka}, so it would be straightforward to project the correlated component out of the temperature and polarization data and thereby remove CMB lensing as a source of contamination for other signals.

\section{Estimators, variance and bias}
\label{estimators-variance-bias}

We now turn to a discussion of optimal estimators for the lensing bispectrum and their variance.
As reviewed in Appendix~\ref{vectorization},
in the case of an isotropic survey (full sky and uniform noise)
the optimal estimator for the amplitude of a bispectrum template $\tB$  is, for small signals, given by
\be
\hat{S} =  \frac{1}{F} \sum_{l_1 \le l_2 \le l_3} \vecp(\tB_{l_1 l_2
  l_3})^\dag \mCov^{-1}_{l_1 l_2 l_3} \vecp(\hat{\tB}_{l_1 l_2 l_3}) =
\frac{1}{6F} \sum_{l_1 l_2 l_3} (B^{ijk}_{l_1 l_2 l_3})^* (\tC_{\text{tot}}^{-1})^{ip}_{l_1}(\tCtot^{-1})^{jq}_{l_2}(\tCtot^{-1})^{kr}_{l_3} \hat{B}^{pqr}_{l_1 l_2 l_3},
\label{opt_est}
\ee
where $\vecp(\tB_{l_1 l_2 l_3})$ is the vector of distinct elements of
$\tB_{l_1 l_2 l_3}$ with covariance matrix $\mCov_{l_1 l_2 l_3}$,
$\tCtot^{ip}{}_{l_1}$ is the total
cross-power spectrum including noise,
and  $F^{-1}=\la \hat{S}^2\ra$ is the inverse of the Fisher error in the limit of no non-Gaussianity, given by
\be
F = \frac{1}{6} \sum_{l_1 l_2 l_3} (B^{ijk}_{l_1 l_2 l_3})^*
(\tC_{\text{tot}}^{-1})^{ip}_{l_1}(\tC_{\text{tot}}^{-1})^{jq}_{l_2}(\tC_{\text{tot}}^{-1})^{kr}_{l_3}
B^{pqr}_{l_1 l_2 l_3} = \sum_{l_1 \le l_2 \le l_3} \Delta_{l_1 l_2
  l_3}^{-1} (B^{ijk}_{l_1 l_2 l_3})^*
(\tC_{\text{tot}}^{-1})^{ip}_{l_1}(\tC_{\text{tot}}^{-1})^{jq}_{l_2}(\tC_{\text{tot}}^{-1})^{kr}_{l_3}
B^{pqr}_{l_1 l_2 l_3}.
\label{eq:fisherB}
\ee
Here $\Delta_{l_1 l_2 l_3} =
6\delta_{(l_1}^{l_1}\delta_{l_2}^{l_2}\delta_{l_3)}^{l_3}$ (with no
implicit sums over ${l}$-labels): $\Delta_{l_1 l_2 l_3}$ is 6 if $l_1=l_2=l_3$, 2 if two of the indices are equal, and 1 otherwise.

The Fisher error in Eq.~(\ref{eq:fisherB}) was calculated for Gaussian
$a^i_{lm}$, but in the presence of lensing the variance is necessarily
larger since there is a guaranteed
signal and this itself has some variance.
We will motivate an expression for this increase in variance by recasting the estimator for the lensing bispectrum as a cross-correlation between
a quadratic estimate of the lensing potential and the CMB itself. We begin for simplicity in Section~\ref{sec:variance_temperature} by considering
the temperature-only case. Polarization is a straightforward
generalization and is presented in Section~\ref{sec:variance_polarization}. Then in
Section~\ref{sec:detection_significance} we combine these results to determine the significance with which the lensing bispectrum may be
detected. In Section~\ref{sec:joint_analysis} we generalize our results further, to the case where the lensing bispectrum is used in a joint
analysis with other bispectra, as a source of bias to be subtracted or marginalized over.

\subsection{Temperature}
\label{sec:variance_temperature}

Equation~(\ref{opt_est}) for the amplitude of the lensing bispectrum
from temperature data can be rewritten as
\begin{equation}
\hat{S} = \frac{1}{F} \sum_{l_1 m_1} \frac{\tilde{T}_{l_1 m_1}}{\tCtot^{TT}{}_{l_1}}
\left[\sum_{l_2 l_3}^{l_1 \leq l_2 \leq l_3} \Delta_{l_1 l_2 l_3}^{-1} B_{l_1 l_2 l_3}
\sum_{m_2 m_3} \threej{l_1}{l_2}{l_3}{m_1}{m_2}{m_3}
\frac{\tilde{T}_{l_2 m_2}}{\tCtot^{TT}{}_{l_2}}
\frac{\tilde{T}_{l_3 m_3}}{\tCtot^{TT}{}_{l_3}} \right].
\end{equation}
The term in square brackets is (proportional to) a quadratic estimator
for the lensing potential $\psi_{l_1 m_1}^*$~\cite{Okamoto03}. To see this, consider
taking the expectation value of this term over noise, small-scale modes of the
unlensed temperature, and the modes of the lensing potential with
$(lm) \neq (l_1 m_1)$. We have that
\ba
 \sum_{m_2 m_3} \threej{l_1}{l_2}{l_3}{m_1}{m_2}{m_3} \la \Ltemp_{l_2 m_2} \Ltemp_{l_3 m_3}\ra_{(lm)\ne (l_1m_1)} &\approx&
 \sum_{m_2 m_3} \threej{l_1}{l_2}{l_3}{m_1}{m_2}{m_3}
\left\la \frac{\delta}{\delta \psi_{l_1 m_1}^*}  \left(\Ltemp_{l_2 m_2} \Ltemp_{l_3 m_3}\right)\right\ra \psi_{l_1 m_1}^*\nonumber\\
&\equiv&
\frac{\cla^{TT}_{l_1 l_2 l_3}}{2l_1+1}  \psi^*_{l_1 m_1} ,
\ea
which is correct to first order in modes of $\psi$ at $l_1$ and
non-perturbatively correct in its other modes. Here,
\be
\cla^{TT}_{l_1 l_2 l_3} = \tilde{C}_{l_3}^{\temp\temp} F^0_{l_2 l_1 l_3} + \tilde{C}_{l_2}^{\temp\temp} F^0_{l_3 l_1 l_2} ,
\ee
which is related to the squeezed limit of the lensing bispectrum by
$B_{l_1 l_2 l_3} = C_{l_1}^{T\psi} \cla^{TT}_{l_1 l_2 l_3}$.
It follows that there is a quadratic estimator for $\psi_{l_1 m_1}$ of
the form
\be
\hat{\psi}_{l_1 m_1}^* = N_{l_1}^{(0)} \sum_{l_2 l_3}^{l_1 \leq l_2 \leq l_3}
\Delta_{l_1 l_2 l_3}^{-1}
\cla^{\temp\temp}_{l_1 l_2 l_3}  \sum_{ m_2 m_3} \threej{l_1}{l_2}{l_3}{m_1}{m_2}{m_3}
\frac{ \Ltemp_{l_2 m_2}\Ltemp_{l_3 m_3}}{\tCtot^{\temp\temp}{}_{l_2}\tCtot^{\temp\temp}{}_{l_3}} ,
\label{psiest}
\ee
where
\be
[N^{(0)}_{l_1}]^{-1} \equiv \frac{1}{2l_1+1}\sum_{l_2 l_3}^{l_1 \leq l_2 \leq l_3}
\Delta_{l_1 l_2 l_3}^{-1}
\frac{
[\cla^{\temp\temp}_{l_1 l_2 l_3}]^2}{\tCtot^{\temp\temp}{}_{l_2}\tCtot^{\temp\temp}{}_{l_3}} .
\ee
This estimator satisfies $\langle \hat{\psi}_{l_1 m_1} \rangle_{(lm)\neq
(l_1 m_1)} = \psi_{l_1 m_1}$ to first-order in $\psi_{l_1 m_1}$ but is
non-perturbatively correct in the $l\neq l_1$ modes of $\psi$.
It is a non-perturbative version of the usual quadratic estimator~\cite{Okamoto03}, avoiding the low-$l$ (`$N^{(2)}$') bias in the standard estimator that was identified by Ref.~\cite{Hanson:2010rp} and generalizing the perturbative corrections of Ref.~\cite{Hanson:2010rp} to a non-perturbative form by simply using the lensed small-scale power spectra in the filter functions.
In the
Gaussian limit, the variance of the estimator is simply $N^{(0)}_l$ and the
weighting in $l_2$ and $l_3$ in Eq.~(\ref{psiest}) can be shown to minimise
this Gaussian variance subject to the estimator being unbiased.

We can now rewrite the estimator of the bispectrum amplitude in
terms of the $\psi$ reconstruction as
\be
\hat{S} = \frac{1}{F} \sum_{l_1 m_1} C_{l_1}^{T\psi}
\frac{\tilde{T}_{l_1 m_1}}{\tCtot^{TT}{}_{l_1}}\frac{\hat{\psi}^*_{l_1 m_1}}{N_{l_1}^{(0)}} ,
\label{eq:standardS}
\ee
and the normalization
\be
F \approx \sum_{l_1} (2l_1 +1)\left(C_{l_1}^{T\psi}\right)^2 \left(\tCtot^{TT}{}_{l_1}\right)^{-1}
\left(N_{l_1}^{(0)}\right)^{-1}.
\ee
Recasting the bispectrum estimator in this form leads directly to an understanding of the contribution to the error from
signal variance. The estimator $\hat{S}$ depends on the empirical
cross-power $\hat{C}_{l_1}^{T\psi}$ between the $\psi$ reconstruction and
the large-scale observed temperature:
\be
\hat{C}_{l_1}^{T\psi} \equiv \frac{1}{2l_1+1}\sum_{m_1} \tilde{T}_{l_1 m_1}
\hat{\psi}^*_{l_1 m_1} \equiv \hat{S}_{l_1} C_{l_1}^{T\psi}.
\ee
Since $\langle \hat{C}_{l_1}^{T\psi} \rangle= C_{l_1}^{T\psi}$, each $\hat{S}_{l_1}$ is
an unbiased estimate of the bispectrum amplitude.
As with any other power spectrum estimator,
$\hat{C}^{\temp\psi}_{l_1}$ has uncertainty both from reconstruction noise and from  signal/cosmic variance:
\be
\text{var}\, \hat{C}_{l_1}^{T\psi} \approx \frac{1}{2l_1+1} \left[
  \tCtot^{TT}{}_{l_1}
(C_{l_1}^{\psi\psi} + N_{l_1}^{(0)}) + \left(C_{l_1}^{T\psi}\right)^2\right] ,
\label{eq:adc4}
\ee
so
\be
\var\, \hat{S}_{l_1} = \frac{1 + r_{l_1}^{-2} }{2l_1+1}  + \frac{1}{F_{l_1}},
\ee
where $r_l\equiv C^{\temp\psi}_{l}/\sqrt{\tCtot^{\temp\temp}{}_{l} C^{\psi\psi}_{l}}$ and the
 usual zero-signal variance term is $1/F_l = \tCtot^{TT}{}_l N_l^{(0)}/[(2l+1)(C_l^{T\psi})^2]$ which comes from the term involving $N_{l_1}^{(0)}$ in Eq.~(\ref{eq:adc4}).
In the standard estimator, Eq.~(\ref{eq:standardS}),
the $\hat{S}_{l_1}$ are weighted with the $F_{l_1}$ and the normalisation
is accordingly $F = \sum_{l_1} F_{l_1}$. In the presence of a non-zero signal,
we can reduce the variance by weighting the $\hat{S}_{l_1}$ with the full
inverse variance. This defines a lower-variance estimator for the
bispectrum amplitude,
\be
\hat{S} = \frac{1}{\clf}\sum_{l_1} \left(\frac{1+r_{l_1}^{-2}}{2l_1+1} + F_{l_1}^{-1}\right)^{-1} \hat{S}_{l_1},
\label{shat}
\ee
which has variance given by $\clf^{-1}$ where
\be
\clf = \sum_{l_1} \left(\frac{1+r_{l_1}^{-2}}{2l_1+1} + F_{l_1}^{-1}\right)^{-1} .
\label{shatf}
\ee
When $F_{l_1}$ is large, so that the lensing modes $\psi_{lm}$ are reconstructed with a small error, the contribution of the signal variance terms $1+r_{l_1}^{-2}$ become important, ensuring that the total signal-to-noise never exceeds that expected from the cosmic-variance limit on the cross-correlation.
Neglect of the signal variance term would lead to an overestimation of the significance for a detection of the lensing bispectrum
(a similar effect happens with primordial
non-Gaussianities\footnote{To account for the signal variance we have
  used an $l_1$-dependent weighting in Eqs.~\eqref{shat} and \eqref{shatf}; Creminelli et al.~\cite{Creminelli:2006gc} use a single realization-dependent change to the overall estimator normalization, which should be less optimal. The argument for lensing here can straightforwardly be generalized for estimation of local non-Gaussianity, using a quadratic estimator for the small-scale primordial power modulation rather than the lensing potential~\cite{Hanson:2009gu}; the corresponding estimator may be a fast nearly-optimal alternative to a fully Bayesian method~\cite{Elsner:2010hb} if the non-Gaussianity were large.} ~\cite{Creminelli:2006gc}).

The optimal Fisher variance in Eq.~\eqref{shatf} can easily be related to that for an optimal measurement of the cross-correlation, giving
\be
\clf =  \sum_{l_1} \frac{(2{l_1}+1)}{1+\tCtot^{TT}{}_{l_1}(C_{l_1}^{\psi\psi}+N^{(0)}_{l_1})/\left(C_{l_1}^{\temp\psi}\right)^2 }.
\label{fisher_variance}
\ee
This is exactly the same result as obtained from an optimal estimator of the amplitude of the cross-correlation $C^{\temp\psi}_l$ using the estimator $\hat{\psi}_{l_1 m_1}$ for the lensing potential.

Note that here we have only discussed the optimal estimator from the measured cross-correlation (bispectrum). If auto-spectra (power spectrum and lensing trispectrum) are also included the variance can be reduced further\footnote{We thank
the referee for raising this point.}; however since the correlation $r$ is always $r<0.5$, and $r\alt 0.2$ where the signal-to-noise peaks even with no noise, the gain from a more optimal joint estimator is rather modest, being $\clo(r)$. We do not discuss joint estimators further here, but a likelihood analysis of actual data should of course properly account for the full covariance structure of the estimators used.

\subsection{Polarization}
\label{sec:variance_polarization}

The arguments above carry over quite directly to polarization.
The original estimator for the bispectrum amplitude,
Eq.~(\ref{opt_est}), can be written as
\begin{equation}
\hat{S} = \frac{1}{F} \sum_{l_1 m_1} C_{l_1}^{a^i \psi}
(\tCtot^{-1})_{l_1}^{ip} \tilde{a}^p_{l_1 m_1} \left(N_l^{(0)}\right)^{-1} \hat{\psi}^*_{l_1 m_1},
\label{eq:newpol1}
\end{equation}
which involves the quadratic estimator
\begin{equation}
\hat{\psi}_{l_1 m_1}^* = N_{l_1}^{(0)} \sum_{l_2 l_3}^{l_1 \leq l_2 \leq l_3}
\Delta_{l_1 l_2 l_3}^{-1} (\cla_{l_1 l_2 l_3}^{jk})^* \sum_{m_2 m_3}
\threej{l_1}{l_2}{l_3}{m_1}{m_2}{m_3} (\tCtot^{-1})_{l_2}^{jq} \tilde{a}^q_{l_2 m_2}
(\tCtot^{-1})_{l_3}^{kr} \tilde{a}^r_{l_3 m_3} ,
\label{eq:pol-psiest}
\end{equation}
where the normalisation
\begin{equation}
[N_{l_1}^{(0)}]^{-1} \equiv \frac{1}{2l_1 +1}\sum_{l_2 l_3}^{l_1 \leq l_2 \leq l_3}
\Delta_{l_1 l_2 l_3}^{-1} (\cla_{l_1 l_2 l_3}^{jk})^* (\tCtot^{-1})_{l_2}^{jq} (\tCtot^{-1})_{l_3}^{kr} \cla_{l_1 l_2 l_3}^{qr} .
\end{equation}
The overall normalization of $\hat{S}$ can be rewritten as
\begin{equation}
F = \sum_{l_1} (2l_1+1)C_{l_1}^{a^i \psi} (\tCtot^{-1})_{l_1}^{ip} C_{l_1}^{a^p \psi}
\left(N_{l_1}^{(0)}\right)^{-1} .
\end{equation}
Here, in the approximations of Sec.~\ref{full-sky},
\begin{equation}
\cla_{l_1 l_2 l_3}^{jk} = \left(\tilde{C}_{l_2}^{a^k a^j} F^{s_k}_{l_3 l_1 l_2}
+ \tilde{C}_{l_3}^{a^j a^k} F^{s_j}_{l_2 l_1 l_3}\right)
+ i \left(\tilde{C}_{l_2}^{\bar{a}^k a^j} F^{-s_k}_{l_3 l_1 l_2}
- \tilde{C}_{l_3}^{\bar{a}^j a^k} F^{-s_j}_{l_2 l_1 l_3}\right) ,
\end{equation}
and is related to the squeezed limit of the bispectrum by
$B^{ijk}_{l_1 l_2 l_3} = C_{l_1}^{a^i \psi} \cla_{l_1 l_2 l_3}^{jk}$.
As for temperature, the quadratic estimator $\hat{\psi}_{l_1 m_1}$
is a non-perturbative generalization of those constructed in~\cite{Okamoto03}.
Averaging over small-scale unlensed
CMB modes and the modes of $\psi$ with $(lm) \neq (l_1 m_1)$ returns
$\psi_{l_1 m_1}$ to first-order since
\ba
 \sum_{m_2 m_3} \threej{l_1}{l_2}{l_3}{m_1}{m_2}{m_3} \la \tilde{a}^j_{l_2 m_2} \tilde{a}^k_{l_3 m_3}\ra_{l\ne l_1} &\approx&
 \sum_{m_2 m_3} \threej{l_1}{l_2}{l_3}{m_1}{m_2}{m_3}
\left\la \frac{\delta}{\delta \psi_{l_1 m_1}^*}  \left(\tilde{a}^j_{l_2 m_2} \tilde{a}^k_{l_3 m_3}\right)\right\ra \psi_{l_1 m_1}^*\nonumber\\
&\equiv& \frac{\cla_{l_1 l_2 l_3}^{jk}}{2l_1+1} \psi^*_{l_1 m_1}.
\label{quadestavpol}
\ea
The weighting in Eq.~(\ref{eq:pol-psiest}) minimises the Gaussian variance
which is simply $N_{l_1}^{(0)}$.

The data enters the estimator, Eq.~(\ref{eq:newpol1}), through
the empirical cross spectrum,
\begin{equation}
\hat{C}_{l_1}^{a^p \psi} \equiv \frac{1}{2l_1+1}\sum_{m_1} \tilde{a}^p_{l_1 m_1}
\hat{\psi}^*_{l_1 m_1} \equiv \hat{S}^p_{l_1} C_{l_1}^{a^p \psi} .
\end{equation}
As with temperature, there is a signal contribution to the (co)variance, since
\ba
\text{Cov}(\hat{S}_{l_1}^{i},\hat{S}_{l_1}^{p}) &\approx&
\frac{\left(C_{l_1}^{a^i \psi} C_{l_1}^{a^p \psi}+\tCtot^{ip}{}_{l_1} C_{l_1}^{\psi\psi}\right)}{(2l_1+1) C_{l_1}^{a^i \psi} C_{l_1}^{a^p \psi}}
+ \frac{\tCtot^{ip}{}_{l_1} N_{l_1}^{(0)}}{(2l_1+1) C_{l_1}^{a^i \psi} C_{l_1}^{a^p \psi}} \nonumber \\
&\equiv& \frac{\left(C_{l_1}^{a^i \psi} C_{l_1}^{a^p \psi}+\tCtot^{ip}{}_{l_1} C_{l_1}^{\psi\psi}\right)}{(2l_1+1) C_{l_1}^{a^i \psi} C_{l_1}^{a^p \psi}}
+ (\bar{F}_{l_1}^{-1})^{ip} .
\ea
The Gaussian contribution, $ (\bar{F}_{l_1}^{-1})^{ip}$, to the covariance
can be rewritten as the inverse of%
\be
\bar{F}^{ip}_{l_1} = \sum_{l_2 l_3}^{l_1 \leq l_2 \leq l_3} \Delta^{-1}_{l_1 l_2 l_3} (B^{ijk}_{l_1 l_2 l_3})^* (\tCtot^{-1})^{ip}_{l_1}(\tCtot^{-1})^{jq}_{l_2}(\tCtot^{-1})^{kr}_{l_3} B^{pqr}_{l_1 l_2 l_3} \qquad \mbox{(no sum on $i$ or $p$)} .
\label{eq:newpol2}
\ee
The original bispectrum estimator
$\hat{S} = F^{-1} \sum_{l_1 i p}\bar{F}^{ip}_{l_1} \hat{S}_{l_1}^p$
and the normalization can be written $F = \sum_{l_1 i p } \bar{F}_{l_1}^{ip}$.
The $\hat{S}_{l_1}^p$ are weighted by their zero-signal inverse covariance;
we can improve on this by using the inverse of the full
$\text{Cov}(\hat{S}_{l_1}^{i},\hat{S}_{l_1}^{p})$. The resulting
estimator has variance $\clf^{-1}$, where
\ba
\clf &=& \sum_{l_1 i p} \text{Cov}^{-1}(\hat{S}_{l_1}^{i},\hat{S}_{l_1}^{p}) \nonumber \\
&=& \sum_{l_1 i p} \left[(\bar{F}_{l_1}^{-1})^{ip} + \frac{\left(C_{l_1}^{a^i \psi} C_{l_1}^{a^p \psi}+\tCtot^{ip}{}_{l_1} C_{l_1}^{\psi\psi}\right)}{(2l_1+1) C_{l_1}^{a^i \psi} C_{l_1}^{a^p \psi}} \right]^{-1} ,
\label{fisher_lens_signal}
\ea
where the matrix inverse is taken of the term in square brackets in the second
line.

\subsection{Detection significance}
\label{sec:detection_significance}

\begin{figure}
\includegraphics[width=10cm]{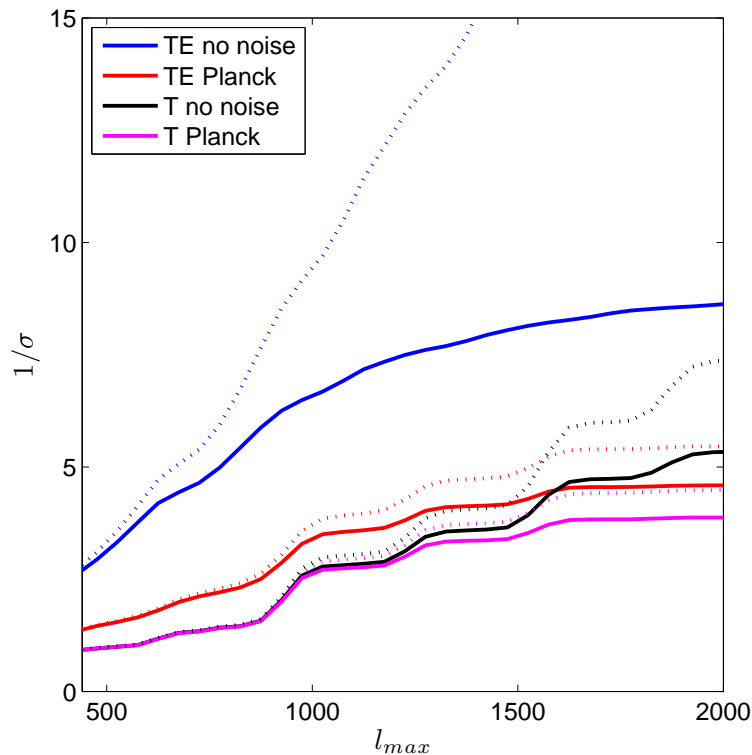}
\caption{Fisher `detection significance' (defined by $1/\sigma = \clf^{1/2}$, Eq.~\eqref{fisher_lens_signal}) of the CMB lensing bispectrum
  as a function of $\lmax$ for no instrument noise and for Planck-like
  noise levels using just the temperature bispectrum (black for no
  noise; magenta for Planck) or using all the $T$ and $E$-polarization
  bispectra
(blue for no noise; red for Planck).
The dotted lines show the (incorrect) results obtained if the signal contribution to the variance is neglected: all results are bounded by the cosmic-variance detection limit on a measurement of the low-$l$ cross-correlation spectra $C_l^{\temp\psi}$ and $C_l^{E\psi}$.
If the cross-correlation part of the signal variance is neglected (as for a null hypothesis test), the `significance' is $\alt 5\%$ larger.
}
\label{detect_sigma}
\end{figure}

Collecting our results for temperature and polarization, in
Fig.~\ref{detect_sigma} we plot the expected detection significance of
the CMB lensing bispectrum as a function of the maximum observed
multipole $\lmax$, assuming noise-free data and Planck-like noise
levels.
Here, we have only included the $T$ and $E$-polarization
bispectra. Including $B$-mode spectra is not expected to improve the
variance significantly for large $l_{\text{max}}$: despite the $E$--$B$ estimator being the most
powerful for reconstructing the lensing potential in the absence of
noise~\cite{Hu:2001kj}, using only noise-free $T$ and $E$ the statistical noise in the
reconstruction is already below cosmic variance on large scales.
The dotted and solid lines in Fig.~\ref{detect_sigma} give the results
using the zero-signal Fisher error and the result including the
additional cosmic variance due to the signal, respectively. The
detection significance is bounded by the cosmic-variance limit on the
detection of the lensing-potential cross-correlation power
spectra. Planck should see the lensing bispectrum at about $5\sigma$,
while a zero-noise experiment should get very close to the cosmic variance limit of about $9\sigma$ using temperature and polarization.

\begin{figure}
\includegraphics[width=10cm]{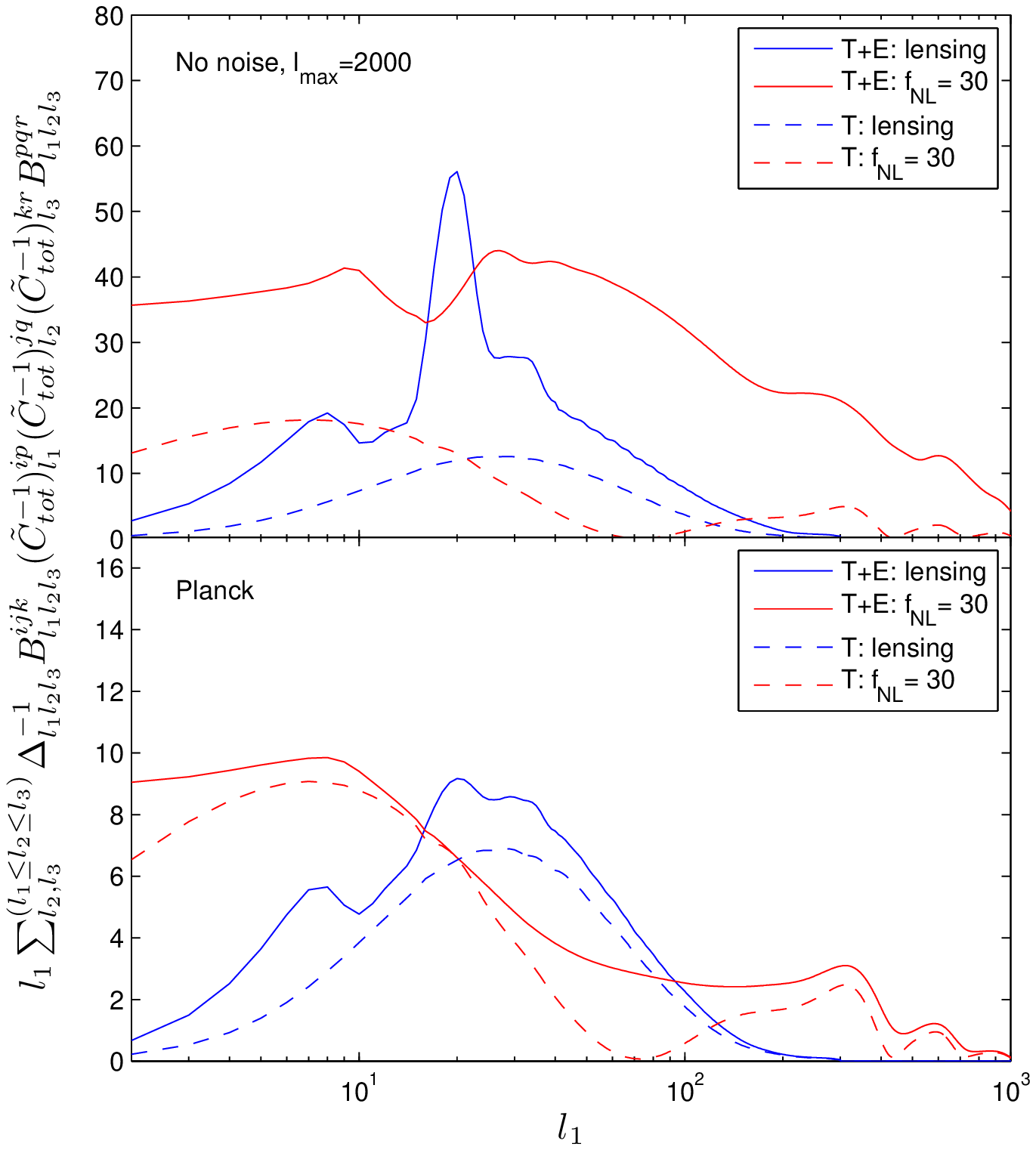}
\caption{Contributions (per $\ln l_1 $) to the Fisher inverse variance
  as a function of $l_1$ assuming no noise (top) and Planck-like noise (bottom).
Blue lines are for the lensing bispectrum using only $T$ (dashed) and
$T$ and $E$-polarization (solid). For comparison, the contributions to the
inverse variance of the local-model bispectrum with $\fnl = 30$ are shown in red.
For such a large $\fnl$ (chosen for clarity in the plot),
signal-variance contributions might be important but are not included
here; however, the lensing results do include signal variance
(in addition to the zero-signal term in the $y$-axis label).
}
\label{FisherContribs}
\end{figure}

Figure~\ref{FisherContribs} shows the contributions to the
signal-to-noise as a function of $l_1$, the largest-scale mode. The distinctly different dependence from the local $\fnl$
contributions is clear. The signal-to-noise peaks for $l_1\sim 20$,
which is a trade-off between the $C_l^{T\psi}$ and $C_l^{E\psi}$ contributions falling rapidly with $l$, and the small number of modes at low $l_1$.

\subsection{Joint analysis of different bispectra}
\label{sec:joint_analysis}

\begin{table}
\centering
\begin{tabular} {| c | c | c | c | c | c |}
\hline
& $\sigma_{\fnl}$ &  $\sigma_{\lens}$ & correlation & bias on $\fnl$ &  $\sigma_{\fnl}^{\text{marge}}$ \\ [0.5ex]
\hline
T & 4.31 & 0.19 & 0.24 & 9.5 & 4.44   \\
T+E & 2.14 & 0.12 & 0.022 & 2.6 & 2.14 \\
Planck T & 5.92 & 0.26 & 0.22 & 6.4 & 6.06\\
Planck T+E & 5.19 & 0.22 & 0.13 & 4.3 &  5.23 \\
\hline
\end{tabular}
\caption{
Errors and biases on CMB lensing and primordial local-model
non-Gaussianity parameterized by $\fnl$ for Planck-like noise
(assuming isotropic coverage over the full sky with
sensitivity $\Delta T = \Delta Q/2 = \Delta U/2 = 50\,\mu\text{K\,arcmin}$ [$N_l^{\temp}=N_l^{\rm{Q/U}}/4=2\times 10^{-4}\muK^2$]
and a beam FWHM of $7\, \rm{arcmin}$) or cosmic-variance limited data
with $\lmax=2000$. From Eq.~\eqref{eq:optfisher} the errors
$\sigma_{\fnl}$ and $\sigma_{\lens}$ are the errors on the amplitudes
of the corresponding bispectrum templates individually when the other
one is fixed;  $\sigma_{\fnl}^{\text{marge}}$ is the Fisher error on
$\fnl$ if the amplitude of the lensing contribution is marginalized
over; and the correlation is that between the two bispectrum shapes.
The bias is the systematic error on $\fnl$ if the CMB lensing
contribution is neglected, i.e.\ Eq.~\eqref{eq:bias}.
}
\label{table:biases}
\end{table}


The primary objective of non-Gaussianity searches is to look for primordial non-Gaussianity, which is expected to be small in many models. If we have a template primordial bispectrum $\tB$, the optimal estimator for small levels of non-Gaussianity would obtain a mean contribution due to CMB lensing given by
\be
\la \hat{f}_{\rm{NL}}\ra_{\text{lens}}= \frac{F_0(\tB^{\text{lens}},\tB)}{ F_0(\tB,\tB) } ,
\label{eq:bias}
\ee
where the Fisher matrix for bispectra with zero expected signal is given by~\cite{Smith:2006ud}
\be
F_0^{ab}\equiv F(\tB^a,\tB^b) = \frac{1}{6}\sum_{l_1 l_2 l_3}(B^{a,ijk}_{l_1 l_2 l_3})^* (\tCtot^{-1})^{ip}_{l_1}(\tCtot^{-1})^{jq}_{l_2}(\tCtot^{-1})^{kr}_{l_3} B^{b,pqr}_{l_1 l_2 l_3}.
\label{fish_correlation}
\ee
This lensing bispectrum contribution to the primordial signal estimator must be carefully distinguished in order not to obtain a spurious detection. Whether the bias is large or not depends on how similar the shapes of the two bispectra are. Since the CMB lensing bispectrum is mostly in squeezed triangles, the bias is expected to be most significant for nearly-local forms of primordial non-Gaussianity:
there is an expected bias of $\la \hat{f}_{\rm{NL}} \ra_{\text{ lens}}=9$ using cosmic-variance limited temperature data to $\lmax=2000$~\cite{Hanson:2009kg}.
Provided that the underlying cosmology is well understood,
this bias may simply be subtracted. Even from current WMAP constraints, $C_l^{T\psi}$
is constrained to $\clo(10\%)$ if we assume a flat, $\Lambda$CDM cosmology with massive neutrinos\footnote{More rigorously, the standard deviation for independent samples of $C_l^{T\psi}$ from a MCMC exploration of the WMAP+SZ+LENS seven-year data likelihood using \textsc{cosmomc}~\cite{Lewis:2002ah} is $\sim10\%$ for $l \leq 200$. The primary uncertainty is in the amplitude of the correlation rather than the shape.}.
As discussed in Refs.~\cite{Babich:2004yc,Hanson:2009kg} the signal
variance of the CMB lensing bispectrum will also slightly increase the
variance of other non-Gaussianity estimators, even if we assume that
the shape and amplitude of the lensing bispectrum are perfectly
known. Suppose that one obtains an unbiased estimate of some other
bispectrum $B$ by subtracting out the expected lensing contribution.
If the estimator is not weighted accounting for the signal variance
from lensing, i.e.\ constructed simply using Eq.~\eqref{opt_est}, then
the variance of the bispectrum amplitude, $f_{\rm{NL}}$, for squeezed
shapes is
\be
\var \,\hat{f}_{\rm{NL}} = \left( \sum_{l_1} F_{l_1}^{B
    B}\right)^{-1} +  \left( \sum_{l_1} F_{l_1}^{B
    B}\right)^{-2}\sum_{l_1} \left(F^{B \cla}_{l_1}\right)^2
\frac{\tCtot^{\temp\temp}{}_{l_1} C^{\psi\psi}_{l_1}+ (C^{\temp\psi}_{l_1})^2}{2l_1+1}
,
\label{eqn:varsubopt}
\ee
where
\be
F_{l_1}^{ab} \equiv \sum_{l_2,l_3}^{(l_1\le l_2 \le l_3)} \Delta^{-1}_{l_1 l_2 l_3} B^a_{l_1 l_2 l_3}
[\tCtot^{\temp\temp}]^{-1}_{l_1}[\tCtot^{\temp\temp}]^{-1}_{l_2}[\tCtot^{\temp\temp}]^{-1}_{l_3}
B^b_{l_1 l_2 l_3}.
\ee
The first term in Eq.~(\ref{eqn:varsubopt}) is the usual zero-signal
variance while the second term gives an additive contribution from CMB lensing.
For cosmic-variance limited temperature data to $\lmax=2000$, the analytical increase in the local $\fnl$ variance is from $17.2$ to $19.4$, in good agreement with the result found by direct simulation in Ref.~\cite{Hanson:2009kg}, corresponding to a $\sim 5\%$ increase in the $\fnl$ error bar. For Planck the increase is smaller, $\sim 1\%$.
CMB lensing also affects the primordial bispectrum itself, by smoothing out the acoustic peaks, which would give some additional correction to the expected constraint that we have neglected here, but which is discussed in Ref.~\cite{Hanson:2009kg}.

If instead we re-weight the bispectrum estimator in $l_1$ to suppress the cosmic variance,  then for squeezed shapes the modified variance will be given by $\clf_{BB}^{-1}$ where
\be
\clf_{BB} = \sum_{l_1} \left(\left[\frac{F^{B\cla}_{l_1}}{F^{BB}_{l_1}}\right]^2
\frac{\tCtot^{\temp\temp}{}_{l_1} C^{\psi\psi}_{l_1}+ [C^{\temp\psi}_{l_1}]^2}{2l_1+1}
+ \frac{1}{F_{l_1}^{BB}}\right)^{-1} .
\ee
For cosmic-variance limited temperature data to $\lmax=2000$, the optimized $\fnl$ variance
variance is $18.6$, rather smaller than the sub-optimal value above. However even with the optimized weighting, lensing still degrades the constraint since the optimized weighting is effectively decreasing the amount information available in the largest-scale modes.

As in the temperature case, CMB lensing also contributes to the variance of other non-Gaussianity estimators for polarization, where Eq.~\eqref{eqn:varsubopt} generalizes to
\be
\var \,\hat{f}_{\rm{NL}} = \var \,\hat{f}_{\rm{NL}}|_{\psi=0} +  \left( \var \,\hat{f}_{\rm{NL}}|_{\psi=0}\right)^{2}\sum_{l_1,ijpq}
\bar{F}^{B\cla (pi)}_{l_1}
\left[\frac{C_{l_1}^{a^i\psi}C_{l_1}^{a^j\psi}  + \tCtot^{a^i a^j}{}_{l_1}
    C_{l_1}^{\psi\psi}}{2l_1+1}\right] \bar{F}^{\cla j, B q}_{l_1}.
\ee
Here,
\be
\bar{F}^{ai,bp}_{l_1} = \sum_{l_2 l_3}^{l_1 \leq l_2 \leq l_3} \Delta^{-1}_{l_1 l_2 l_3} (B^{a,ijk}_{l_1 l_2 l_3})^* (\tCtot^{-1})^{ip}_{l_1}(\tCtot^{-1})^{jq}_{l_2}(\tCtot^{-1})^{kr}_{l_3} B^{b,pqr}_{l_1 l_2 l_3} \qquad \mbox{(no sum on $i$ or $p$)},
\ee
generalizes the Gaussian inverse covariance of Eq.~(\ref{eq:newpol2}) to
analyses of joint bispectra, with an $\cla$ component making the replacement $B^{ijk}_{l_1 l_2 l_3} \rightarrow \cla^{jk}_{l_1 l_2 l_3}$.
The fractional increase in the $\fnl$ error bar is similar to the temperature case.

As the detailed shape of the lensing bispectrum is quite different from that of primordial non-Gaussianity, the correlation between the two estimators is small and the amplitudes of the lensing and primordial bispectra may also be fit for simultaneously, rather than simply assuming a fiducial lensing bispectrum. This is a safer method to correct for the lensing bispectrum, although it may result in slightly larger error bars (one could view this as a trade-off between systematic and experimental errors).
%
If the amplitudes of multiple bispectra are estimated simultaneously,
then the optimal estimator for the amplitude of $B^a$ with the first
field at scale $l_1$ and of type $h$ can be written (in the approximation in which the inverse-variance filtering is given by the lensed power spectrum) as
\be
\hat{S}^{ah}_{l_1} = \sum_{l_2l_3}^{(l_1\le l_2 \le l_3)} [\bar{F}^{-1}_{l_1}]^{ah,bi}  \Delta^{-1}_{l_1 l_2 l_3} (B^{b,ijk}_{l_1 l_2 l_3})^*
[\tCtot^{-1}]^{ip}_{l_1}[\tCtot^{-1}]^{jq}_{l_2}[\tCtot^{-1}]^{kr}_{l_3}
\hat{B}^{pqr}_{l_1 l_2 l_3},
\ee
and the covariance will have contributions from the lensing variance
\be
\cov(\hat{S}^{ai}_{l_1},\hat{S}^{bj}_{l_1})  =
[\bar{F}_{l_1}^{-1}]^{ai,bj} +  \delta^{a,\lens}\delta^{b,\lens}
\left[\frac{C_{l_1}^{a^i\psi}C_{l_1}^{a^j\psi}  + \tCtot^{ij}{}_{l_1}
    C_{l_1}^{\psi\psi}}{(2l_1+1)C_{l_1}^{a^i\psi}C_{l_1}^{a^j\psi}}\right] ,
\ee
in agreement with the previous result when just estimating the lensing
bispectrum amplitude. Combining the information in all $l_1$ and in all the fields, the overall optimal joint estimators therefore have Fisher matrix
\ba
F^{ab} &=& \sum_{l_1,ij}\left([\bar{F}_{l_1}^{-1}]^{ai,bj} +  \delta^{a,\lens}\delta^{b,\lens} \left[\frac{C_{l_1}^{a^i\psi}C_{l_1}^{a^j\psi}  + \tCtot^{ij}{}_{l_1} C_{l_1}^{\psi\psi}}{(2l_1+1)C_{l_1}^{a^i\psi}C_{l_1}^{a^j\psi}}\right]\right)^{-1} ,
\label{eq:optfisher}
\ea
where the inversion is of a large matrix with lumped indices $ai$ and $bj$.
This can be used to calculate the correlation between estimators, and assess the errors on individual bispectrum amplitudes with and without marginalization over uncertainty in the amplitudes of the other bispectra: $1/F^{aa}$ gives the variance on bispectrum $B^a$ if the other bispectra are fixed (known amplitude); $[F^{-1}]^{aa}$ gives the variance if the other bispectrum amplitudes are marginalized over. In practice Eq.~\eqref{eq:optfisher} involves inversion of singular matrices, and should actually be evaluated using matrices with $C_l^{a^i \psi}$ factored out; lensing alters the error on other forms of non-Gaussianity even when $C_l^{a^i \psi}$ (and hence the corresponding lensing bispectrum) is zero.
\dmh{I have difficulty understanding this-- why is the inversion inside the $l_1$, $ij$ sum?}

With only temperature data the correlation between primordial local non-Gaussianity and lensing is $\sim 0.22$ for Planck, which means that the error on $\fnl$ increases only by a factor $\sqrt{[F^{-1}]^{BB}/[F^{BB}]^{-1}}\sim 2\%$ if
the amplitude of the lensing bispectrum is marginalized over (assuming the template is fixed). If polarization data is included the bispectrum shapes are even more different, so the correlation is even smaller. Table~\ref{table:biases} contains a summary of results for cosmic-variance limited and Planck data, using temperature-only or temperature and $E$-polarization.




\section{Conclusions}
\label{conclusions}

We can summarize our current understanding of the lensing bispectra as follows.
\begin{enumerate}
\item The late-time integrated Sachs-Wolfe effect gives a significant
  CMB temperature bispectrum. The signal is mostly in squeezed
  triangles, and strong enough to bias estimators of local
  non-Gaussianity if unaccounted for. The signal is thought to be well
  constrained even with current data, however, and also has a
  distinctive shape and scale dependence compared to that expected from any
  adiabatic primordial non-Gaussianity model, which allows the two
  bispectra also to be separated in data.
\item The $E$-polarization from reionization is correlated with the lensing potential and gives a contribution to the CMB polarization bispectra. This is in principle detectable at $\sim 2.5\sigma$, and, like the temperature lensing bispectrum, can bias polarization estimators for primordial non-Gaussianity if not accounted for. The signal is mostly on very large scales, falling rapidly with decreasing size of the largest mode.
\item The leading-order perturbative calculation for the CMB lensing bispectrum is inaccurate at the $\sim 10\%$ level. There is a simple non-perturbative approximate calculation that agrees well with simulations, and is will approximated by using lensed rather than unlensed small-scale power spectra in the standard lowest-order result.
\item The covariance of the CMB lensing bispectrum has significant
  contributions from the signal variance, which is easily understood
  by thinking of the bispectrum as the empirical cross-correlation of
  the temperature with a quadratic estimator for the lensing
  potential. The detection significance of the lensing bispectra is limited by the cosmic-variance limit on the detection of the cross-correlation power spectra $C^{T\psi}_l$ and $C^{E\psi}_l$, about $9\sigma$ in total.
\end{enumerate}
We have developed a numerical code to calculate the CMB lensing
potential cross-correlation power spectra and the lensing and
local-model primordial bispectra, which is publicly available as part
of CAMB\footnote{\url{http://camb.info}}. The public
LensPix\footnote{\url{http://cosmologist.info/lenspix/}} code can be
used to simulate the properly correlated unlensed fields as well as the lensed CMB temperature and polarization.

Note that although we have focused in this work on a CMB-only analysis of the lensing temperature and polarization bispectra, it is also possible to form bispectra involving mixtures of the CMB and other observables, for example $B^{\Delta\temp\temp}$ where $\Delta$ is some tracer of large-scale structure. CMB-large-scale-structure bispectra have already been detected, giving the first signatures of the CMB lensing effect~\cite{Smith:2007rg,Hirata:2008cb}, and a joint analysis with the other bispectra can slightly improve constraints on primordial non-Gaussianity~\cite{Mead:2010bv}. If $\Delta$ is only used on relatively large scales and can be treated as a Gaussian field, then our previous results immediately apply to these other forms of bispectra, for example on the flat sky the combination of Eq.~\eqref{eqn:short_leg_bispec} and Eq.~\eqref{eqn:lensedcovresponse} becomes
\be
\la \Delta(\vl_1) \tilde{\temp}(\vl_2) \tilde{\temp}(\vl_3)\ra \approx
-\frac{1}{2\pi} \delta(\vl_1+\vl_2+\vl_3) C^{\Delta\psi}_{l_1} \left[ (\vl_1\cdot \vl_2)  \Cgrads_{l_2} + (\vl_1\cdot \vl_3)  \Cgrads_{l_3}\right].
\ee
Thus non-perturbative corrections can also be important for the
cross-correlation bispectra. As mentioned in the context of
Eq.~\eqref{psiest}, the corrections can easily be incorporated into
the quadratic estimator framework for lens reconstruction by suitably
substituting $\Cgrads$ for $C^{\temp\temp}$ in the weighting of the
observed temperature fields and the normalisation of the estimator.
As shown in Ref.~\cite{Hanson:2010rp} by a direct calculation, this modification to the estimator also
improves the bias properties of the estimated lensing power spectrum
on large scales.

We have focussed here on the linear contributions to the CMB-lensing potential correlation. Future work should also incorporate an accurate model of Rees-Sciama and SZ contributions~\cite{Verde:2002mu,Mangilli:2009dr,Calabrese:2009bu}, where the `unlensed short-leg' approximation may be less accurate. In addition to CMB lensing there are also other non-linear effects at recombination that can give rise to non-Gaussianity even in the absence of a primordial signal~\cite{Pitrou:2010sn}; future precision constraints on primordial non-Gaussianity should also model these, and there may also be some overlap between the shape of non-Gaussianities at recombination and CMB lensing (for example modulation of the sound horizon at recombination by large-scale modes looks rather similar to lensing magnification of the last-scattering surface).


\section{Acknowledgements}
AL thanks James Fergusson for help with Fig.~\ref{threedcontour}
and David Seery for discussion.
AL was supported by the Science and Technology Facilities Council (grant numbers ST/F002858/1 and PP/C001214/2).
Some of the results in this paper have been derived using HealPix~\cite{Gorski:2004by}.
Part of the research described in this paper was carried out at the Jet Propulsion Laboratory, California Institute of Technology, under a contract with the National Aeronautics and Space Administration.

\appendix

\section{$E$--$\psi$ correlation from reionization}
\label{app:Epsi}

For the calculations in the main text we compute the $E$--$\psi$ correlation
accurately using CAMB~\cite{Lewis:1999bs}. However, to build intuition,
we give here a simple analytic treatment under the assumption that
scattering at reionization is instantaneous.

Consider an electron at reionization in a direction $\hat{\vn}$; its position
relative to the origin (the observation point) is $\vx = \chire \hat{\vn}$
where $\chire \sim 10000\,\mathrm{Mpc}$ is the comoving distance to
reionization. Linear polarization is generated at reionization by scattering
of the local temperature quadrupole there. Reionization occurs well
before dark energy becomes dynamically relevant and so the temperature
quadrupole at reionization is simply a projection of the perturbations
on the electron's last-scattering surface (which has a comoving radius
$\Delta \chi \equiv \chi_* - \chire \approx 4200\, \mathrm{Mpc}$, where $\chi_* \approx 14200\,\mathrm{Mpc}$ is the radius of our last scattering
surface~\cite{Larson:2010gs}). The modes
that contribute most to the quadrupole at reionization have
wavenumber $k \approx 2/4200\, \mathrm{Mpc}^{-1}$ and are therefore
well outside the sound horizon at last scattering (where the conformal
age is $264 \,\mathrm{Mpc}$). For adiabatic perturbations, the
temperature quadrupole at reionization is therefore well approximated
by the simple Sachs-Wolfe result, giving
\begin{equation}
\Theta_{2m}(\chire \vnhat;\etare) = -\frac{4\pi}{3} \int \frac{d^3 \vk}{(2\pi)^{3/2}} \, \Psi(\vk;\eta_*)e^{i\vk\cdot \chire \vnhat} j_2(k\Delta \chi) Y_{2m}^*(\hat{\vk}) ,
\label{eq:appa1}
\end{equation}
where the (dimensionless) temperature anisotropy for radiation propagating in direction $\ve$ at position $\vx$ and conformal time $\eta$ is $\Theta(\eta,\vx,\ve) = \sum_{lm} \Theta_{lm}(\vx;\eta) Y_{lm}(\ve)$, and $\Psi$ is the
gravitational potential. The polarization that we observe from reionization
along the line of sight $\hat{\vn}$ is~\cite{Hu:1997hp}
\begin{equation}
(Q \pm i U)(\hat{\vn}) \approx -\frac{\sqrt{6}\tau}{10} \sum_m
\Theta_{2m}(\chire \vnhat;\etare) {}_{\pm 2} Y_{2m}(\hat{\vn}) ,
\end{equation}
where $\tau$ is the optical depth to reionization.
Using Eq.~(\ref{eq:appa1}), we have
\begin{equation}
(Q \pm i U)(\hat{\vn}) \approx \frac{\sqrt{6}\tau}{10} \frac{4\pi}{3}
\int  \frac{d^3 \vk}{(2\pi)^{3/2}} \, \Psi(\vk;\eta_*)e^{i\vk\cdot \chire \vnhat} j_2(k\Delta \chi) \sum_{m} Y_{2m}^*(\hat{\vk}) {}_{\pm 2} Y_{2m}(\hat{\vn}).
\end{equation}
This is a pure $E$-mode signal, where\footnote{%
Our polarization conventions throughout this paper follow~\cite{Lewis:2006fu}
so the $E$ and $B$ multipoles
have opposite sign to~\cite{Seljak:1997gy} and to the output of
\CMBFAST~\cite{Seljak:1996is} and \CAMB~\cite{Lewis:1999bs}.
The $E$--$\psi$ correlation we compute here therefore has opposite sign
to that displayed in Fig.~\ref{tpecorrelations}.
}
\begin{equation}
(Q\pm i U)(\hat{\vn}) = \sum_{lm} (E_{lm} \pm i B_{lm})
{}_{\pm 2}Y_{lm}(\hat{\vn}) ,
\end{equation}
with $E$-mode multipoles
\begin{equation}
E_{lm} =- \pi \tau i^l \int \frac{d^3 \vk}{(2\pi)^{3/2}}
\, \Psi(\vk;\eta_*) j_2(k\Delta \chi) \left[\sqrt{\frac{(l+2)!}{(l-2)!}}
\frac{j_l(k\chire)}{(k\chire)^2}\right] Y_{lm}^*(\hat{\vk}) .
\end{equation}

In the flat universe we are assuming here, the lensing potential of
Eq.~(\ref{psin}) has multipoles
\begin{equation}
\psi_{lm} = -8\pi i^l \int \frac{d^3 \vk}{(2\pi)^{3/2}}
\left(\int_0^{\chi_*} d\chi\, \frac{\chi_*-\chi}{\chi_* \chi} \Psi(\vk;
\eta_0 - \chi) j_l(k\chi)\right) Y_{lm}^*(\hat{\vk}) ,
\end{equation}
so the $E$--$\psi$ power spectrum is given by
\begin{equation}
C_l^{E\psi} = 2\pi \tau \sqrt{\frac{(l+2)!}{(l-2)!}}
\int d\ln k\, j_2(k\Delta \chi) \frac{j_l(k\chire)}{(k\chire)^2}
\left(\int_0^{\chi_*} d \chi \, \frac{\chi_*-\chi}{\chi_*\chi} j_l(k\chi)
\mathcal{P}_\Psi(k;\eta_*,\eta_0 - \chi)\right) ,
\end{equation}
where the unequal-time power spectrum of $\Psi$ is defined by
\begin{equation}
\langle \Psi(\vk;\eta) \Psi^*(\vk';\eta') \rangle = \frac{2\pi^2}{k^3}
\mathcal{P}_\Psi(k;\eta,\eta') \delta(\vk-\vk').
\end{equation}
To make further analytic progress, we neglect the late-time evolution of the
gravitational potential which is a reasonable approximation since CMB
lensing is most efficient around $z \approx 2$ and, furthermore, low-redshift
lenses will be only weakly correlated with the polarization from reionization.
We therefore have $\mathcal{P}_\Psi(k;\eta_*,\eta_0 - \chi) \approx
\mathcal{P}_\Psi(k;\eta_*)$ and, on the (large) scales of interest,
this is directly related to the power spectrum, $\mathcal{P}_{\mathcal{R}}(k)$,
of the primordial curvature perturbation and so is nearly scale-invariant.
Defining the integral
\begin{equation}
\mathcal{I}_l \equiv \int d\ln k\, j_2(k\Delta \chi) \frac{j_l(k\chire)}{(k\chire)^2}
\left(\int_0^{\chi_*} d \chi \, \frac{\chi_*-\chi}{\chi_*\chi} j_l(k\chi)
\right) ,
\end{equation}
we can approximate
\begin{equation}
C_l^{E\psi} \approx \tau A_s \frac{18\pi}{25} \sqrt{\frac{(l+2)!}{(l-2)!}}
\mathcal{I}_l ,
\end{equation}
where $\mathcal{P}_\Psi(k;\eta_*) = 9\mathcal{P}_{\mathcal{R}}(k)/25 \approx
9A_s/25$. The integral $\mathcal{I}_l$ can be (very) roughly approximated by
Limber-approximating (e.g.~\cite{Hu:1995sv}) in the integral over $\chi$ and then again
in the integral over $k$:
\begin{equation}
\mathcal{I}_l \sim \frac{\pi}{2l^5} \frac{\chi_*-\chire}{\chi_*} j_2(l\Delta\chi/\chire).
\end{equation}
This approximation is only good to around 50\% (away from zero crossings)
but does capture the basic shape of the $C_l^{E\psi}$ correlation in
Fig.~\ref{tpecorrelations}. In particular, we expect $[l(l+1)]^{3/2}C_l^{E\psi}
\propto j_2(l\Delta \chi/\chire)$ which gives rise to the oscillations seen
in Fig.~\ref{tpecorrelations}. At a given $l\gg 2$, most of the correlation
is from wavenumbers $k \sim l/ \chire$; the angular projection of
plane-wave fluctuations in $\Psi$ at wavenumber $k$ over the electron's
last scattering surface gives a local temperature quadrupole going
as $j_2(k \Delta \chi)$.
For $l=2$ where the signal is largest, the dominant contributions are actually from $k \sim 4/ \chire$ as shown in Fig.~\ref{waves}.

On large scales, $C_l^{E\psi} > 0$ ($<0$ in the \CMBFAST/\CAMB\ conventions
adopted in Fig.~\ref{tpecorrelations}), corresponding to radial
polarization around large overdense lenses. The part of the $T$--$E$
correlation from reionization coming from temperature anisotropies sourced
at last scattering has the opposite sign to $C_l^{E\psi}$ on large scales
since large-scale overdensities on the last-scattering surface appear cold.
However, the contribution of the late-time ISW effect to the $T$--$E$
correlation from reionization is positive on large scales since the
decay of large-scale potential wells at late times produces positive
temperature fluctuations.

\section{Perturbative temperature lensing bispectrum on the flat sky}
\label{third_order}

Expanding $\tilde{T}(\vx) = T(\vx+\vgrad\psi)$ to third order in $\psi$ we have
\ba
\tilde{\temp}(\vl) =\temp(\vl) &-&\int \frac{\ud^2 \vl_1}{2\pi}\frac{\ud^2 \vL}{2\pi} \temp(\vL) \psi(\vl_1)\vl_1\cdot\vL (2\pi)\delta(\vl_1+\vL-\vl)\nonumber\\
&+&\frac{1}{2} \int \frac{\ud^2 \vl_1}{2\pi}\frac{\ud^2 \vl_2}{2\pi}\frac{\ud^2 \vL}{2\pi}
\temp(\vL) \psi(\vl_1)\psi(\vl_2)\vl_1\cdot\vL \vl_2\cdot\vL (2\pi)\delta(\vl_1+\vl_2+\vL-\vl)\nonumber \\
&-&\frac{1}{6}
\int \frac{\ud^2 \vl_1}{2\pi}\frac{\ud^2 \vl_2}{2\pi}\frac{\ud^2 \vl_3}{2\pi}\frac{\ud^2 \vL}{2\pi}
\temp(\vL) \psi(\vl_1)\psi(\vl_2)\psi(\vl_3)\vl_1\cdot\vL \vl_2\cdot\vL \vl_3\cdot\vL (2\pi)\delta(\vl_1+\vl_2+\vl_3+\vL-\vl).
\ea

For $l_1\le l_2\le l_3$ and $l_1 \ll l_2,l_3$ (assuming $C^{T\psi}_l\approx 0$ for large $l=l_2,l_3$, and $\temp(\vl_1)=\tilde{\temp}(\vl_1)$) to third order in $\psi$ we have
\ba
b_{l_1 l_2 l_3} &\approx& - C_{l_1}^{T\psi}\left[ \vl_1\cdot \vl_2 C^{\temp\temp}_{l_2} (1-l_2^2 R^\psi) +
\int \frac{\ud^2 \vL}{(2\pi)^2} \vl_1\!\cdot\! \vL C^{\temp\temp}_L C^{\psi\psi}_{|\vl_2-\vL|} [(\vl_2-\vL)\cdot\vL]^2
\right]
+(\vl_2\leftrightarrow \vl_3).
\label{flat_squeezed1}
 \\
&=& - \vl_1\cdot \vl_2 C_{l_1}^{T\psi}\left[  C^{\temp\temp}_{l_2} (1-l_2^2 R^\psi) +
\int \frac{\ud^2 \vL}{(2\pi)^2} \frac{\vl_2\!\cdot\!\vL}{l_2^2}  C^{\temp\temp}_L C^{\psi\psi}_{|\vl_2-\vL|} [(\vl_2-\vL)\cdot\vL]^2
\right]
+ (\vl_2\leftrightarrow \vl_3).
\label{flat_squeezed}
\ea
Here we have defined
\begin{equation}
R^\psi\equiv \frac{1}{2} \la |\grad\psi|^2\ra = \frac{1}{4\pi}\int \frac{\ud l}{l} \,l^4 C^\psi_l, \label{R_def}
\end{equation}
which is half the total deflection angle power. The term in square brackets in Eq.~\eqref{flat_squeezed} is just the second-order result for $\Cgrads_{l_2}$.  If we approximate $\vl_1\cdot \vL \approx \vl_1\cdot \vl_2$ (corresponding to a `large-lens' approximation, where $l^2 C^{\psi\psi}_l$ falls rapidly at high $l$) and
use the second-order result for the lensed power spectrum~\cite{Hu:2000ee},
\begin{equation}
\tC^{\temp\temp}_l
\approx  (1-l^2 R^\psi) C_l^{\temp\temp} +
 \int \frac{\ud^2 \vl'}{(2\pi)^2} \left[ \vl'\cdot(\vl-\vl')\right]^2 C_{|\vl-\vl'|}^\psi C_{l'}^{\temp\temp} ,
\label{TT_lensed_series}
\end{equation}
Eq.~\eqref{flat_squeezed} is then equivalent to the first-order result but using the lensed power spectrum:
\be
b_{l_1 l_2 l_3} \approx -\left[(\vl_1\cdot \vl_2) C_{l_1}^{\temp\psi} \tC^{\temp\temp}_{l_2} +
(\vl_1\cdot \vl_3) C_{l_1}^{\temp\psi} \tC^{\temp\temp}_{l_3} \right].
\ee


In Eq.~\eqref{flat_squeezed} (and the main text) we assumed that $\tilde{\temp}(\vl_1) = \temp(\vl_1)$. If we relax this approximation there are additional third-order terms. Keeping only terms involving $C^{\temp\psi}_l$ at low $l$ where it is non-zero, there are two contributions: from $\clo(\psi)\times \clo(\psi)\times \clo(\psi)$ and $\clo(\psi) \times\clo(\psi^2) \times \clo(1)$,
\begin{multline}
\la \tilde{\temp}(\vl_1)\tilde{\temp}(\vl_2)\tilde{\temp}(\vl_3) \ra \approx
\la \temp(\vl_1)\tilde{\temp}(\vl_2)\tilde{\temp}(\vl_3) \ra
-\frac{1}{2\pi} \delta(\vl_1+\vl_2+\vl_3)
 \int \frac{\ud^2 \vL}{(2\pi)^2}
C^{\temp\psi}_L C^{\psi\psi}_{|\vl_1-\vL|}
\times \\
\biggl\{
[(\vl_1-\vL)\cdot \vL][(\vl_2+\vL)\cdot \vL][(\vL-\vl_1)\cdot (\vL+\vl_2)]C^{\temp\temp}_{|\vl_2+\vL|}
-[(\vl_1-\vL)\cdot \vL][\vl_2\cdot \vL][(\vL-\vl_1)\cdot \vl_2] C^{\temp\temp}_{|\vl_2|}
\biggr\}
 + (\vl_2\leftrightarrow \vl_3),
 \label{mixed_terms}
\end{multline}
%
and an additional small term from $\clo(\psi^2)\times \clo(\psi)\times \clo(1)$ which is down by a power of $l_1/l_2$.
Individually the separate terms in Eq.~\eqref{mixed_terms} are significant (if not large), but since $\vL$ is small, for large $|\vl_2|$ we have $\vl_2+\vL\approx \vl_2$, and the terms nearly cancel.
Using $\tilde{\temp}(\vl_1) = \temp(\vl_1)$ is therefore a good approximation, as expected on physical grounds for small $\vl_1$
and verified with simulations in the main text. Corrections from Eq.~\eqref{mixed_terms} are fractionally most important for
triangles where $\vl_1$ is orthogonal to $\vl_2$, where the signal is small anyway, and remain sub-dominant to the correction obtained by using
the lensed rather than unlensed power spectrum.

\section{Calculation of the gradient power spectra}
\label{Cgards_correlation}
In this appendix, we calculate the power spectrum of the $\tilde{T} \tilde{\nabla T}$ correlation, and the equivalent results for the polarization.
We will assume here that we can neglect $\temp$--$\psi$ correlations for this calculation. As we only require
$\Cgrads_l$ on small scales where the ISW contribution is small this should be a reliable approximation.
We start with the flat-sky limit.
Following~\cite{Seljak:1996ve,Lewis:2006fu} we construct the correlation function as a function of $\vr\equiv \vx -\vx'$:
\ba
\chi(r)\equiv \vr \cdot \la \widetilde{\vgrad \temp}(\vx) \tilde{\temp}(\vx')\ra &=& \vr\cdot \la [\vgrad \temp](\vx+\valpha) \temp(\vx'+\valpha')\ra \nonumber \\
&=&\int \frac{\ud \vl}{2\pi} \frac{\ud \vl'}{2\pi} \la e^{i\vl\cdot(\vx+\valpha)} e^{-\vl'\cdot(\vx'+\valpha')}\ra
\la (i\vr\cdot \vl) T(\vl) T(\vl')^*\ra \nonumber \\
&=& \int \frac{\ud^2\vl}{(2\pi)^2} C^{\temp\temp}_l (i\vl\cdot\vr) e^{i\vl\cdot\vr} e^{-l^2[\sigma^2(r)+\cos 2\phi \Cgtwo(r)]/2} \nonumber \\
&=&
\int \frac{\ud l}{l} \frac{l^2 C^{\temp\temp}_l}{2\pi} e^{-l^2\sigma^2(r)/2} \frac{lr}{2} \sum_{n=-\infty}^\infty
\left[ J_{2n-1}(lr) - J_{2n+1}(lr)\right] I_n[l^2\Cgtwo(r)/2].
\label{eq:AppB1}
\ea
Here, we have defined $\phi$ as the angle between $\vl$ and $\vr$,
$\sigma^2(r) \equiv \langle (\valpha - \valpha')^2 \rangle /2$ as half
the variance of the relative deflection, and $\Cgtwo(r)$ as the non-isotropic
part of the correlation function of $\valpha$: $\Cgtwo(r) \equiv
-2 \hat{r}^i \hat{r}^j \langle \alpha_i \alpha'_j \rangle$ where
$\hat{\vr}=\vr / r$ and angle brackets around indices denote the symmetric,
trace-free part. Bessel functions and modified Bessel functions
are denoted by $J_n(x)$ and $I_n(x)$ respectively.
Note that $\vr \cdot \la \widetilde{\vgrad \temp}(\vx) \tilde{\temp}(\vx')\ra$
has no component perpendicular to $\vr$ so the correlation is fully
described by $\chi(r)$.
Expanding Eq.~(\ref{eq:AppB1}) gives the leading terms
\be
\chi(r) = -\int \frac{\ud l}{l} \frac{l^2 C^{\temp\temp}_l}{2\pi} e^{-l^2\sigma^2(r)/2} lr
\left( J_1(lr) + \frac{l^2\Cgtwo(r)}{4}\left[J_{3}(lr)- J_{1}(lr)\right] + \cdots \right).
\ee
Transforming the correlation function we then have
\be
\Cgrads_l = - 2\pi \int r\ud r \frac{J_1(lr)}{lr} \chi(r).
\ee
See Fig.~\ref{Cgrads} for numerical comparison with the lensed power spectrum.

For the polarization we have\footnote{%
Note that $P^*(\vl)$ is \emph{not} the complex conjugate $[P(\vl)]^*$ of
$P(\vl)$; rather $P^*(\vl)=[P(-\vl)]^*$ so that $E(\vl)$ and
$B(\vl)$ are the Fourier transforms of real fields with e.g.\
$[E(\vl)]^* = E(-\vl)$.}
\ba
P(\vl) &\equiv& E(\vl)+iB(\vl) = - \int \frac{\ud^2 \vx}{2\pi} P(\vx) e^{-2i\phi_\vl} e^{-i\vl\cdot\vx} , \\
P^*(\vl) &\equiv& E(\vl)-iB(\vl) = - \int \frac{\ud^2 \vx}{2\pi} P^*(\vx) e^{2i\phi_\vl} e^{-i\vl\cdot\vx},
\ea
and
\ba
 \frac{\delta}{\delta \psi(\vl_1)^*}  \tilde{P}(\vl_2) &=& -\frac{i}{2\pi} e^{2i(\phi_{\vl_3}-\phi_{\vl_2})} \vl_1\cdot \widetilde{\vgrad P}(\vl_1+\vl_2),\\
 \frac{\delta}{\delta \psi(\vl_1)^*}  \tilde{P}^*(\vl_2) &=& -\frac{i}{2\pi}  e^{-2i(\phi_{\vl_3}-\phi_{\vl_2})} \vl_1\cdot \widetilde{\vgrad P^*}(\vl_1+\vl_2),
\ea
where $\vl_3 = -\vl_1-\vl_2$.

We define $\phi\equiv \phi_\vl -\phi_\vr$ and $\vr^\pm_i = r_i \pm i \epsilon^j{}_i r_j$, so that
\be
\vr^\pm \cdot \vl = rl e^{\pm i\phi}
\qquad
i \epsilon^{jk}  l_j r^\pm_k = \mp \, r l e^{\pm i\phi},
\ee
and introduce the correlation functions
\ba
\chi_+(r) &\equiv& \vr \cdot \la \widetilde{\vgrad P^*}(\vx) \tilde{P}(\vx')\ra = \vr\cdot \la \vgrad P^*(\vx+\valpha) P(\vx'+\valpha')\ra\nonumber\\
&=&\int \frac{\ud l}{l} \frac{l^2 (C^{EE}_l+C^{BB}_l) }{2\pi} e^{-l^2\sigma^2(r)/2} \frac{lr}{2} \sum_{n=-\infty}^\infty
\left[ J_{2n-1}(lr) - J_{2n+1}(lr)\right] I_n[l^2\Cgtwo(r)/2] \nonumber\\
&=& -\int \frac{\ud l}{l} \frac{l^2 (C^{EE}_l+C^{BB}_l)}{2\pi} e^{-l^2\sigma^2(r)/2} lr
\left( J_1(lr) + \frac{l^2\Cgtwo(r)}{4}\left[J_{3}(lr)- J_{1}(lr)\right] + \cdots \right) ,
\ea
\ba
\chi^\pm_-(r) &\equiv& \vr^\pm \cdot \la e^{-4i\phi_\vr} \widetilde{\vgrad P}(\vx) \tilde{P}(\vx')\ra
= \vr^\mp \cdot \la e^{4i\phi_\vr} \widetilde{\vgrad P^*}(\vx) \tilde{P}^*(\vx')\ra
\nonumber \\
&=& \int \frac{\ud^2\vl}{(2\pi)^2} (C^{EE}_l-C^{BB}_l) e^{4i\phi} (i\vr^\pm \cdot \vl) e^{i\vl\cdot\vr} e^{-l^2[\sigma^2(r)+\cos 2\phi \Cgtwo(r)]/2}\nonumber\\
&=&
\mp \int \frac{\ud l}{l} \frac{l^2  (C^{EE}_l-C^{BB}_l) }{2\pi} e^{-l^2\sigma^2(r)/2} lr \sum_{n=-\infty}^\infty
J_{2n+4\pm 1}(lr) I_n[l^2\Cgtwo(r)/2]\nonumber\\
&=&\mp \int \frac{\ud l}{l} \frac{l^2 (C^{EE}_l-C^{BB}_l)}{2\pi} e^{-l^2\sigma^2(r)/2} lr
\left( J_{4\pm 1}  + \frac{l^2\Cgtwo(r)}{4}\left[  J_{2\pm 1}(lr) + J_{6\pm 1}(lr)\right] + \cdots \right).
\ea

\ba
\chi^\pm_\times(r) &\equiv& \vr^\pm \cdot \la e^{-2i\phi_\vr} \widetilde{\vgrad T}(\vx) \tilde{P}(\vx')\ra
=\vr^\mp \cdot \la e^{2i\phi_\vr} \widetilde{\vgrad T}(\vx) \tilde{P}^*(\vx')\ra \nonumber\\
&=& \vr^\pm \cdot \la e^{-2i\phi_\vr} \widetilde{\vgrad P}(\vx) \tilde{T}(\vx')\ra
= \vr^\mp \cdot \la e^{2i\phi_\vr} \widetilde{\vgrad P^*}(\vx) \tilde{T}(\vx')\ra
\nonumber\\
&=& - \int \frac{\ud^2\vl}{(2\pi)^2} C^{TE}_l e^{2i\phi} (i\vr^\pm \cdot \vl) e^{i\vl\cdot\vr} e^{-l^2[\sigma^2(r)+\cos 2\phi \Cgtwo(r)]/2}\nonumber\\
&=&
\mp \int \frac{\ud l}{l} \frac{l^2 C^{TE}_l }{2\pi} e^{-l^2\sigma^2(r)/2} lr \sum_{n=-\infty}^\infty
J_{2n+2\pm 1}(lr) I_n[l^2\Cgtwo(r)/2]\nonumber\\
&=& \mp \int \frac{\ud l}{l} \frac{l^2 C^{TE}_l}{2\pi} e^{-l^2\sigma^2(r)/2} lr
\left( J_{2\pm 1}(lr) + \frac{l^2\Cgtwo(r)}{4}\left[ J_{4\pm 1}(lr) \pm  J_1(lr) \right]
 + \cdots \right).
\ea

The relevant correlations in Fourier space are
\ba
\la \widetilde{\vgrad P}(\vl) \tilde{P}^*(\vl') \ra = \la \widetilde{\vgrad P^*}(\vl) \tilde{P}(\vl') \ra
&=&
\int \frac{\ud^2 \vx}{2\pi} \frac{\ud^2 \vx'}{2\pi} \la \widetilde{\vgrad P}(\vx) \tilde{P}^*(\vx')\ra e^{-2i\phi_\vl} e^{-i\vl\cdot\vx}e^{2i\phi_{\vl'}} e^{-i\vl'\cdot\vx'} \nonumber
\\
&=&\delta(\vl+\vl') \int \ud^2 \vr \la \widetilde{\vgrad P}(\vx) \tilde{P}^*(\vx')\ra  e^{-i\vl\cdot\vr} \nonumber
\\
&=&
\delta(\vl+\vl') \int \frac{\ud^2 \vr}{r^2} \vr \chi_+(r) e^{-i\vl\cdot\vr} ,
\ea
\ba
\la \widetilde{\vgrad P}(\vl) \tilde{P}(\vl') \ra = \la \widetilde{\vgrad P^*}(-\vl) \tilde{P}^*(-\vl') \ra^*
&=&
\int \frac{\ud^2 \vx}{2\pi} \frac{\ud^2 \vx'}{2\pi} \la \widetilde{\vgrad P}(\vx) \tilde{P}(\vx')\ra e^{-2i\phi_\vl} e^{-i\vl\cdot\vx}e^{-2i\phi_{\vl'}} e^{-i\vl'\cdot\vx'} \nonumber
\\
&=&\delta(\vl+\vl') \int \ud^2 \vr \la e^{-4i\phi_\vr} \widetilde{\vgrad P}(\vx) \tilde{P}(\vx')\ra e^{-4i\phi} e^{-i\vl\cdot\vr} \nonumber
\\
&=&
\delta(\vl+\vl') \int \frac{\ud^2 \vr}{2r^2}  \left(\chi_-^-(r) \vr^+ + \chi_-^+(r)  \vr^-\right)  e^{-4i\phi} e^{-i\vl\cdot\vr} ,
\ea
\ba
\la \widetilde{\vgrad T}(\vl) \tilde{P}(\vl') \ra = [\la \widetilde{\vgrad T^*}(-\vl) \tilde{P}^*(-\vl') \ra]^*
&=&
\la \widetilde{\vgrad P}(\vl) \tilde{T}(\vl') \ra = \la \widetilde{\vgrad P^*}(-\vl) \tilde{T}^*(-\vl') \ra^* \nonumber \\
&=&
- \int \frac{\ud^2 \vx}{2\pi} \frac{\ud^2 \vx'}{2\pi} \la \widetilde{\vgrad T}(\vx) \tilde{P}(\vx')\ra e^{-i\vl\cdot\vx}e^{-2i\phi_{\vl'}} e^{-i\vl'\cdot\vx'} \nonumber
\\
&=& -\delta(\vl+\vl') \int \ud^2 \vr \la e^{-2i\phi_\vr} \widetilde{\vgrad T}(\vx) \tilde{P}(\vx')\ra e^{-2i\phi} e^{-i\vl\cdot\vr} \nonumber
\\
&=&
- \delta(\vl+\vl') \int \frac{\ud^2 \vr}{2r^2}  \left( \chi_\times^-(r) \vr^+ + \chi_\times^+(r) \vr^-\right)  e^{-2i\phi} e^{-i\vl\cdot\vr} .
\ea
We can express these in terms of power spectra as
\ba
-i \la \widetilde{\vgrad P}(\vl) \tilde{P}^*(\vl') \ra &=& \delta(\vl+\vl') (\tilde{C}_l^{E\grad E} + \tilde{C}_l^{B\grad B}) \vl ,
\\
-i \la \widetilde{\grad_i P}(\vl) \tilde{P}(\vl') \ra &=& \delta(\vl+\vl')\left[(\tilde{C}_l^{E\grad E} - \tilde{C}_l^{B\grad B}) l_i - i \epsilon_{ji} l^j \tilde{C}_l^{PP\perp}\right] ,
\\
-i \la \widetilde{\grad_i P}(\vl) \tilde{T}(\vl') \ra &=& \delta(\vl+\vl')\left[ \tilde{C}_l^{T\grad E} l_i - i \epsilon_{ji} l^j \tilde{C}_l^{TP\perp}\right] ,
\ea
where
\ba
\tilde{C}_l^{E\grad E} + \tilde{C}_l^{B\grad B} &\equiv& -2\pi \int \frac{r \ud r}{l r}  J_1(lr)  \chi_+(r) ,
\\
\tilde{C}_l^{E\grad E} - \tilde{C}_l^{B\grad B} &\equiv& \int 2\pi \frac{r \ud r}{2l r} \left\{
\chi_-^-(r) J_3(lr) - \chi_-^+(r) J_5(lr)
\right\},
\\
\tilde{C}_l^{PP\perp} &\equiv & \int 2\pi \frac{r \ud r}{2l r} \left\{
\chi_-^-(r) J_3(lr) + \chi_-^+(r) J_5(lr)
\right\},
\\
\tilde{C}_l^{T\grad E}  &\equiv& -2\pi \int \frac{r \ud r}{2l r} \left\{
\chi_\times^+(r) J_3(lr) - \chi_\times^-(r) J_1(lr)
\right\},
\\
\tilde{C}_l^{TP\perp} &\equiv & - 2 \pi \int \frac{r \ud r}{2l r} \left\{ \chi_\times^+(r) J_3(lr) + \chi_\times^-(r) J_1(lr)\right\}.
\ea
In the absence of lensing $\tilde{C}_l^{E\grad E}\rightarrow C_l^{EE}$,
$\tilde{C}_l^{B\grad B}\rightarrow C_l^{BB}$ and $\tilde{C}_l^{T\grad E} \rightarrow
C_l^{TE}$, but $\tilde{C}_l^{PP\perp} \rightarrow 0$ and $\tilde{C}_l^{TP\perp} \rightarrow 0$.
Results required for the bispectrum then follow. For example, using
\ba
\left\la\left(\frac{\delta}{\delta \psi(\vl_1)^*} \tilde{B}(\vl_2)\right)\tilde{E}(\vl_3)\right\ra\! &=&\!
\frac{1}{2\pi} \delta(\vl_1+\vl_2+\vl_3) \left[\vl_1 \cdot \vl_3
 \tilde{C}^{E\grad E}_{l_3} \sin 2(\phi_{\vl_2}-\phi_{\vl_3}) + \frac{1}{2}
\epsilon_{ji} l_3^j l_1^i \tilde{C}^{PP\perp}_{l_3}
\cos 2(\phi_{\vl_2}-\phi_{\vl_3}) \right], \quad\quad\\
\left\la \tilde{B}(\vl_2)\left(\frac{\delta}{\delta \psi(\vl_1)^*} \tilde{E}(\vl_3)\right)\right\ra \!&=&\!
\frac{1}{2\pi} \delta(\vl_1+\vl_2+\vl_3) \left[\vl_1 \cdot \vl_2
 \tilde{C}^{B\grad B}_{l_2} \sin 2(\phi_{\vl_2}-\phi_{\vl_3}) + \frac{1}{2}
\epsilon_{ji} l_2^j l_1^i \tilde{C}^{PP\perp}_{l_2}
\cos 2(\phi_{\vl_2}-\phi_{\vl_3}) \right], \quad\quad
\ea
we have\footnote{%
Note that the flat-sky bispectrum of an odd-parity product of fields, like
$T$, $B$ and $E$, depends not only on the lengths $l_1$, $l_2$ and $l_3$
but also which of the two parity-related configurations of the three
vectors is being considered. The bispectra for the two configurations
have opposite signs.
}
\ba
\langle T(\vl_1)\tilde{B}(\vl_2) \tilde{E}(\vl_3) \rangle &=&
\frac{1}{2\pi} \delta(\vl_1+\vl_2+\vl_3) C_{l_1}^{T\psi}
\Biggl[\vl_1 \cdot (\tilde{C}_{l_2}^{B\grad B}\vl_2 +\tilde{C}_{l_3}^{E\grad E}\vl_3)
\sin  2(\phi_{\vl_2}-\phi_{\vl_3}) \nonumber \\
&& \hspace{0.25\textwidth} + \frac{1}{2} \epsilon_{ji}l_1^i(
\tilde{C}^{PP\perp}_{l_2}l_2^j + \tilde{C}^{PP\perp}_{l_3}l_3^j )\cos  2(\phi_{\vl_2}-\phi_{\vl_3})
\Biggr].
\label{eq:AppB2}
\ea
The leading-order result sets the $\tilde{C}_l^{PP\perp}$ terms to zero and
replaces $\tilde{C}_l^{E\grad E}$ and $\tilde{C}_l^{B\grad B}$ with $C_l^{EE}$ and $C_l^{BB}$ respectively. The $C_l^{BB}$ contribution vanishes if there
are no unlensed $B$ modes.
The approximation made in the text to capture the main,
non-perturbative corrections to the leading-order result is to replace
$\tilde{C}_l^{E\grad E}$ with $\tilde{C}_l^{EE}$. This neglects the
$\tilde{C}_l^{B\grad B}$ and $\tilde{C}_l^{PP\perp}$ terms in
Eq.~(\ref{eq:AppB2}). These are of similar magnitude to the lensed
$B$-mode spectrum, $\tilde{C}_l^{BB}$, and much smaller than the change
in $C_l^{EE}$ due to lensing which therefore dominates the corrections to
the leading-order bispectrum.

\subsection{Full sky}
\label{app:full_sky}
In the semi-squeezed limit of most interest we can accurately approximate the bispectrum of the lensed fields as
\ba
B^{ijk}_{l_1 l_2 l_3} &\approx & \sum_{m_1 m_2 m_3} \threej{l_1}{l_2}{l_3}{m_1}{m_2}{m_3} \la a^i_{l_1 m_2} \tilde{a}^j_{l_2 m_2} \tilde{a}^k_{l_3 m_3} \ra \nonumber \\
&=& C^{a^i\psi}_{l_1}
\sum_{m_1 m_2 m_3} \threej{l_1}{l_2}{l_3}{m_1}{m_2}{m_3}  \left\la \left(\frac{\delta}{\delta \psi_{l_1 m_1}^*} \tilde{a}^j_{l_2 m_2}\right) \tilde{a}^k_{l_3 m_3} \right \ra + [ (j, l_2) \leftrightarrow (k,l_3) ].
\label{fullbi}
\ea
We shall pursue a non-perturbative curved-sky analysis in the approximation in which sky-curvature effects on the arcminute-scale of the deflection angles can be neglected, similar to the approximations used when calculating the full-sky lensed power spectra~\cite{Lewis:2006fu}.
Acting on a lensed field of any spin we write
\be
 \frac{\delta }{\delta \psi_{l_1 m_1}^*}
 = -\frac{1}{2}\left( \frac{\delta ({}_1\alpha)}{\delta \psi_{l_1 m_1}^*}  \beth +  \frac{\delta ({}_{-1}\alpha )}{\delta \psi_{l_1 m_1}^*} \edth\right)
 = -\frac{\sqrt{l_1 (l_1+1)}}{2}\left( {}_{-1}Y_{l_1 m_1}^* \beth - {}_{-1}Y_{l_1 m_2}^* \edth\right)
\ee
where we used spin $\pm 1$ components of the deflection field $\valpha = \vgrad \psi$ expanded in terms of spin $\pm 1$ spherical harmonics
\be
{}_1\alpha = -\sum_{lm} \sqrt{l(l+1)}  \,{}_1 Y_{lm} \psi_{lm}\qquad
{}_{-1}\alpha = \sum_{lm} \sqrt{l(l+1)}  \, {}_{-1} Y_{lm} \psi_{lm}.
\ee
The combination $ -\frac{1}{2}\left( {}_1\alpha \beth + {}_{-1}\alpha \edth\right)$ is the spin-weight analogue of $\valpha\cdot \vgrad$
acting on a rank $|s|$ tensor field~\cite{Challinor:2002cd}.

For a lensed spin $s_2$ field $\XstwoL = {}_{-s_2} \tilde{X}^{*}$ we then have
\ba
\frac{\delta}{\delta \psi_{l_1 m_1}^*} \XstwoL_{l_2 m_2}
&=&
\int \ud\Omega_{\vnhat} \frac{\delta }{\delta \psi_{l_1 m_1}^*} \Xstwo(\vnhat+\valpha)  {}_{s_2}Y_{l_2 m_2}^*(\vnhat)
\\
&=& -\frac{\sqrt{l_1(l_1+1)}}{2}\int \ud\Omega_{\vnhat}
\left\{ {}_{-1}Y_{l_1 m_1}^*(\vnhat) [\beth \Xstwo](\vnhat+\valpha) - {}_1 Y_{l_1 m_1}^*(\vnhat) [\edth \Xstwo](\vnhat+\valpha)\right\} {}_{s_2} Y_{l_2 m_2}^*(\vnhat) \nonumber\\
&=& \sum_{l'm'}\frac{\sqrt{l_1(l_1+1)}}{2} \sqrt{\frac{ (2l_1+1)(2l_2+1)(2l'+1)}{4\pi}}\threej{l_1}{l_2}{l'}{m_1}{m_2}{m'} \nonumber\\
&&\qquad\qquad\times \left[
\threej{l_1}{l_2}{l'}{-1}{s_2}{1-s_2}{}_{-s_2}  \tilde{X}^{-*}_{l'm'}\kappa^-_{l's_2} -\threej{l_1}{l_2}{l'}{1}{s_2}{-s_2-1} {}_{-s_2}\tilde{X}^{+*}_{l'm'}\kappa^+_{l's_2}
\right]
\ea
where
\ba
\,[\edth \Xstwo](\vnhat+\valpha)  &=& \sum_{l' m'} {}_{s_2}\tilde{X}^-_{l'm'}{}_{s_2+1}Y_{l'm'}(\vnhat) \kappa^+_{l' s_2}   =\sum_{l' m'}  {}_{s_2+1} Y_{l'm'}(\vnhat+\valpha) \Xstwo_{l'm'} \kappa^+_{l' s_2} \nonumber \\
\, [\beth \Xstwo] (\vnhat+\valpha)  &=& \sum_{l' m'} {}_{s_2}\tilde{X}^+_{l'm'}{}_{s_2-1}Y_{l'm'}(\vnhat) \kappa^-_{l' s_2} =\sum_{l' m'}  {}_{s_2-1} Y_{l'm'}(\vnhat+\valpha) \Xstwo_{l'm'} \kappa^-_{l' s_2}
\ea
and
$$
\kappa^+_{ls} \equiv \sqrt{ l(l+1)-s(s+1)} \qquad
\kappa^-_{ls} \equiv -\sqrt{l(l+1) -s(s-1)}.
$$
For the polarization we can then write
\ba
\frac{\delta}{\delta \psi_{l_1 m_1}^*}\left( {}_2 P_{l_2 m_2}\pm  {}_{-2} P_{l_2 m_2}\right)
&=& \sum_{l'm'}\frac{1}{2} \sqrt{\frac{ (2l_1+1)(2l_2+1)(2l'+1)}{4\pi}}\threej{l_1}{l_2}{l'}{m_1}{m_2}{m'} \nonumber\\
&&\times \biggl[
{}_{2}M^\parallel_{l_1 l_2 l'}\left( {}_{-2}\tilde{P}^{-*}_{l'm'} + {}_{-2}\tilde{P}^{+*}_{l'm'} \pm (-1)^p[ {}_{2}\tilde{P}^{-*}_{l'm'} + {}_{2}\tilde{P}^{+*}_{l'm'} ]\right)  \nonumber\\
&&\qquad+{}_{2}M^\perp_{l_1 l_2 l'} \left( {}_{-2}\tilde{P}^{-*}_{l'm'} - {}_{-2}\tilde{P}^{+*}_{l'm'}   \mp (-1)^p[ {}_{2}\tilde{P}^{-*}_{l'm'} - {}_{2}\tilde{P}^{+*}_{l'm'} ] \right)
\biggr]
\ea
where
\ba
{}_{2}M^\parallel_{l_1 l_2 l'} &\equiv&
\frac{\sqrt{l_1(l_1+1)}}{2}\left[\threej{l_1}{l_2}{l'}{-1}{2}{-1}  \kappa^-_{l'2} -\threej{l_1}{l_2}{l'}{1}{2}{-3}\kappa^+_{l'2}\right]
\nonumber \\
&=& \frac{1}{2} \left[l_1(l_1+1) + l'(l'+1)  - l_2(l_2+1)\right] \threej{l_1}{l_2}{ l'}{0}{2}{-2},
\ea
\ba
{}_{2}M^\perp_{l_1 l_2 l'} &\equiv&
\frac{\sqrt{l_1(l_1+1)}}{2}\left[\threej{l_1}{l_2}{l'}{-1}{2}{-1}  \kappa^-_{l'2} +\threej{l_1}{l_2}{l'}{1}{2}{-3}\kappa^+_{l'2}\right].
\ea
and the parity is determined by $p=l_1 + l_2 + l_3$.

The required power spectra are of the general form
\ba
\tilde{C}^{VW}_{l} &=& \la {}_{s}\tilde{V}_{l m }^* \,{}_{s'}\tilde{W}_{l m} \ra = \frac{1}{2l+1} \sum_{m}\la {}_{s}\tilde{V}_{l m }^* \,{}_{s'}\tilde{W}_{l m} \ra \\
&=& \frac{1}{2l+1} \sum_{m}\int \ud\Omega_{\vnhat}\ud \Omega_{\vnhat'} \la {}_{s}\tilde{V}(\vnhat)^* {}_sY_{lm}(\vnhat) \,{}_{s'}\tilde{W}(\vnhat') {}_{s'}Y_{lm}^*(\vnhat') \ra\\
&=& 2\pi \int_{-1}^1 \ud \cos\beta \,d^l_{ss'}(\beta) \chi^{VW}(\beta)
\ea
where the last line follows easily from rotation invariance if  $\vnhat$ is chosen to lie along the $z$-axis and $\vnhat'$ is in the $x$-$z$ plane at an angle $\beta$ to the $z$-axis (using $d^l_{sm}(0)=\delta_{s m}$ and the relationship between spin-weight spherical harmonics and Wigner functions).
The correlation functions can be calculated following the method in Refs.~\cite{Challinor:2005jy,Lewis:2006fu} where we make the same choice of $\vnhat$ and $\vnhat'$:
\ba
\chi^{VW}(\beta) &\equiv& \la {}_{s}\tilde{V}(\vnhat)^*  \,{}_{s'}\tilde{W}(\vnhat') \ra \nonumber \\
&=&
\sum_{l m} C^{VW}_l \la {}_{s}Y_{lm}^*(\vnhat+\valpha) {}_{s'}Y_{lm}(\vnhat'+\valpha') \ra \nonumber \\
&=&
\sum_{l m m'} C^{VW}_l d^l_{mm'}(\beta) \la e^{-s i\psi} {}_{s}Y_{lm}^*(\alpha,\psi) {}_{s'}Y_{lm'}(\alpha',\psi') e^{is' \psi'}\ra \nonumber \\
&\approx&
\sum_l  \frac{2l+1}{4\pi} C^{VW}_{l} e^{-l(l+1)\sigma^2/2} \sum_{n=-l}^l I_n[l(l+1)\Cgtwo/2] d^l_{n+s,-n+s'}.\quad
\ea

\section{Simulating the correlated unlensed fields}
\label{simulation}
Given a method for generating independent Gaussian normal variates with unit variance $g_T$, $g_E$ and $g_\psi$,
the unlensed full-sky temperature, polarization and lensing potential multipoles at a given $l$ and $m$ can be generated with the correct correlations using the
Cholesky decomposition of the covariance matrix between the three fields. Explicitly, this is given by
\ba
T_{lm} &=& \sqrt{C_l^{TT}} g_T \\
E_{lm} &=& \frac{C_l^{TE}}{\sqrt{C_l^{TT}}} g_T + \left(C_l^{EE} -
  \frac{(C_l^{TE})^2}{C_l^{TT}}\right)^{1/2}g_E \\
\psi_{lm} &=& \frac{C_l^{T\psi}}{\sqrt{C_l^{TT}}} g_T +
 \frac{\left(C_l^{\psi E} - \frac{C_l^{\psi T} C_l^{T E} }{C_l^{TT}}\right) }
 {\left(C_l^{E E} - \frac{(C_l^{T E})^2 }{C_l^{TT}}\right)^{1/2}} g_E
 +\left(C_l^{\psi\psi} - \frac{(C_l^{\psi T})^2}{C_l^{TT}} -
  \frac{\left(C_l^{\psi E} - \frac{C_l^{\psi T} C_l^{T E} }{C_l^{TT}}\right)^2 }
 {\left(C_l^{E E} - \frac{(C_l^{T E})^2 }{C_l^{TT}}\right)}
\right)^{1/2} g_\psi.
  \label{psi_sim}
\ea
The lensed field can then be obtained by re-mapping points by the lensing deflection angle; this is implemented in the public LensPix\footnote{\url{http://cosmologist.info/lenspix/}} code~\cite{Lewis:2005tp,Hamimeche:2008ai}.

As written above, the temperature realization is generated first and
correlations to the other fields are generated by using $g_T$ when
they are simulated. Permuting fields, e.g. $T\leftrightarrow \psi$,
similar expressions can be obtained where other fields are generated first.

\subsection*{Simulation for lensing bispectrum estimation}

For simulation of a bispectrum it can be useful to reduce the variance by using an estimator that subtracts off most of the random realization-dependent scatter in the bispectrum estimator, while leaving the estimator unbiased~\cite{Hanson:2009kg}, e.g.
 \ba
 \hat{B}_{l_1 l_2 l_3}^{ijk} &=& \sum_{m_1 m_2}  \threej{l_1}{l_2}{l_3}{m_1}{m_2}{m_3} \left[
 \tilde{a}_{l_1 m_1}^i \tilde{a}_{l_2 m_2}^j \tilde{a}_{l_3 m_3}^k -  \bar{a}_{l_1 m_1}^{i} \bar{a}_{l_2 m_2}^{j} \bar{a}_{l_3 m_3}^{k} \right],
 \ea
 where $m_3=-m_1-m_2$ and $\la\bar{a}_{l_1 m_1}^{i} \bar{a}_{l_1 m_1}^{j} \bar{a}_{l_1
   m_1}^{k} \ra=0$. Note that, here, the overbar denotes counterterms
 in the bispectrum estimator rather than the different usage in
Eq.~(\ref{fullspherbispec}).
When simulating the lensing bispectrum the simplest method would be to
take $\bar{a}_{lm}$ to be the unlensed field. However since the lensed
and unlensed multipoles decorrelate at small scales (lenses shift the
field around, mixing up the $l,m$), this does not help very much on
small scales. It is therefore preferable to use a lensed realization
for $\bar{a}_{lm}$ where the fields are as close as possible to the
$\tilde{a}_{lm}$ realization but constructed to have zero bispectrum
in the mean. For a given random seed, one way to achieve this is to construct $\psi_{lm}$ from
Eq.~\eqref{psi_sim} with the correlation terms omitted, i.e.\ set $g_T
= g_E = 0$. This does succeed in reducing the variance, but it is still
significant since the large-scale lenses are modified by removing the
correlations and this alters the small-scale lensed $\bar{a}_{lm}$.
If we are only interested in simulating $B_{l_1 l_2 l_3}$ for a
particular range of $\{l_1\}$, we can instead generate counter-term
realizations as above but with $\psi_{lm}$ only modified for $l \in \{
l_1 \}$, and $T_{lm},E_{lm}$ modified for $l \notin \{
l_1 \}$
This significantly reduces the change in the $\psi$ field, and hence
the difference between the small-scale $\tilde{a}_{lm}$ and
$\bar{a}_{lm}$ reducing the variance even further. In the main text,
when simulating $B_{l_1 l_2 l_3}$ for $l_1 = 4$,
we simulate the $l=4$ unlensed fields using a `$T$-first' algorithm as
described above, but generate all other multipoles using a
`$\psi$-first' algorithm, with $\bar{a}_{lm}$ generated using the same
random seed but terms correlating $\psi$ to $T$ and $E$ set to zero. This trick works better for higher $l_1$ (our $l_1=50$ results) where the correlations are small, but gains less for $l_1=4$ where the correlations are significant so the zero-bispectrum field is still significantly different.


\section{Tensor vectorization and forms of the bispectrum covariance}
\label{vectorization}

Here we generalize some results for matrix vectorization to tensors, aiming to derive the general result that gives the equivalence of the various forms of the bispectrum covariance. For further details and references for matrix results see e.g. Refs.~\cite{Gupta99,Hamimeche:2008ai}.

The elements of a general matrix $\mA$ can be assigned column-wise into a vector $\vec(\mA)$. We can extend this to a 3-tensor $\tB$ so that
\be
\vec(\tB) = ( B_{111}, B_{211}, B_{311}\dots B_{n11}, B_{121},B_{221},B_{321}\dots )^T,
\ee
and a 3-tensor contraction can be written as
\be
B_{ijk} D_{ijk} = \vec(\tB)^T \vec(\tD).
\ee

The Kronecker product of an $m\times n$ matrix $\mA$ with a $p\times q$ matrix $\mB$ is defined to be the $mp \times nq$ matrix
\be
\mA\otimes\mB = \begm A_{11}\mB & A_{12}\mB &\dots & A_{1n}\mB \\ A_{21}\mB & A_{22}\mB & \dots & A_{2n}\mB
\\ \vdots & \vdots & & \vdots
\\ A_{m1}\mB & A_{m2}\mB & \dots & A_{mn}\mB
\enm.
\ee
We can write a matrix-tensor contraction e.g. as $\tB(\mF,\mG,\mH)$, so $[\tB(\mF,\mG,\mH)]_{ijk} = F_{ip} G_{jq} H_{kr} B_{pqr}$.
We can then write
\be
\vec( \tB(\mF,\mG,\mH)) = (\mH \otimes \mG\otimes \mF) \vec(\tB).
\ee
Using this we have
\be
B^{ijk}_{l_1 l_2 l_3}
(C^{-1})^{ip}_{l_1}(C^{-1})^{jq}_{l_2}(C^{-1})^{kr}_{l_3} B^{pqr}_{l_1
  l_2 l_3} =  \vec(\tB_{l_1 l_2 l_3})^T (\mC^{-1}_{l_3} \otimes
\mC^{-1}_{l_2} \otimes \mC^{-1}_{l_1}) \vec(\tB_{l_1 l_2 l_3}) .
\ee
This result extends straightforwardly to higher-rank tensors (e.g. for higher $n$-point functions), following the generalization below, where we also consider the case where the tensor may be symmetric on one or more indices as when $l_1,l_2$ and $l_3$ are not distinct.

\subsection*{Results for $k$-tensor and symmetric tensor vectorization}
The vector $\vec(\mA)$ is the elements of a general matrix $\mA$ assigned column-wise; we can extend this to a
$k$-tensor $\tB$ so that
\be
\vec(\tB) = ( B_{11\dots 1}, B_{21\dots 1}, B_{31\dots 1}\dots B_{n1\dots 1}, B_{12\dots 1},B_{22\dots 1},B_{32\dots 1}\dots
B_{nn\dots n})^T.
\ee
Full contraction of $k$-tensors can then be written as
\be
B_{a_1 a_2\dots a_k} D_{a_1 a_2\dots a_k} = \vec(\tB)^T \vec(\tD).
\ee
We can write a matrix-tensor contraction e.g. as
$\tB(\mM^1,\mM^2,\dots, \mM^k)$, so
\be
\tB(\mM^1,\mM^2,\dots, \mM^k)]_{a_1 a_2\dots a_k} = M^1_{a_1 b_1}
M^2_{a_2 b_2} \dots M^k_{a_k b_k} B_{b_1 b_2\dots b_k}
\ee
and
\be
\vec(\tB(\mM^1,\mM^2,\dots, \mM^k)) = (\mM^k \otimes \dots\otimes\mM^2\otimes \mM^1) \vec(\tB).
\ee

For a fully-symmetric $n$-dimensional symmetric rank-$k$ tensor there are only $S_{nk}\equiv \frac{(n+k-1)!}{k!(n-1)!}$ distinct elements: for $k=3$ there are $n(n+1)(n+2)/6$.
A tensor which has $s$ of $k$ symmetric indices has $S_{nks}\equiv \frac{(n+s-1)!}{s!(n-1)!}n^{k-s}$ distinct elements;
we define
$\vecp(\tB)$ to be the corresponding vector of distinct components of $\tB$. In general $\vecp(\tB)$ is the vector of
distinct components of the symmetrized tensor, for example
if $\tB$ is symmetric in its last $s$ indices,
\be
\vecp(\tB) = \begm B_{11 1\dots 11}, & B_{1\dots 1(1\dots 12)},& B_{1\dots 1(1\dots 22)}, &\dots, & B_{1\dots 1 2\dots 22}, & \dots & B_{1\dots 1(1 \dots n n)}, & \dots, & B_{n\dots n n\dots nn} \enm^T.
\ee
In terms of components, for a fully-symmetric tensor $[\vecp(\tB)]_i
= B_{(a_1 a_2\dots a_k)}$  where $a_{j}\le a_{j+1}$ and
\be
i =a_1 + \frac{a_2(a_2-1)}{2} + \frac{(a_3-1)a_3(a_3+1)}{6} +\dots + \frac{(a_k+k-2)!}{k!(a_k-2)!}.
\ee
It is sometimes useful to write the vector index in the form
$[\vecp(\tB)]_{a_1 a_2 \dots a_k}$.
In general a lumped index can be used with additional non-symmetric indices.

The rank $n^k\times S_{nks}$ matrix $\mBp$ is defined so that for a tensor $\tA$
\be
\vecp(\tA) = \mBp^T \vec(\tA).
\ee
For example, for $k=s$ a $2\times 2\times 2$ tensor $\tA$ has
\be
\mBp^T\vec(\tA) = \begm A_{111}, & (A_{211}+A_{121}+A_{112})/3, & (A_{221}+A_{212}+A_{122})/3,  & A_{222}\enm^T.
\ee
Explicitly
\be
[\mBp]_{a_1 a_2\dots a_s A_{s+1}\dots A_k , b_1 b_2 \dots b_s B_{s+1}\dots B_k} = \delta_{a_1}^{(b_1} \delta_{a_2}^{b_2}\dots \delta_{a_s}^{b_s)} \delta_{A_{s+1}}^{B_{s+1}}\dots \delta_{A_k}^{B_k} \qquad [\text{for }b_i\le b_{i+1}]
\ee
(we are putting the non-symmetric indices last for convenience).
The pseudo-inverse $\mBp^+ \equiv (\mBp^T \mBp)^{-1}\mBp^T$ can be used to construct $\vec(\tD)$ from $\vecp(\tD)$ when $\tD$ is symmetric:
\be
\vec(\tD) = (\mBp^+)^T \vecp(\tD),
\ee
and in general gives the vectorization of the symmetrized tensor.
The symmetric matrix $\mBp\mBp^+$ is the $n^k\times n^k$ matrix for symmetrizing vectorized tensors, with components
\be
(\mBp\mBp^+)_{a_1 a_2\dots a_s A_{s+1}\dots A_k , b_1 b_2\dots b_s B_{s+1}\dots B_k} = \delta_{(a_1}^{(b_1}  \delta_{a_2}^{b_2}\dots  \delta_{a_s)}^{b_s)} \delta_{A_{s+1}}^{B_{s+1}}\dots\delta_{A_{k}}^{B_{k}}.
\ee
Since
\be
(\mM^k \otimes \dots \otimes \mM^2 \otimes \mM^1) _{a_1 a_2\dots a_k,
  b_1 b_2\dots b_k} =  M^1_{a_1 b_1} M^2_{a_2 b_2}\dots M^k_{a_k b_k} ,
\ee
we have
\be
\mBp\mBp^+ (\mM^{k} \otimes \dots \mM^{s+1} \otimes \mM \otimes \dots \otimes \mM) =   (\mM^{k} \otimes \dots \mM^{s+1} \otimes \mM \otimes \dots \otimes \mM)  \mBp\mBp^+.
\ee
Since $(\mA\otimes\mB)^{-1}=\mA^{-1}\otimes \mB^{-1}$ (for non-singular matrices) it follows that
\ba
[\mBp^T (\mM_{k} \otimes \dots\otimes \mM_{s+1}\otimes \mM \otimes \dots \otimes \mM) \mBp ]^{-1}  =\mBp^+ (\mM_{k}^{-1} \otimes \dots \otimes \mM_{s+1}^{-1} \otimes \mM^{-1} \otimes \dots \otimes \mM^{-1}) (\mBp^+)^T.
\ea
Hence for symmetric $\tD$ and $\tB$ symmetric on $s$ indices we have
\begin{multline}
\vec(\tD)^T \vec( \tB(\mC_{l}^{-1},\mC_l^{-1},\dots, \mC_l^{-1},\mC_{l_{s+1}},\dots,\mC_{l_k}^{-1})) \\=  \vecp(\tD)^T [\mBp^T (\mC_{l_k} \otimes \dots \otimes \mC_{l_{s+1}} \otimes \mC\otimes\dots\otimes \mC) \mBp ]^{-1} \vecp(\tB).
\label{gensymmresult}
\end{multline}


\subsection*{Bispectra}
A bispectrum estimator using a vector of different fields $\va_{lm}$ for a specific set of $\{l\}$ is
\be
\hat{B}_{l_1 l_2 l_3}^{ijk} = \sum_{m_1 m_2 m_3} \threej{l_1}{l_2}{l_3}{m_1}{m_2}{m_3} a^i_{l_1 m_1} a^j_{l_2 m_2} a^k_{l_3 m_3}
\ee
or equivalently
\be
\vec(\hat{\tB}_{l_1 l_2 l_3}) =\sum_{m_1 m_2 m_3} \threej{l_1}{l_2}{l_3}{m_1}{m_2}{m_3} \va_{l_3 m_3} \otimes \va_{l_2 m_2} \otimes \va_{l_1 m_1}.
\ee
If we restrict $l_1\le l_2 \le l_3$ then $\vec(\hat{\tB}_{l_1 l_2 l_3})$ is uncorrelated to other vectors with different $\{l\}$-labels.

Using $(\mA\otimes\mB)^T = \mA^T\otimes \mB^T$  and the general result (for appropriately sized matrices) that $(\mA\otimes\mB)(\mC\otimes\mD) = (\mA\mC)\otimes(\mB\mD)$ gives
\be
(\va_{l_3 m_3} \otimes \va_{l_2 m_2} \otimes \va_{l_1 m_1})(\va_{l_3 m_3'} \otimes \va_{l_2 m_2'} \otimes \va_{l_1 m_1'})^\dag
=(\va_{l_3 m_3} \va_{l_3 m_3'}^\dag)\otimes (\va_{l_2 m_2} \va_{l_2 m_2'}^\dag) \otimes (\va_{l_1 m_1} \va_{l_1 m_1'}^\dag).
\ee
Hence using $\la \va_{l m} \va_{l' m'}^\dag\ra =
\delta_{ll'}\delta_{mm'}\mC_{l}$, where $\mC_l$ here is the total
covariance matrix including the effects of instrument noise,
gives
\ba
\la \vec(\hat{\tB}_{l_1 l_2 l_3}) \vec(\hat{\tB}_{l_1 l_2 l_3})^\dag \ra
=
\Delta_{l_1 l_2 l_3} \mC_{l_3}
\otimes \mC_{l_2} \otimes \mC_{l_1} ,
\ea
where $\Delta_{l_1 l_2 l_3} = 6\delta_{(l_1}^{l_1}\delta_{l_2}^{l_2}\delta_{l_3)}^{l_3}$ (no implicit sums over ${l}$-labels: $\Delta_{l_1 l_2 l_3}$ is 6 if $l_1=l_2=l_3$, 2 if two of the indices are equal, and 1 otherwise). Note that this matrix is not invertible if any of the $l$ indices are the same since the tensor is then (partially) symmetric, $\hat{B}_{ll l'}^{ijk}=\hat{B}_{lll'}^{jik}$, so $\vec(\hat{\tB}_{l_1 l_2 l_3})$ is a redundant set (some items are perfectly correlated, so the covariance is singular). For duplicate indices we therefore use $\vecp(\hat{\tB}_{l_1 l_2 l_3})$ instead, so get a vector of only the distinct components, and define
\ba
\mCov_{l_1 l_2 l_3} \equiv \la \vecp(\hat{\tB}_{l_1 l_2 l_3}) \vecp(\hat{\tB}_{l_1 l_2 l_3})^\dag \ra
=
\Delta_{l_1 l_2 l_3} \mBp^T \mC_{l_3}
\otimes \mC_{l_2} \otimes \mC_{l_1}\mBp,
\ea
and it is understood that $\vecp=\vec$ (and $\mBp=\mI$) if $l_1,l_2,
l_3$ are all distinct.

Following Ref.~\cite{Babich:2004yc} we can find a set of weights to get the estimator $\hat{S}=\sum_{l_1\le l_2 \le l_3} \vW_{l_1 l_2 l_3}^\dag \vecp(\hat{\tB}_{l_1 l_2 l_3})$, which
has Gaussian variance
\be
\la \hat{S}^2\ra = \sum_{l_1\le l_2 \le l_3}  \vW_{l_1 l_2 l_3}^\dag \mCov_{l_1 l_2 l_3} \vW_{l_1 l_2 l_3}.
\ee
Minimizing subject to unit response to $\tB$ gives $\vW_{l_1 l_2 l_3}=  \la \hat{S}^2\ra \mCov_{l_1 l_2 l_3}^{-1} \vecp(\tB_{l_1 l_2 l_3})$.
Using Eq.~\eqref{gensymmresult} the Fisher inverse error variance is then
\ba
\sum_{l_1\le l_2 \le l_3} \vecp({\tB}_{l_1 l_2 l_3})^\dag \mCov^{-1}_{l_1 l_2 l_3} \vecp({\tB}_{l_1 l_2 l_3})
&=&
\sum_{l_1\le l_2 \le l_3} \Delta_{l_1 l_2 l_3}^{-1}\vecp({\tB}_{l_1 l_2 l_3})^\dag  [\mBp^T \mC_{l_3}
\otimes \mC_{l_2} \otimes \mC_{l_1}\mBp]^{-1} \vecp({\tB}_{l_1 l_2 l_3})
\nonumber
\\
&=& \sum_{l_1\le l_2 \le l_3} \Delta_{l_1 l_2 l_3}^{-1} \vec({\tB}_{l_1 l_2 l_3})^\dag  \mC_{l_3}^{-1}
\otimes \mC_{l_2}^{-1} \otimes \mC_{l_1}^{-1} \vec({\tB}_{l_1 l_2 l_3})
\nonumber
\\
&=& \frac{1}{6}\sum_{l_1 l_2 l_3}\vec({\tB}_{l_1 l_2 l_3})^\dag  \mC_{l_3}^{-1}
\otimes \mC_{l_2}^{-1} \otimes \mC_{l_1}^{-1} \vec({\tB}_{l_1 l_2 l_3}) \nonumber\\
&=&\frac{1}{6}\sum_{l_1 l_2 l_3}(B^{ijk}_{l_1 l_2 l_3})^*
(C^{-1})^{ip}_{l_1}(C^{-1})^{jq}_{l_2}(C^{-1})^{kr}_{l_3} B^{pqr}_{l_1
  l_2 l_3} .
\ea
This establishes the correspondence between the (zero-signal) variance
in terms of the covariance of the elements, the result of
Ref.~\cite{Yadav:2007rk} and that obtained from an Edgeworth-expansion
of the non-Gaussian likelihood (note the appendix [Sec. 6] of Ref.~\cite{Yadav:2007rk} is somewhat misleading: the vector of components of bispectra using $T$ and $E$-polarization have to include all eight possible terms with distinct $l_1, l_2, l_3$).

\bibliography{../antony,../cosmomc}

\end{document}